{}

\documentclass[10pt,a4paper]{article}

\newif\ifpublic\publictrue

\ifpublic\else\usepackage{showkeys}\fi
\usepackage[bookmarks=true,hyperfigures=true,hidelinks,bookmarksnumbered,unicode]{hyperref}
\usepackage[nosort]{cite}
\usepackage[parsep]{collref}
\usepackage[english]{babel}
\usepackage[T1]{fontenc}
\usepackage[utf8]{inputenc}
\usepackage{lmodern}
\usepackage{amsfonts,amsbsy,dsfont,euscript,mathrsfs,fixmath}
\usepackage{amsmath,amssymb}
\usepackage{mathtools}
\usepackage{scalerel}
\usepackage{enumerate}
\usepackage{graphicx}
\usepackage{tabularx} 
\usepackage{longtable} 
\usepackage{multirow}
\usepackage{tikz,pgfplots}
\usetikzlibrary{arrows,calc,chains,decorations.pathmorphing,decorations.pathreplacing,fadings,matrix,positioning,shapes,shapes.geometric,shapes.symbols,trees}
\pgfplotsset{compat=1.9}

\def\showkeysrefformat#1{{\normalfont\tiny\ttfamily#1}}
\makeatletter
\def\SK@@ref#1>#2\SK@{{\@inlabelfalse\leavevmode\vbox to\z@{\vss\SK@refcolor\rlap{\vrule\raise .75em \hbox{\showkeysrefformat{#2}}}}}}
\makeatother

\usepackage[a4paper,text={160mm,247mm},centering]{geometry}
\allowdisplaybreaks[3]
\numberwithin{equation}{section}
\def\[{\begin{equation}\begin{aligned}}
\def\]{\end{aligned}\end{equation}}

\usepackage[font=small,labelfont=bf,width=0.85\textwidth]{caption}

\expandafter\def\expandafter\bfseries\expandafter{\bfseries\ifmmode\else\boldmath\fi}
\expandafter\def\expandafter\mdseries\expandafter{\mdseries\ifmmode\else\unboldmath\fi}
\expandafter\def\expandafter\normalfont\expandafter{\normalfont\ifmmode\else\unboldmath\fi}

\RequirePackage{verbatim}

\makeatletter
\newwrite\bibinl@out
\newenvironment{bibtex}[1][\jobname]{%
\immediate\openout\bibinl@out #1.bib%
\immediate\write\bibinl@out{\@percentchar generated from `\jobname' starting line \the\inputlineno^^J}%
\def\verbatim@processline{\immediate\write\bibinl@out{\the\verbatim@line}}%
\@bsphack\let\do\@makeother\dospecials\catcode`\^^M\active\verbatim@start%
}
{\immediate\closeout\bibinl@out\@esphack}
\makeatother

\let\barefrac=\frac
\renewcommand{\frac}[2]{\mathinner{\barefrac{#1}{#2}}}

\let\baresqrt=\sqrt
\makeatletter
\renewcommand{\sqrt}{\@ifnextchar[\@sqrt@space@a\@sqrt@space@b}
\def\@sqrt@space@a[#1]#2{\mathinner{\mathchoice{\mkern-3mu}{\mkern-3mu}{}{}\baresqrt[#1]{#2}}}
\def\@sqrt@space@b#1{\mathinner{\mathchoice{\mkern-3mu}{\mkern-3mu}{}{}\baresqrt{#1}}}
\makeatother

\makeatletter
\let\per@dot@old=\.
\def\.{\ifmmode\def\per@dot@sel{\mkern3mu}\else\def\per@dot@sel{\per@dot@old}\fi\per@dot@sel}
\makeatother

\let\barefootnote=\footnote
\renewcommand{\footnote}[1]{\barefootnote{#1\vspace{3pt}}}

\ExplSyntaxOn
\NewDocumentEnvironment{sequation}{O{\small}b}
{
\yufip_sequation:nnn {equation}{#1}{#2}
}{}
\NewDocumentEnvironment{sequation*}{O{\small}b}
{
\yufip_sequation:nnn {equation*}{#1}{#2}
}{}
\cs_new_protected:Nn \yufip_sequation:nnn
{
\begin{#1}
\mbox{#2$\displaystyle#3$}
\end{#1}
}
\ExplSyntaxOff

\newcommand{\vfrac}[2]{\ifmmode\mathinner{\textstyle^{#1}\!/\!_{#2}}\else$^{#1}\!/\!_{#2}$\fi}

\DeclareMathOperator{\diag}{diag}
\DeclareMathOperator{\tr}{tr}
\DeclareMathOperator{\Tr}{Tr}

\newcommand{\Real}{\mathds{R}}
\newcommand{\Complex}{\mathds{C}}
\newcommand{\Integer}{\mathds{Z}}

\newcommand{\Unit}{\mathds{U}}

\let\Re\relax\DeclareMathOperator{\Re}{Re}

\DeclareMathOperator{\csch}{csch}

\newcommand{\ind}[1]{{\scriptscriptstyle{#1}}}

\makeatletter
\newcommand*\bigcdot{\mathpalette\bigcdot@{.5}}
\newcommand*\bigcdot@[2]{\mathbin{\vcenter{\hbox{\scalebox{#2}{$\m@th#1\bullet$}}}}}
\makeatother

\newcommand{\alg}[1]{\mathfrak{#1}}
\newcommand{\grp}[1]{\mathrm{#1}}
\DeclareMathOperator{\rank}{rank}
\DeclareMathOperator{\Lie}{Lie}
\DeclareMathOperator{\ad}{ad}
\DeclareMathOperator{\Ad}{Ad}

\newcommand{\Lax}{\mathcal{L}}

\def\<{\big\langle}
\def\>{\big\rangle}

\newcommand{\geom}[1]{\mathrm{#1}}

\newcommand{\AdS}{\geom{AdS}}

\newcommand{\CP}{\Complex\mathbf{P}}

\newcommand{\Act}{\mathcal{S}}

\DeclareFontEncoding{LS1}{}{}
\DeclareFontSubstitution{LS1}{stix}{m}{n}
\DeclareSymbolFont{stixsymbols}{LS1}{stixscr}{m}{n}
\SetSymbolFont{stixsymbols}{bold}{LS1}{stixscr}{b}{n}
\DeclareMathSymbol{\kay}{\mathalpha}{stixsymbols}{"6B}
\DeclareMathSymbol{\hay}{\mathalpha}{stixsymbols}{"68}
\DeclareMathAlphabet{\mathdsl}{U}{bbm}{m}{sl}
\newcommand{\gdsl}{\mathdsl{g}}
\newcommand{\udsl}{\mathdsl{u}}

\newcommand{\interval}[1]{(#1]}

\usepackage{pifont}
\providecommand{\href}[2]{#2}

\makeatletter
\def\mr@ignsp#1 {\ifx\:#1\@empty\else #1\expandafter\mr@ignsp\fi}
\newcommand{\multiref}[1]{\begingroup%
\xdef\mr@no@sparg{\expandafter\mr@ignsp#1 \: }%
\def\mr@comma{}\def\mr@dash{-}%
\@for\mr@refs:=\mr@no@sparg\do{%
\ifx\mr@refs\mr@dash\def\mr@comma{}--\else%
\mr@comma\def\mr@comma{,}\ref{\mr@refs}\fi}%
\endgroup}
\renewcommand{\eqref}[1]{(\multiref{#1})}
\makeatother

\makeatletter
\newcommand{\namedref}[2]{\hyperref[#2]{#1~\ref*{#2}}}
\newcommand{\namedrefs}[3]{#1~\hyperref[#2]{\ref*{#2}}~and~\hyperref[#3]{\ref*{#3}}}
\newcommand{\secref}{\@ifstar{\namedref{Section}}{\namedref{sec.}}}
\newcommand{\secsref}{\@ifstar{\namedrefs{Sections}}{\namedrefs{secs.}}}
\newcommand{\appref}{\@ifstar{\namedref{Appendix}}{\namedref{app.}}}
\newcommand{\tabref}{\@ifstar{\namedref{Table}}{\namedref{tab.}}}
\newcommand{\figref}{\@ifstar{\namedref{Figure}}{\namedref{fig.}}}
\newcommand{\foottref}{\@ifstar{\namedref{Footnote}}{\namedref{foot.}}}
\makeatother

\let\oldbib=\thebibliography
\def\thebibliography{\phantomsection\addcontentsline{toc}{section}{\refname}\oldbib}

\let\oldtoc=\tableofcontents
\def\tableofcontents{\phantomsection\addcontentsline{toc}{section}{\contentsname}\oldtoc}

\providecommand{\hypersetup}[1]{}
\providecommand{\texorpdfstring}[2]{#1}
\hypersetup{plainpages=false}
\hypersetup{pdfpagemode=UseNone}
\hypersetup{bookmarksnumbered=true}
\hypersetup{pdfstartview=FitH}
\hypersetup{colorlinks=false}
\hypersetup{citebordercolor={1 1 1}}
\hypersetup{urlbordercolor={1 1 1}}
\hypersetup{linkbordercolor={1 1 1}}

\makeatletter
\let\@keywords\@empty
\let\@subject\@empty
\providecommand{\keywords}[1]{\gdef\@keywords{#1}}
\providecommand{\subject}[1]{\gdef\@subject{#1}}
\def\thetitle{\@title}
\def\theauthor{\@author}
\def\thesubject{\@subject}
\def\thedate{\@date}
\def\thekeywords{\@keywords}
\makeatother
\AtBeginDocument{
\hypersetup{pdftitle={\thetitle}}
\hypersetup{pdfauthor={\theauthor}}
\hypersetup{pdfsubject={\thesubject}}
\hypersetup{pdfkeywords={\thekeywords}}}

\newif\ifshownote
\shownotetrue

\ifpublic\shownotefalse\fi

\ifshownote

\ifpdf\else\RequirePackage[active]{srcltx}\fi
\RequirePackage{xcolor}
\RequirePackage{ulem}

\newcommand{\remark}[2][]{{\normalfont\sffamily\hspace{1ex}
\def\emph{\textsl}\def\textbullet{$\bullet$}
\def\tmparga{#1}
\def\tmpargb{BH}\ifx\tmparga\tmpargb\color[rgb]{0.7,0,0}\fi
\def\tmpargb{RH}\ifx\tmparga\tmpargb\color[rgb]{0,0.7,0}\fi
\def\tmpargb{}\ifx\tmparga\tmpargb\normalfont\color{red}\fi
\def\tmpargb{}\ifx\tmparga\tmpargb\else \textbf{#1:}\fi
#2\hspace{1ex}}}

\else

\newcommand{\remark}[2][]{\ignorespaces}

\fi

\title{Twists of trigonometric sigma models}
\author{Rashad Hamidi and Ben Hoare}

\begin{document}

\pdfbookmark[1]{Title Page}{title}
\thispagestyle{empty}

\vspace*{2cm}
\begin{center}
\begingroup\Large\bfseries\thetitle\par\endgroup
\vspace{1cm}

\renewcommand{\thefootnote}{\roman{footnote}}
\begingroup\theauthor\par\endgroup
\vspace{1cm}

\textit{Department of Mathematical Sciences \\ Durham University \\ Durham DH1 3LE, UK}

\vspace{0.5cm}

\texttt{\{rashad.m.hamidi,ben.hoare\}@durham.ac.uk}

\vspace{5mm}

\vfill

\textbf{Abstract}\vspace{5mm}

\small
\begin{minipage}{12.5cm}
We introduce the $\Integer_N$-twisted trigonometric sigma models, a new class of integrable deformations of the principal chiral model.
Starting from 4d Chern-Simons theory on a cylinder, the models are constructed by introducing a $\Integer_N$ branch cut running along the non-compact direction.
As we pass through the branch cut we apply a $\Integer_N$ automorphism to the algebra-valued and group-valued fields of the theory.
Two instances of models in this class have appeared in the literature and we explain how these fit into our general construction.
We generalise both to the case of $\Integer_N$-twistings, and for each we construct two further deformations, leading to four doubly-deformed models.

\medskip

Mapping the cylinder to a sphere, a novel feature of the construction is that the branch points are at the simple poles corresponding to the ends of the cylinder.
As a result, the untwisted and twisted models have the same number of degrees of freedom.
While twisting does not always lead to inequivalent models, we show that if we use an outer automorphism to twist then the untwisted and twisted models have different symmetries, hence the latter are new integrable sigma models.
\end{minipage}
\normalsize

\vspace*{4cm}

\end{center}

\setcounter{footnote}{0}
\renewcommand{\thefootnote}{\arabic{footnote}}

\newpage

\tableofcontents

\section{Introduction}\label{sec:introduction}

The introduction of 4d Chern-Simons theory \cite{Costello:2013zra,Costello:2013sla} and its extensive development \cite{Costello:2017dso,Costello:2018gyb,Costello:2019tri} has provided new pathways to construct, classify and quantise integrable sigma models in two dimensions.
Many new 2d integrable sigma models have been introduced in recent years, see~\cite{Hoare:2021dix} for a review, and 4d Chern-Simons has vastly improved our understanding of this landscape, see~\cite{Lacroix:2021iit} for a review.
4d Chern-Simons theory is defined on a product of the 2d space-time and a Riemann surface.
Depending on whether the Riemann surface is a sphere, cylinder or torus, the resulting 2d integrable sigma models can be understood as rational, trigonometric or elliptic respectively.
There are close relationships between these surfaces, e.g.,~the cylinder can be mapped to a sphere with two simple poles, hence the same integrable sigma model can have different descriptions.
See \cite{Kawaguchi:2011pf,Kawaguchi:2012ve} for early discussions of the different descriptions of the trigonometric deformation of the $\grp{SU}(2)$ principal chiral model~\cite{Cherednik:1981df}.
In this paper we investigate trigonometric models, but we often exploit the closely related rational description.

To construct 2d integrable sigma models from 4d Chern-Simons we introduce defects.
There are different types of defect and our focus is on the so-called disorder defects.
To write down the action of 4d Chern-Simons, the Chern-Simons 3-form needs to be paired with a 1-form $\omega$, which is chosen to be a meromorphic 1-form on the Riemann surface.
Disorder defects are the zeroes of this meromorphic 1-form, and have been extensively studied in the case that the Riemann surface is a sphere, leading to the construction of new 2d integrable sigma models and a better understanding of the relations between them~\cite{Delduc:2019whp,Bassi:2019aaf,Benini:2020skc,Lacroix:2020flf}.
Moreover, their integrability structure is well-understood through their relation to affine Gaudin models~\cite{Vicedo:2017cge,Vicedo:2019dej,Lacroix:2020flf}, their Hamiltonian integrability and their conserved charges~\cite{Lacroix:2017isl,Lacroix:2018njs,Lacroix:2023gig}.
In general, when the Riemann surface is a cylinder, disorder defects are less well-studied.
However, this is primarily because, as mentioned above, the cylinder can be mapped to a sphere with two simple poles.

The zeroes in $\omega$ imply that the Chern-Simons gauge field $A$ is allowed to have poles at these locations, while we interpret the poles in $\omega$ as boundaries where the behaviour of $A$ should be specified.
Once the meromorphic 1-form is given, along with regularity conditions for $A$ at its zeroes and boundary conditions at its poles, there is a well-established procedure for constructing 2d integrable sigma models from 4d Chern-Simons with disorder defects, which we summarise in \secref{sec:general-theory}.

\medskip

In this paper we introduce a new class of 2d integrable sigma models constructed from 4d Chern-Simons, which we call $\Integer_N$-twisted trigonometric sigma models.
These are most naturally understood as coming from 4d Chern-Simons on the group $\grp{G}^\Complex$ when the Riemann surface is a cylinder.
The novel feature is the introduction of a branch cut running along the non-compact direction of the cylinder.
For a $\Integer_N$ branch cut we return to the original sheet after going through the cut $N$ times.
Twisting then corresponds to relating algebra-valued and group-valued fields on different sheets by acting with a $\Integer_N$ automorphism $\sigma$ of the Lie algebra $\alg{g}^\Complex$, or its lift to the Lie group $\grp{G}^\Complex$, as we pass through the branch cut.
This procedure can be understood as a consistent truncation, guaranteeing the integrability of the resulting 2d integrable sigma model.

\begin{figure}[t]
\begin{center}
\begin{tikzpicture}[scale=1]
\draw[dashed,very thick,gray!25] (-3,0) ellipse (1 and 0.5);
\draw[dashed,very thick,gray!25] (3,0) ellipse (1 and 0.5);
\draw[style={decorate, decoration=snake},very thick,gray] (-3,-2.5)--(-3,1.5);
\draw[-,very thick] (-4,-2)--(-4,2);
\draw[-,very thick] (-2,-2)--(-2,2);
\draw[very thick] (-3,2) ellipse (1 and 0.5);
\draw[very thick] (-3,-2) ellipse (1 and 0.5);
\draw[-,very thick] (2,-2)--(2,2);
\draw[-,very thick] (4,-2)--(4,2);
\draw[very thick] (3,2) ellipse (1 and 0.5);
\draw[very thick] (3,-2) ellipse (1 and 0.5);
\draw[-,very thick] (-2.15,0.1)--(-1.95,-0.1);
\draw[-,very thick] (-2.15,-0.1)--(-1.95,0.1);
\draw[-,very thick] (-2.05,0.1)--(-1.85,-0.1);
\draw[-,very thick] (-2.05,-0.1)--(-1.85,0.1);
\draw[very thick] (-2.4,0.4) circle (0.1);
\draw[very thick] (-2.4,-0.4) circle (0.1);
\draw[-,very thick] (2.15,0.1)--(1.95,-0.1);
\draw[-,very thick] (2.15,-0.1)--(1.95,0.1);
\draw[-,very thick] (2.05,0.1)--(1.85,-0.1);
\draw[-,very thick] (2.05,-0.1)--(1.85,0.1);
\draw[very thick] (2.4,0.4) circle (0.1);
\draw[very thick] (2.4,-0.4) circle (0.1);
\draw[-,very thick] (4.15,0.1)--(3.95,-0.1);
\draw[-,very thick] (4.15,-0.1)--(3.95,0.1);
\draw[-,very thick] (4.05,0.1)--(3.85,-0.1);
\draw[-,very thick] (4.05,-0.1)--(3.85,0.1);
\draw[very thick] (3.6,0.4) circle (0.1);
\draw[very thick] (3.6,-0.4) circle (0.1);
\draw[dashed,very thick,gray!25] (-3,-6) circle (1);
\draw[dashed,very thick,gray!25] (3,-6) circle (1);
\draw[style={decorate, decoration=snake},very thick,gray] (-3,-6)--(-1,-4);
\draw[-,very thick] (-1,-6)--(-5,-6);
\draw[-,very thick] (-3,-8)--(-3,-4);
\draw[-,very thick] (-2.9,-6.1)--(-3.1,-5.9);
\draw[-,very thick] (-2.9,-5.9)--(-3.1,-6.1);
\draw[-,very thick] (-0.9,-3.9)--(-1.1,-4.1);
\draw[-,very thick] (-0.9,-4.1)--(-1.1,-3.9);
\draw[-,very thick] (-2.15,-5.9)--(-1.95,-6.1);
\draw[-,very thick] (-2.15,-6.1)--(-1.95,-5.9);
\draw[-,very thick] (-2.05,-5.9)--(-1.85,-6.1);
\draw[-,very thick] (-2.05,-6.1)--(-1.85,-5.9);
\draw[very thick] (-2.09,-5.6) circle (0.1);
\draw[very thick] (-2.09,-6.4) circle (0.1);
\draw[-,very thick] (1,-6)--(5,-6);
\draw[-,very thick] (3,-8)--(3,-4);
\draw[-,very thick] (2.9,-6.1)--(3.1,-5.9);
\draw[-,very thick] (2.9,-5.9)--(3.1,-6.1);
\draw[-,very thick] (4.9,-3.9)--(5.1,-4.1);
\draw[-,very thick] (4.9,-4.1)--(5.1,-3.9);
\draw[-,very thick] (2.15,-5.9)--(1.95,-6.1);
\draw[-,very thick] (2.15,-6.1)--(1.95,-5.9);
\draw[-,very thick] (2.05,-5.9)--(1.85,-6.1);
\draw[-,very thick] (2.05,-6.1)--(1.85,-5.9);
\draw[very thick] (2.09,-5.6) circle (0.1);
\draw[very thick] (2.09,-6.4) circle (0.1);
\draw[-,very thick] (4.15,-5.9)--(3.95,-6.1);
\draw[-,very thick] (4.15,-6.1)--(3.95,-5.9);
\draw[-,very thick] (4.05,-5.9)--(3.85,-6.1);
\draw[-,very thick] (4.05,-6.1)--(3.85,-5.9);
\draw[very thick] (3.91,-5.6) circle (0.1);
\draw[very thick] (3.91,-6.4) circle (0.1);
\end{tikzpicture}
\end{center}
\caption{Introducing a $\Integer_2$ branch cut along the non-compact direction of the cylinder and taking the $N$-fold cover leads to a $\Integer_N$-equivariant distribution of poles and zeroes on the cylinder as depicted here in the top two diagrams for $N=2$.
The bottom two diagrams show the same setup, but after mapping the cylinder to a sphere, drawn as $\Complex \cup \{\infty\}$, with two simple poles.}\label{fig0}
\end{figure}
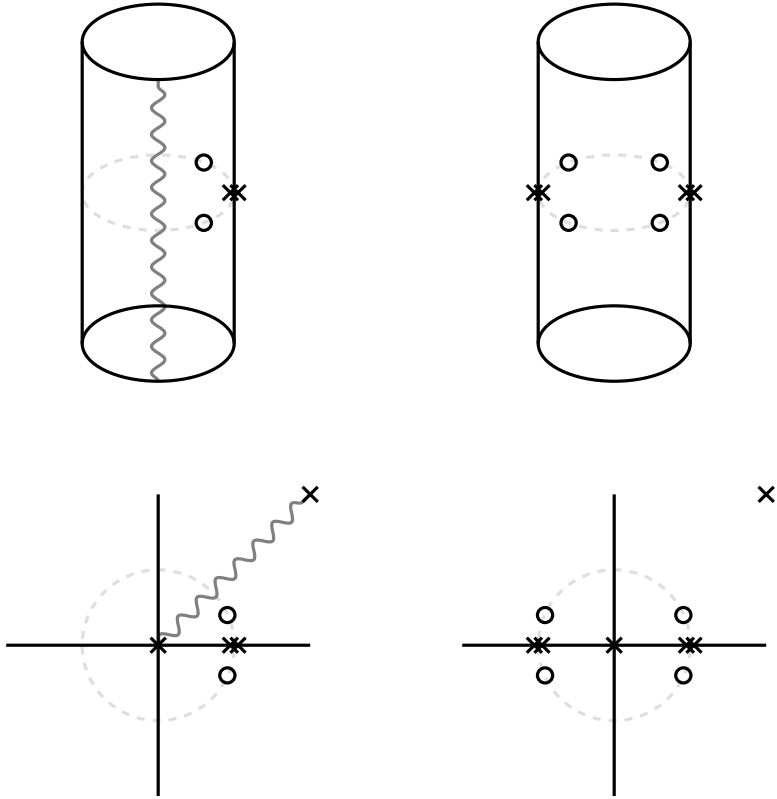

Practically, we work with $N$-fold cover of the cylinder, which has a $\Integer_N$-equivariant distribution of poles and zeroes, as depicted in \figref{fig0}.
This means that there is a $\Integer_N$ action on the cylinder leaving the sets of poles and zeroes invariant.
It is also be useful to map the cylinder to a sphere with two simple poles.
In this picture, again depicted in \figref{fig0}, the branch cut runs between the two simple poles and, similarly, the $N$-fold cover is a sphere with two simple poles that are fixed points of a $\Integer_N$ action.
Again the distribution of poles and zeroes is $\Integer_N$-equivariant.

The models we find before twisting, i.e.~before we introduce the branch cuts, are the same as those found by taking the $\Integer_N$ automorphism to be the identity.
Therefore, we refer to these as untwisted models.
Such models are well-known in the literature and include the Yang-Baxter deformation of the principal chiral model~\cite{Klimcik:2002zj}, also known as the $\eta$-model, the current-current deformation of the WZW model~\cite{Sfetsos:2013wia}, also known as the $\lambda$-model, and their deformations~\cite{Klimcik:2008eq,Sfetsos:2014cea,Sfetsos:2015nya,Sfetsos:2017sep}.
Their 4d Chern-Simons constructions are given in~\cite{Delduc:2019whp}.
Other than the untwisted models, two instances of $\Integer_N$-twisted trigonometric sigma models have appeared in the literature.
The construction in~\cite{Fukushima:2020kta} takes a specific $\Integer_2$-equivariant configuration of poles and zeroes in 4d Chern-Simons and consistently truncates the resulting 2d sigma model.
An interpretation in terms of branch cuts and covering spaces is briefly proposed in \cite{Tian:2020ryu}.
On the other hand, the construction in \cite{Borsato:2024alk} is formulated directly in 2d using the language of spectators.
In both cases the $\Integer_2$-twisting is implemented in 2d and we explain in detail how these models fit into our general construction from 4d Chern-Simons in \secref{sec:general-theory}.
We then generalise both of these to the case of $\Integer_N$-twistings, and for each we consider two further deformations, leading us to four doubly-deformed models whose actions we explicitly construct in the $\Integer_2$ case.
To summarise, the models we explicitly construct are:
\begin{itemize}
\item $\Integer_N$-twisted $\eta$-model,
\item $\Integer_N$-twisted $\lambda$-model,
\item YB-deformed $\Integer_2$-twisted $\eta$-model,
\item CC-deformed $\Integer_2$-twisted $\eta$-model,
\item YB-deformed $\Integer_2$-twisted $\lambda$-model,
\item CC-deformed $\Integer_2$-twisted $\lambda$-model,
\end{itemize}
where we explain the nomenclature in detail in \secref{sec:general-theory}.

In contrast to other discussions of branch cuts in 4d Chern-Simons~\cite{Costello:2019tri,Tian:2020ryu,Fukushima:2020dcp,Tian:2020pub,Costello:2020lpi,Berkovits:2024reg,Cole:2024skp} where the cuts run between two marked points that are not poles or zeroes leading to coset models, the $\Integer_N$-twisted trigonometric sigma models are particularly interesting since they have the same number of degrees of freedom as their untwisted counterparts.
It is instructive to briefly consider the simplest case, the $\Integer_N$-twisted $\eta$-model, which corresponds to the setup shown in \figref{fig0}.
For particular boundary conditions this leads to the sigma model
\begin{equation}\label{eq:actionetatwistintro}
\Act_{\eta}^{(N)}(g) = \frac{4\hay\eta}{1+\eta^{2}} \int d^{2}\sigma \, \tr\Big( J_{+}^{(0)}\frac{1+\eta^{2}}{1-\eta\tilde{\mathcal{R}}}J_{-}^{(0)} + \sum_{a\in\Integer_N\backslash\{0\}} \Big(1+i\eta \Big(1-\frac{2a}{N}\Big) \Big) J_{+}^{(a)}J_{-}^{(N-a)} \Big) ~,
\end{equation}
where $J_\pm^{(a)} = P_a(g^{-1}\partial_\pm g)$ and $g \in \grp{G}$, a particular real form of $\grp{G}^\Complex$, $\sigma^\pm$ are light-cone coordinates, $\hay$ and $\eta$ are parameters, $\tilde{\mathcal{R}}$ is an antisymmetric solution to the non-split modified classical Yang-Baxter equation on $\alg{g} = \Lie(\grp{G})$, and $P_a$ are projectors onto the $N$ eigenspaces of $\sigma$.
The untwisted case corresponding to $\sigma = 1$ yields the well-known Yang-Baxter deformation of the principal chiral model~\cite{Klimcik:2002zj}.
Rescaling $\hay$ and taking $\eta \to 0$, we see that the dependence on the $\Integer_N$ automorphism drops out from the action~\eqref{eq:actionetatwistintro} and this model is a deformation of the principal chiral model for any $N$.
\unskip\footnote{In terms of the meromorphic 1-form, this corresponds to mapping the cylinder to a sphere and using an $\grp{SL}(2;\Complex)$ transformation to fix the positions of the simple zeroes and double pole, such that $\eta \to 0$ corresponds to the two simple poles merging to a double pole.}
This suggests that the twisting procedure is intrinsic to the trigonometric setup and that there are many integrable deformations of symmetric sigma models with potentially interesting applications still to construct.

\medskip

The outline of this paper is as follows.
In \secref{sec:general-theory} we review the construction of integrable sigma models from 4d Chern-Simons and give an extended overview of the general theory of the $\Integer_N$-twisted trigonometric sigma models.
In \secref{sec:bcgs} we discuss the different regularity conditions, boundary conditions and gauge fixings that we use to construct the models in \secref{sec:sigmamodels}.
Explicit backgrounds are computed for $\Integer_2$-twisted $\grp{SU}(2)$ models in \secref{sec:examples}.
While we show that these are equivalent to the backgrounds of untwisted models, in \secref{sec:equivalence} we give setups where the untwisted and $\Integer_2$-twisted models are inequivalent.
These are $\grp{SU}(n)$ models where we use the $\Integer_2$ outer automorphism to implement the twisting.
Our conventions are detailed in \appref{app:conv}, and we conclude in \secref{sec:conclusions} with a summary of our results, open directions and a discussion of possible applications of the newly constructed $\Integer_N$-twisted trigonometric sigma models.

\section{General theory of \texorpdfstring{$\Integer_N$}{ZN}-twisted trigonometric sigma models}\label{sec:general-theory}

\subsection{4d Chern-Simons with disorder defects}

Our starting point is the 4d Chern-Simons action on $\CP^1 \times \Sigma$ with disorder defects \cite{Costello:2019tri,Delduc:2019whp,Benini:2020skc,Lacroix:2021iit}
\begin{equation}\label{eq:4dCS}
\Act = \frac{i}{4\pi}\int_{\CP^1 \times \Sigma} \omega \wedge \mathrm{CS}_3(A) ~,
\end{equation}
where $\Sigma$ is the 2d space-time, $\mathrm{CS}_3(A)$ is the Chern-Simons 3-form for the 1-form gauge field $A$ valued in the simple Lie algebra $\alg{g}^\Complex$
\begin{equation}
\mathrm{CS}_3(A) = \tr(A \, dA) + \frac{1}{3} \tr(A \, [A,A] ) ~,
\end{equation}
and $[,]$ and $\tr()$ are the Lie bracket and normalised ad-invariant bilinear form on $\alg{g}^{\Complex}$, extended to take forms as arguments as described in \appref{app:conv1}.
We introduce a complex coordinate $z$ on $\CP^1$ and parametrise the meromorphic 1-form $\omega$ in terms of the twist function $\varphi$ as
\begin{equation}
\omega = \varphi(z) dz ~.
\end{equation}
The zeroes of $\omega$ are known as disorder defects, while we interpret the poles as 2d boundaries.
The dynamical degrees of freedom of the effective 2d integrable sigma model live at these poles.
Importantly, the presence of $\omega$ in the action means that $\iota_z A$ drops out of the action and we assume that $A$ takes the form
\begin{equation}
A = A_\mu d\sigma^\mu + A_{\bar z} d\bar z ~,
\end{equation}
where $\mu = \pm$ and $\sigma^\pm$ are light-cone coordinates on $\Sigma$.

We will not give a detailed account of how to extract 2d integrable sigma models from 4d Chern-Simons here, or review the different methods that are described in the literature.
Instead we summarise the key steps that are needed for our construction.

The presence of poles in $\omega$ means that the 4d Chern-Simons action~\eqref{eq:4dCS} has a non-trivial boundary variation.
To define our theory we need to impose boundary conditions to ensure that these boundary terms vanish.
Varying the action~\eqref{eq:4dCS} we find
\begin{equation}\label{Eq:4dCSvariation}
\Act = \frac{i}{2\pi}\int_{\CP^1\times \Sigma} \omega \wedge \tr(\delta A \, F(A))
-\frac{i}{4\pi}\int_{\CP^1 \times \Sigma} d\omega \wedge \tr(A \, \delta A) ~,
\end{equation}
where
\begin{equation}
F(A) = dA + \frac12 [A,A] ~,
\end{equation}
is the usual field strength for the connection $A$.
The precise details of the allowed boundary conditions leading to an effective 2d integrable sigma model can be subtle and depend on the choice of $\omega$.
To address this question it is useful to introduce the concept of a defect Lie algebra, $\mathfrak{d}$, together with an ad-invariant bilinear form $\langle\!\langle,\rangle\!\rangle$ \cite{Lacroix:2020flf}.
The defect Lie algebra and its bilinear form are built from $d\omega$ and $\tr$, and has dimension $\dim\alg{g}\times\sum_{i=1}^{n} m_i$, where $n$ is the number of distinct poles and $m_i$ are their multiplicity.
From $A$ and its $\CP^1$-derivatives evaluated at the positions of the poles we can define, $\mathds{A} \in \mathfrak{d}$, which is a 1-form on $\Sigma$, such that, up to a factor of $\frac12$, the boundary condition becomes
\begin{equation}\label{eq:boundaryterm}
\frac{1}{2\pi i}\int_{\CP^1 \times \Sigma} d\omega \wedge \tr(A \, \delta A) = \int_\Sigma \langle\!\langle \mathds{A} , \delta \mathds{A} \rangle\!\rangle = 0 ~.
\end{equation}
If $\mathfrak{d}$ is even-dimensional, it follows that a large class of boundary conditions is given by demanding $\mathds{A}$ is valued in a Lagrangian subalgebra of $\mathfrak{d}$.
While these are not the most general allowed boundary conditions~\cite{Costello:2019tri,Ashwinkumar:2023zbu,Cole:2023umd}, they have been extensively studied and are known to lead to 2d sigma models that are Hamiltonian integrable in the sense that the Poisson bracket of the Lax matrix takes the non-ultralocal $r$/$s$-form~\cite{Vicedo:2017cge,Vicedo:2019dej,Lacroix:2020flf} of Maillet~\cite{Maillet:1985ek,Maillet:1985fn}.
In this paper, we restrict to this setup when constructing models.
Moreover, we focus on 1-forms $\omega$ that have pairs of simple poles with equal and opposite residues and double poles with vanishing residues.
The precise from of the defect Lie algebra and the boundary conditions for these cases is discussed in~\secref{sec:bcgs}.

Once we have fixed our boundary conditions, we proceed by defining
\unskip\footnote{We restrict to configurations for which such a gauge choice exists.}
\begin{equation}\label{eq:Laxparam}
A = \hat{g} \Lax \hat{g}^{-1} - d\hat{g} \hat{g}^{-1} = \Lax^{\hat{g}} ~, \qquad \hat{g} \in \grp{G}^{\Complex} ~.
\end{equation}
such that
\begin{equation}
\Lax = \Lax_\mu d\sigma^\mu ~, \qquad \Lax_{\bar z} = 0 ~.
\end{equation}
Up to a boundary term, which we drop, the resulting action is written
\begin{equation}\label{eq:4dCSL}
\Act = \frac{i}{4\pi}\int_{\CP^1 \times \Sigma} \omega \wedge \mathrm{CS}_3(\Lax) +
\frac{i}{4\pi} \int_{\CP^1 \times \Sigma} d\omega \wedge \tr(\Lax \, \hat{g}^{-1} d\hat{g})
+ \frac{i}{4\pi} \int_{\CP^1 \times \Sigma} \omega \wedge \tr(\hat{g}^{-1} d\hat{g} \, [\hat{g}^{-1} d\hat{g} , \hat{g}^{-1} d\hat{g} ]) ~.
\end{equation}
The bulk equations of motion for $\Lax$ are
\begin{equation}
\omega \wedge \tr(\delta \Lax F(\Lax) ) = 0 ~.
\end{equation}
Since $\Lax$ only has two non-vanishing components $\Lax_\mu$, this yields two equations
\begin{equation}
\omega \wedge \partial_{\bar z} \Lax = 0 ~.
\end{equation}
We solve these equations, which imply that $\Lax$ is a meromorphic function of $z$ and is allowed poles at the zeroes of $\omega$.
The bulk equations of motion for $A$, $\omega \wedge \tr(\delta A F(A))$, imply a third equation that comes from the variation of $A_{\bar z}$.
In terms of $\Lax$, this is given by
\begin{equation}
\omega \wedge (d_\Sigma \Lax + \frac12[\Lax,\Lax]) = 0 ~.
\end{equation}
This equation becomes the zero-curvature of the Lax connection, which is then equivalent to the equations of motion of the 2d integrable sigma model.

With a suitable choice of the pole structure of $\Lax$, which we refer to as regularity conditions, it is then possible to use meromorphicity and the boundary conditions to solve for the Lax connection $\Lax$ in terms of the 2d edge modes, which are given by $\hat{g}$ and its $\CP^1$-derivatives evaluated at the poles of $\omega$.
We consider 1-forms $\omega$ that only have simple zeroes and discuss the regularity conditions in more detail in \secref{sec:bcgs}.
Finally, substituting the expression for $\Lax$ back into the full action~\eqref{eq:4dCSL} we find that the bulk terms vanish and we are left with a 2d integrable sigma model for the edge modes with the Lax connection given by $\Lax$.

There are two types of residual gauge symmetries that can survive in the 2d integrable sigma model and it can be simpler to fix these prior to solving for the Lax connection.
These are the external and internal gauge symmetries.
The external gauge symmetries originate from the transformation of the 1-form $A$
\begin{equation}\label{eq:extgauge}
A \to u A u^{-1} - du u^{-1} ~, \qquad \hat{g} \to u \hat{g} ~, \qquad
\end{equation}
In the same way that from $A$ and its $\CP^1$-derivatives evaluated at the poles we can construct $\mathds{A} \in \alg{d}$, from $\hat{g}$ and its $\CP^1$-derivatives, we can construct $\gdsl \in \exp(\alg{d})$, a group-valued field on $\Sigma$.
Similarly, we can also construct $\udsl \in \exp(\alg{d})$ on $\Sigma$, such that under the transformations~\eqref{eq:extgauge} we have
\begin{equation}\label{eq:external_gauge_symmetry}
\mathds{A} \to \udsl \mathds{A} \udsl^{-1} - d \udsl \udsl^{-1} ~, \qquad
\gdsl \to \udsl \gdsl ~.
\end{equation}
This tells us how the transformations~\eqref{eq:extgauge} act on the edge modes.
Not all these transformations survive as gauge symmetries in the 2d integrable sigma model.
In particular, those that do must preserve the boundary conditions that we impose.
If we take $\mathds{A}$ to be valued in a Lagrangian subalgebra of $\alg{d}$, then those $\udsl$ in the corresponding Lagrangian subgroup of $\exp(\alg{d})$ survive as gauge symmetries.

The internal gauge symmetries originate from the redundancy in the parametrisation~\eqref{eq:Laxparam} and act as
\begin{equation}\label{eq:internal_gauge_symmetry}
\Lax \to v^{-1} \Lax v + v^{-1} d v~, \qquad \hat g \to \hat g v ~,
\end{equation}
where $v \in \grp{G}^\Complex$ is a group-valued field on $\Sigma$.
It is $\CP^1$-independent to ensure that the transformations preserve the condition $\Lax_{\bar z} = 0$.

\subsection{Trigonometric models in 4d Chern-Simons}\label{sec:trig_mod_4dcs}

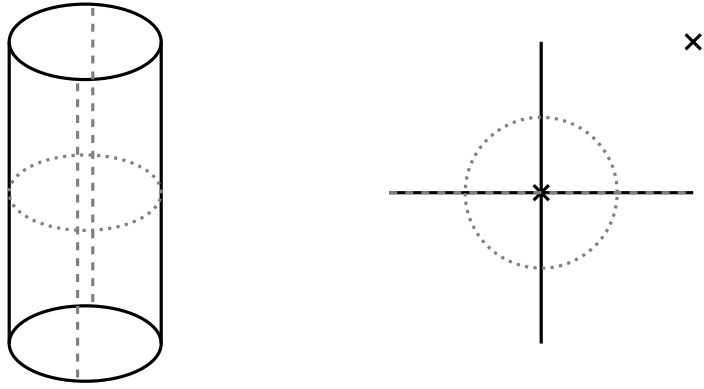
\begin{figure}
\begin{center}
\begin{tikzpicture}
\draw[-,very thick] (-4,-2)--(-4,2);
\draw[-,very thick] (-2,-2)--(-2,2);
\draw[very thick] (-3,2) ellipse (1 and 0.5);
\draw[very thick] (-3,-2) ellipse (1 and 0.5);
\draw[dotted,very thick,gray] (-3,0) ellipse (1 and 0.5);
\draw[dashed,very thick,gray] (-3.1,-2.45)--(-3.1,1.45);
\draw[dashed,very thick,gray] (-2.9,-1.45)--(-2.9,2.45);
\draw[-,very thick] (1,0)--(5,0);
\draw[-,very thick] (3,-2)--(3,2);
\draw[-,very thick] (2.9,-0.1)--(3.1,0.1);
\draw[-,very thick] (2.9,0.1)--(3.1,-0.1);
\draw[-,very thick] (4.9,2.1)--(5.1,1.9);
\draw[-,very thick] (4.9,1.9)--(5.1,2.1);
\draw[dotted,very thick,gray] (3,0) circle (1);
\draw[dashed,very thick,gray] (1,0)--(5,0);
\end{tikzpicture}
\end{center}
\caption{Map between the cylinder and $\CP^1 = \Complex \cup \{\infty\}$ with poles at $0$ and $\infty$.
The dotted line is mapped to the unit circle, while the dashed line is mapped to the horizontal axis.
We consider two classes of reality conditions for which conjugation corresponds to reflection in these two lines.}\label{fig1}
\end{figure}

Let us now turn to the trigonometric models, for which we replace $\CP^1$ by a cylinder $\Complex^\times$.
It is convenient to think of the cylinder in two equivalent ways, either as a cylinder or as a $\CP^1$ with simple poles at $z=0$ and $z=\infty$.
This is represented in~\figref{fig1}.
The map between these is well-known and is given by
\begin{equation}\label{eq:maptrigplane}
z= e^{i w} ~,
\end{equation}
where $w$, with $\Re w$ with $\Re w \in \interval{-\pi,\pi}$, is the coordinate on the cylinder and $z$ is the coordinate on $\CP^1$.
The trigonometric dependence on $w$ is the reason for the name of these models, however, we largely use the rational description and write twist functions and Lax connections in terms of $z$.

To construct non-trivial 2d integrable sigma models, we introduce the explicit form of the meromorphic 1-form $\omega$
\begin{equation}\label{eq:omegatrigi}
\omega_{\mathrm{trig}} = c_0 \varphi_{\mathrm{trig}}(z) \frac{dz}{z} ~,
\end{equation}
where the trigonometric twist function $\varphi_{\mathrm{trig}}(z)$ has an equal number of poles and zeroes, taking multiplicities into account, away from $0$ and $\infty$.
Here $c_0$ is a constant equal to $1$ or $i$ depending on our choice of conjugation on $\Complex^{\times}$, as we will explain shortly.
It is worth noting that any twist function on $\CP^1$ that has at least two simple poles can be mapped to our setup by an $\grp{SL}(2;\Complex)$ transformation
\begin{equation}
z \to \frac{az+b}{cz+d} ~, \qquad a,b,c,d \in \Complex ~, \quad ad - bc = 1 ~.
\end{equation}
This does not include, e.g.,~the principal chiral model for which $\omega$ has two double poles on $\CP^1$, but it does include many deformed models, such as the Yang-Baxter and current-current deformations \cite{Delduc:2019whp}.
Note that once the positions of the two simple poles have been fixed at $z=0$ and $z=\infty$, the residual $\grp{SL}(2;\Complex)$ transformations connected to the identity are given by
\begin{equation}\label{eq:residsl2c}
z \to a^2 z ~, \qquad a \in \Complex ~.
\end{equation}

Before we introduce the $\Integer_N$-twisted trigonometric models, let us briefly discuss conjugation and reality conditions.
There are two different choices of conjugation on $\Complex^{\times}$ that are of interest.
These correspond to whether the ends of the cylinder are conjugate to themselves (reflection in the dashed line in \figref{fig1}) or to each other (reflection in the dotted line in \figref{fig1}).
The two setups are distinguished by the value of $c_0$.
Denoting standard complex conjugation by $*$, for $c_0 = 1$ we have $\bar w = - w^*$, while for $c_0 = i$, $\bar w = w^*$.
Using the map~\eqref{eq:maptrigplane} it follows that
\begin{equation}
\begin{aligned}
& c_0 = 1~: \qquad && \bar z = z^* ~,
\\ & c_0 = i~: \qquad && \bar z = 1/z^* ~.
\end{aligned}
\end{equation}
We define the real line, which we denote $\mathfrak{R}$, to correspond to the line of reflection, which is left invariant.
For $c_0 = 1$, the real line on $\CP^1$ is the real numbers, which we denote $\Real$, while for $c_0 = i$, the real line on $\CP^1$ is the unit circle, which we denote $\Unit$.
Importantly, assuming $\varphi_{\mathrm{trig}}^*(z) = \varphi_{\mathrm{trig}}(\bar z)$, the 1-form $\omega_{\mathrm{trig}}$ in eq.~\eqref{eq:omegatrigi} satisfies the following conjugation property
\begin{equation}
\omega_{\mathrm{trig}}^*(z) = \omega_{\mathrm{trig}} (\bar z) ~,
\end{equation}
highlighting the role that the parameter $c_0$ is playing.
This property allows us to ensure that the 2d integrable sigma models we construct are real.
To satisfy $\varphi_{\mathrm{trig}}^*(z) = \varphi_{\mathrm{trig}}(\bar z)$ we demand that zeroes and double poles lie on the real line, while simple poles come in pairs that are either real or complex conjugate.
Note that, demanding the residual $\grp{SL}(2;\Complex)$ transformations preserve the conjugation properties implies that $a^2 \in \mathfrak{R}$, that is $a^2 \in \Real$ for $c_0 = 1$ and $a^2 \in \Unit$ for $c_0 = i$.

The final step for constructing real theories is to introduce an involutive antilinear automorphism $\tau$ of the complex Lie algebra $\alg{g}^\Complex$.
We denote the fixed point subalgebra of $\tau$, i.e.~the real form of $\alg{g}^\Complex$, by $\alg{g}$.
We then impose that algebra-valued fields evaluated at conjugate points, i.e.~at $z$ and $\bar z$, are related by the action of $\tau$, and group-valued fields by its lift to the Lie group.
The boundary conditions and regularity conditions, as well as any gauge fixings should all be compatible with this restriction.

\subsection{Twists of trigonometric models}\label{sec:twisttrigmod}

In order to twist the trigonometric models we start from a given trigonometric twist function $\varphi^{(1)}_{\mathrm{trig}}(\tilde z)$ and 1-form $\omega_{\mathrm{trig}}^{(1)} = c_0 \varphi^{(1)}_{\mathrm{trig}}(\tilde z) \frac{d\tilde z}{\tilde z}$, and introduce a $\Integer_N$ branch cut that runs from $\tilde z=0$ to $\tilde z = \infty$ on $\CP^1$, or equivalently cuts the cylinder along its non-compact direction.
As we go through the branch cut we apply a $\Integer_N$ automorphism of the Lie algebra $\alg{g}^\Complex$, which we denote $\sigma$
\begin{equation}\label{eq:sigmaN=1}
[\sigma(X),\sigma(Y)] = \sigma([X,Y]) ~, \quad X,Y \in \alg{g}^\Complex ~, \qquad \sigma^N = 1 ~,
\end{equation}
to the fields in our theory.
The goal of this paper is to construct the resulting twisted theories and start to investigate their properties.

For a $\Integer_N$ branch cut we come back to the original sheet after going through the cut $N$ times.
To avoid dealing explicitly with branch cuts, we instead work with the $N$-fold cover of the cylinder, which is the cylinder itself.
The trigonometric twist function $\varphi^{(N)}_{\mathrm{trig}}(z)$ on the $N$-fold cover is then given by
\begin{equation}
\varphi^{(N)}_{\mathrm{trig}} (z) = \varphi^{(1)}_{\mathrm{trig}}(z^N) ~,
\end{equation}
which we immediately see satisfies the $\Integer_N$-equivariance condition
\begin{equation}\label{eq:trigtwistaut}
\varphi^{(N)}_{\mathrm{trig}}(e^{\frac{2i\pi}{N}} z) = \varphi^{(N)}_{\mathrm{trig}}(z) ~.
\end{equation}
The fixed points of the $\Integer_N$ action on $\CP^1$
\begin{equation}\label{eq:varsigma}
\varsigma: z \to e^{\frac{2i\pi}{N}} z ~,
\end{equation}
are at $z=0$ and $z=\infty$, which correspond to the branch points as expected.
The trigonometric 1-form $\omega^{(N)}_{\mathrm{trig}}$ on the $N$-fold cover is then defined as
\unskip\footnote{Note that there is a factor of $N$ in the relation between the 1-forms, $\omega^{(N)}_{\mathrm{trig}}(z) = \frac{1}{N} \omega^{(1)}_{\mathrm{trig}}(z^N)$.}
\begin{equation}\label{eq:omegatrig}
\omega^{(N)}_{\mathrm{trig}} = c_0 \varphi^{(N)}_{\mathrm{trig}}(z) \frac{dz}{z} ~,
\end{equation}
which we see satisfies the $\Integer_N$-equivariance condition~\eqref{eq:trigtwistaut}.
Note that above and from now on, we take $\tilde z$ to be the $\CP^1$ coordinate on the original sheet, i.e.~before taking the $N$-fold cover, and $z$ to be the $\CP^1$ coordinate on the $N$-fold cover.

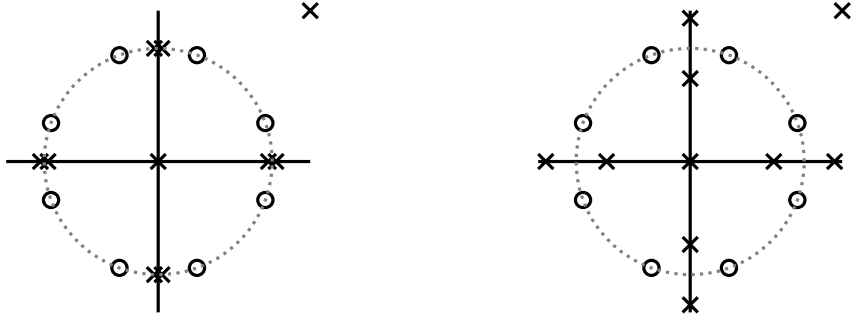
\begin{figure}
\begin{center}
\begin{tikzpicture}
\draw[-,very thick] (1,0)--(5,0);
\draw[-,very thick] (3,-2)--(3,2);
\draw[-,very thick] (2.9,-0.1)--(3.1,0.1);
\draw[-,very thick] (2.9,0.1)--(3.1,-0.1);
\draw[-,very thick] (4.9,2.1)--(5.1,1.9);
\draw[-,very thick] (4.9,1.9)--(5.1,2.1);
\draw[-,very thick] (4.35,0.1)--(4.55,-0.1);
\draw[-,very thick] (4.35,-0.1)--(4.55,0.1);
\draw[-,very thick] (4.45,0.1)--(4.65,-0.1);
\draw[-,very thick] (4.45,-0.1)--(4.65,0.1);
\draw[very thick] (4.41,0.51) circle (0.1);
\draw[very thick] (4.41,-0.51) circle (0.1);
\draw[-,very thick] (1.65,0.1)--(1.45,-0.1);
\draw[-,very thick] (1.65,-0.1)--(1.45,0.1);
\draw[-,very thick] (1.55,0.1)--(1.35,-0.1);
\draw[-,very thick] (1.55,-0.1)--(1.35,0.1);
\draw[very thick] (1.59,0.51) circle (0.1);
\draw[very thick] (1.59,-0.51) circle (0.1);
\draw[-,very thick] (2.95,1.4)--(3.15,1.6);
\draw[-,very thick] (2.95,1.6)--(3.15,1.4);
\draw[-,very thick] (2.85,1.4)--(3.05,1.6);
\draw[-,very thick] (2.85,1.6)--(3.05,1.4);
\draw[very thick] (3.51,1.41) circle (0.1);
\draw[very thick] (2.49,1.41) circle (0.1);
\draw[-,very thick] (2.95,-1.4)--(3.15,-1.6);
\draw[-,very thick] (2.95,-1.6)--(3.15,-1.4);
\draw[-,very thick] (2.85,-1.4)--(3.05,-1.6);
\draw[-,very thick] (2.85,-1.6)--(3.05,-1.4);
\draw[very thick] (3.51,-1.41) circle (0.1);
\draw[very thick] (2.49,-1.41) circle (0.1);
\draw[dotted,very thick,gray] (3,0) circle (1.5);
\draw[-,very thick] (8,0)--(12,0);
\draw[-,very thick] (10,-2)--(10,2);
\draw[-,very thick] (9.9,-0.1)--(10.1,0.1);
\draw[-,very thick] (9.9,0.1)--(10.1,-0.1);
\draw[-,very thick] (11.9,2.1)--(12.1,1.9);
\draw[-,very thick] (11.9,1.9)--(12.1,2.1);
\draw[-,very thick] (11,0.1)--(11.2,-0.1);
\draw[-,very thick] (11,-0.1)--(11.2,0.1);
\draw[-,very thick] (11.8,0.1)--(12,-0.1);
\draw[-,very thick] (11.8,-0.1)--(12,0.1);
\draw[very thick] (11.41,0.51) circle (0.1);
\draw[very thick] (11.41,-0.51) circle (0.1);
\draw[-,very thick] (8.2,0.1)--(8,-0.1);
\draw[-,very thick] (8.2,-0.1)--(8,0.1);
\draw[-,very thick] (9,0.1)--(8.8,-0.1);
\draw[-,very thick] (9,-0.1)--(8.8,0.1);
\draw[very thick] (8.59,0.51) circle (0.1);
\draw[very thick] (8.59,-0.51) circle (0.1);
\draw[-,very thick] (9.9,1)--(10.1,1.2);
\draw[-,very thick] (9.9,1.2)--(10.1,1);
\draw[-,very thick] (9.9,1.8)--(10.1,2);
\draw[-,very thick] (9.9,2)--(10.1,1.8);
\draw[very thick] (10.51,1.41) circle (0.1);
\draw[very thick] (9.49,1.41) circle (0.1);
\draw[-,very thick] (9.9,-1.8)--(10.1,-2);
\draw[-,very thick] (9.9,-2)--(10.1,-1.8);
\draw[-,very thick] (9.9,-1)--(10.1,-1.2);
\draw[-,very thick] (9.9,-1.2)--(10.1,-1);
\draw[very thick] (10.51,-1.41) circle (0.1);
\draw[very thick] (9.49,-1.41) circle (0.1);
\draw[dotted,very thick,gray] (10,0) circle (1.5);
\end{tikzpicture}
\end{center}
\caption{Two examples of $\Integer_N$-equivariant twist functions with (i) double poles and (ii) pairs of simple poles.
Poles are denoted by crosses and zeroes by circles.
Here we have taken $N=4$ and the real line $\mathfrak{R}$ to be the unit circle $\Unit$.
In the nomenclature of this paper, for Dirichlet boundary conditions at the double poles and $\eta$-type boundary conditions at the simple poles, these give the $\Integer_4$-twisted undeformed $\eta$-model and the $\Integer_4$-twisted YB-deformed $\eta$-model.}\label{fig2}
\end{figure}

To construct the $\Integer_N$-twisted trigonometric sigma models we consider trigonometric twist functions on the $N$-fold cover that satisfy eq.~\eqref{eq:trigtwistaut}.
Two examples of such twist functions for $N=4$ are given in \figref{fig2}.
To impose the twisting we then demand that fields in the theory satisfy the $\Integer_N$-equivariance condition
\begin{equation}\label{eq:fieldtwisting}
\Phi(\varsigma(z)) = \Phi(e^{\frac{2i\pi}{N}}z ) = \sigma (\Phi(z)) ~,
\end{equation}
where $\Phi$ represents an algebra-valued field.
A similar relation holds for group-valued fields, with $\sigma$ replaced by $\hat \sigma$, its lift to the Lie group.
Let us note that the twisting can be understood as a consistent truncation, guaranteeing the integrability of the twisted models.

\medskip

In the following sections we give more technical details and construct explicit examples.
Before we do so let us make some comments.
The first comment is on the properties of fields evaluated at the branch points, or equivalently the fixed points of the $\Integer_N$ action on $\CP^1$~\eqref{eq:varsigma}, which we denote $z_{\mathrm{f.p.}}=0,\infty$.
It follows that $\varsigma(z_{\mathrm{f.p.}}) = z_{\mathrm{f.p.}}$, hence, from eq.~\eqref{eq:fieldtwisting}, we find
\begin{equation}
\Phi(z_{\mathrm{f.p.}}) = \sigma(\Phi(z_{\mathrm{f.p.}})) ~,
\end{equation}
which implies that fields evaluated at these points are restricted to lie in the fixed point subalgebra or subgroup of the $\Integer_N$ automorphisms $\sigma$ or $\hat\sigma$, which we denote $\alg{g}_0^\Complex$ and $\grp{G}_0^\Complex$.
The projector onto the fixed point subalgebra can be built from the $\Integer_N$ automorphism as
\begin{equation}\label{eq:p0}
P_0 = \frac{1}{N}\sum_{a\in\Integer_N} \sigma^a ~.
\end{equation}

\medskip

Our second comment is on the compatibility of reality conditions and the $\Integer_N$-twisting.
Let us introduce the projectors onto $\alg{g}_a$, the eigenspaces of the $\Integer_N$ automorphism $\sigma$ with eigenvalue $e^{\frac{2i\pi a}{N}}$,
\unskip\footnote{
We can check that these are projectors explicitly by noting that
\begin{equation*}
P_aP_b = \frac{1}{N^{2}} \sum_{c,d\in\Integer_{N}} e^{-\frac{2i\pi (ca+db)}{N}} \sigma^{c+d} = \frac{1}{N^{2}} \sum_{m\in\Integer_{N}} \sum_{n\in\Integer_{N}+m} e^{-\frac{2i\pi (na+m(b-a))}{N}} \sigma^{n} ~.
\end{equation*}
If $a = b$ we have
\begin{equation*}
P_a^2 = \frac{1}{N^{2}} \sum_{m\in\Integer_{N}} \sum_{n\in\Integer_{N}+m} e^{-\frac{2i\pi na}{N}} \sigma^{n} = \frac{1}{N} \sum_{n\in\Integer_{N}} e^{-\frac{2i\pi na}{N}} \sigma^{n} = P_a ~,
\end{equation*}
where we have used $\sum_{m\in\Integer_N} 1 = N$.
On the other hand, if $a \neq b$ then
\begin{equation*}
P_aP_b = \frac{1}{N^{2}} \sum_{m\in\Integer_{N}} \sum_{n\in\Integer_{N}+m} e^{-\frac{2i\pi (na+m(b-a))}{N}} \sigma^{n} = 0 ~,
\end{equation*}
since $\sum_{m\in\Integer_N} e^{-\frac{2i\pi m(b-a)}{N}} = 0$ for $b\neq a \mod N$.}
\begin{equation}\label{eq:pa}
P_a = \frac{1}{N} \sum_{b\in\Integer_N} e^{-\frac{2i\pi a b}{N}}\sigma^b ~.
\end{equation}
For $c_0 = 1$ we find the $\Integer_N$-twisted trigonometric sigma models are real if $\tau$ commutes with the projectors
\begin{equation}\label{eq:realc01}
[\tau,P_a] = 0 ~, \qquad \tau\sigma = \sigma^{-1}\tau ~,
\end{equation}
while for $c_0 = i$ $\tau$ should commute with the $\Integer_N$ automorphism
\begin{equation}\label{eq:realc0i}
[\tau,\sigma] = 0 ~, \qquad \tau P_a = P_{N-a} \tau ~, \qquad P_N = P_0 ~.
\end{equation}
The conditions~\eqref{eq:realc01} and \eqref{eq:realc0i} are equivalent for $N=1,2$, but differ for $N > 2$.
In both cases we have that the projector $P_0$~\eqref{eq:p0} commutes with $\tau$, and the fixed point subalgebra of both $\tau$ and $\sigma$ is the real form $\alg{g}_0$ of $\alg{g}_0^{\Complex}$.
We specify boundary conditions, regularity conditions and gauge fixings that are compatible with reality conditions on the original sheet, i.e.~before taking the $N$-fold cover, and extend to these to the remaining $N-1$ sheets using $\Integer_N$-equivariance.
Equivalently, we specify the conditions on the sector of the $N$-fold cover corresponding to the original sheet, e.g.~$-\frac{\pi}{N}<\arg z \leq \frac{\pi}{N}$, and use $\Integer_N$-equivariance to extend these to the rest of $\CP^1$.

\medskip

This procedure also ensures that the boundary conditions, regularity conditions and gauge fixings for poles and zeroes away from $z_{\mathrm{f.p.}}$ are compatible with $\Integer_N$-equivariance, which brings us to our third comment.
The compatibility of $\Integer_N$-equivariance with boundary conditions and gauge fixings at $z_{\mathrm{f.p.}}$ is more subtle since these are fixed points.
This is because once we impose these conditions on one sheet, or for one sector of the $N$-fold cover, they are automatically specified for all sheets or sectors.
To ensure compatibility, we consider boundary conditions and gauge fixings at $z_{\mathrm{f.p.}}$ that commute with imposing $\Integer_N$-equivariance.
Explicitly, the boundary conditions should commute with the restriction $(1-P_0) A_\pm\vert_{z=z_{\mathrm{f.p.}}} = 0$.

In this paper we consider two types of boundary conditions at $z_{\mathrm{f.p.}}$, which we refer to as $\eta$-type and $\lambda$-type.
These boundary conditions are applicable to setups where the sum of the residues at $z=0$ and $z=\infty$ vanishes, i.e.~$\lim_{z\to\infty}\varphi_{\mathrm{trig}}(z) = \varphi_{\mathrm{trig}}(0)$.
The $\eta$-type boundary conditions are given by
\begin{equation}\label{eq:bcetac0}
(\tilde{\mathcal{R}}+c_0) A_\pm\vert_{z=0} = (\tilde{\mathcal{R}}-c_0) A_\pm \vert_{z=\infty} ~,
\end{equation}
where the R-matrix $\tilde{\mathcal{R}}$ is an antisymmetric solution to the modified classical Yang-Baxter equation on $\alg{g}$
\begin{equation}\begin{gathered}
[\tilde{\mathcal{R}}X,\tilde{\mathcal{R}}Y]-\tilde{\mathcal{R}}[X,\tilde{\mathcal{R}}Y] - \tilde{\mathcal{R}}[\tilde{\mathcal{R}}X,Y] + c_0^2 [X,Y] = 0 ~,
\\ \tr(X\tilde{\mathcal{R}} Y) + \tr(\tilde{\mathcal{R}} X Y) = 0 ~, \qquad X,Y\in\alg{g} ~.
\end{gathered}
\end{equation}
To ensure that these boundary conditions are compatible with $\Integer_N$-equivariance we require that
\begin{equation}
[\sigma,\tilde{\mathcal{R}}] = 0 ~,
\end{equation}
which implies $[P_0,\tilde{\mathcal{R}}] = 0$.
The $\lambda$-type boundary conditions are
\begin{equation}\label{eq:bclambdac0}
A_\pm\vert_{z=0} = A_\pm\vert_{z=\infty} ~,
\end{equation}
which are immediately compatible with $\Integer_N$-equivariance.

Returning to poles away from $z_{\mathrm{f.p.}}$, we similarly consider different types of boundary conditions.
For double poles we take Dirichlet boundary conditions $A_\pm\vert_{z = z_{\mathrm{pole}}} = 0$, while at pairs of simple poles, we again take either $\eta$-type or $\lambda$-type boundary conditions.
Let us emphasise that since we have guaranteed $\Integer_N$-equivariance by specifying boundary conditions on a single sheet, or in a single sector, there is no additional restriction.
In particular, for $\eta$-type boundary conditions at simple poles away from $z_{\mathrm{f.p.}}$ we do not require the associated R-matrix, which can be different to $\tilde{\mathcal{R}}$, to commute with $\sigma$.

Regarding gauge symmetries, the internal gauge symmetries~\eqref{eq:internal_gauge_symmetry} are restricted to $v\in\grp{G}_0$ upon imposing $\Integer_N$-equivariance.
This follows since $v$ is $\CP^1$-independent, hence the restriction at $z=z_{\mathrm{f.p.}}$ implies this holds over the whole of $\CP^1$.
The external gauge symmetries~\eqref{eq:extgauge} evaluated at the fixed points $z_{\mathrm{f.p.}}$ are similarly restricted to $u\vert_{z=0}, u\vert_{z=\infty} \in \grp{G}_0$.
As discussed above, only those external gauge transformations that preserve the boundary conditions survive.
We will return to this in \secref{sec:bcgs}.

\medskip

The models we find before twisting, i.e.~before we introduce the branch cuts, are the same as the those that come from twisting with the identity automorphism, i.e.~$\sigma = 1$.
Therefore, we refer to these as the untwisted models, which have $\grp{G}_0 = \grp{G}$.
Our fourth comment is that the untwisted and twisted models have the same number of degrees of freedom.
This is due to the fact that the branch points are simple poles.
A sketch of why this is the case for the setups we consider goes as follows.
After fixing the external gauge symmetries, each $N$ double poles or $N$ pairs of simple poles away from the fixed points $z=z_{\mathrm{f.p.}}$ on the $N$-fold cover gives $\dim \grp{G}$ degrees of freedom in the twisted model.
On the other hand the pair of simple poles at $z=z_{\mathrm{f.p.}}$ gives $\dim \grp{G}_0$ degrees of freedom.
The remaining gauge symmetries are the internal ones, which remove $\dim \grp{G}_0$ degrees of freedom.
If the total number of double poles and pairs of simple poles, excluding the pair of simple poles at $z=z_{\mathrm{f.p.}}$, is $N N_{\mathrm{f}}$, we conclude that the total number of degrees of freedom is $N_{\mathrm{f}} \dim \grp{G}$, which does not depend on $\grp{G}_0$ and the choice of automorphism.
\unskip\footnote{Note that, assuming that we gauge fix in a way compatible with $\Integer_N$-equivariance, gauge fixing commutes with the $\Integer_N$-twisting.
Before the $\Integer_N$-twisting, after fixing the external gauge symmetries, there are $(N N_{\mathrm{f}} + 1) \dim \grp{G}$ degrees of freedom, less $\dim\grp{G}$ to account for the internal gauge symmetries.
This leaves $N N_{\mathrm{f}}\dim \grp{G}$ degrees of freedom, which after $\Integer_N$-twisting becomes the expected $N_{\mathrm{f}} \dim \grp{G}$.}

The majority of the literature on $\Integer_N$ branch cuts and $\Integer_N$-equivariant models, see e.g.~\cite{Costello:2019tri,Tian:2020ryu,Fukushima:2020dcp,Tian:2020pub,Costello:2020lpi,Berkovits:2024reg,Cole:2024skp}, has focused on setups that are qualitatively different.
In these setups, the branch cuts are introduced between two marked points on $\CP^1$ that are not poles or zeroes.
In this case, the internal gauge symmetries are restricted to $\grp{G}_0$, but there are no restrictions on the degrees of freedom of the 2d integrable sigma model.
Therefore, these setups define models with $N_{\mathrm{f}} \dim \grp{G} - \dim \grp{G}_0$ degrees of freedom.
This includes coset models such as the symmetric space sigma model, for which $N_{\mathrm{f}} = 1$.

\medskip

There are two instances of $\Integer_N$-twisted trigonometric sigma models in the literature.
In the language introduced above, these are both $\Integer_2$-twistings starting with a single double pole at $\tilde z = 1$.
$\eta$-type boundary conditions at the fixed points $z_{\mathrm{f.p.}} = 0,\infty$ were considered in~\cite{Fukushima:2020kta} from the perspective of 4d Chern-Simons and a consistent truncation implemented in 2d.
We call this model the $\Integer_2$-twisted $\eta$-model.
A brief interpretation in terms of branch cuts is given in \cite{Tian:2020ryu}.
On the other hand, the model that comes from using the $\lambda$-type boundary conditions at $z_{\mathrm{f.p.}} = 0,\infty$ is constructed in~\cite{Borsato:2024alk} directly in 2d using the language of spectators.
We call this model the $\Integer_2$-twisted $\lambda$-model.
In this paper we give an account of the origin of both these models from the perspective of 4d Chern-Simons, their $\Integer_N$ generalisation and integrable deformations.
The deformations we construct come from splitting the double pole at $\tilde z=1$ and imposing either $\eta$-type boundary conditions (YB-deformed) or $\lambda$-type boundary conditions (CC-deformed).
To summarise, we will explicitly construct the following models:
\begin{itemize}
\item $\Integer_N$-twisted $\eta$-model,
\item $\Integer_N$-twisted $\lambda$-model,
\item YB-deformed $\Integer_2$-twisted $\eta$-model,
\item CC-deformed $\Integer_2$-twisted $\eta$-model,
\item YB-deformed $\Integer_2$-twisted $\lambda$-model,
\item CC-deformed $\Integer_2$-twisted $\lambda$-model.
\end{itemize}
Here $\eta$-model or $\lambda$-model refers to the boundary conditions we are choosing at the ends of the cylinder, $z_{\mathrm{f.p.}} = 0,\infty$, while YB-deformed and CC-deformed refer to boundary conditions away from these points.

The untwisted versions of all these models have previously been constructed in the literature.
To help the reader, we conclude this overview by briefly summarising the various names by which these are known.
The untwisted $\eta$-model is often called the (inhomogeneous) Yang-Baxter deformation of the principal chiral model, or the $\eta$ deformation, and was first constructed for $c_0 = i$ in \cite{Klimcik:2002zj}.
Similarly, for $c_0 = 1$ the untwisted $\lambda$-model is normally known as the $\lambda$ deformation \cite{Sfetsos:2013wia}, and can also be understood as the current-current deformation of the WZW model~\cite{Witten:1983ar}.
Of the remaining four models, the YB-deformed untwisted $\eta$-model is most well-known as the bi-Yang-Baxter deformation~\cite{Klimcik:2008eq}.
The untwisted versions of the CC-deformed $\eta$-model and the YB-deformed $\lambda$-model are the same and this model was first written down in~\cite{Sfetsos:2015nya}.
Finally, the CC-deformed untwisted $\lambda$-model is a particular double deformation of the WZW model initially constructed in \cite{Sfetsos:2014cea,Sfetsos:2017sep}.
A summary of these models, including a more detailed review of the literature and their relation via Poisson-Lie duality is given in \cite{Hoare:2022vnw}, while a detailed derivation of the first three untwisted models from 4d Chern-Simons was first given in \cite{Delduc:2019whp}.

\section{Regularity conditions, boundary conditions, gauge symmetries and actions}\label{sec:bcgs}

Having outlined the general construction of the $\Integer_N$-twisted trigonometric sigma models in \secref{sec:general-theory}, we now give more details about the regularity conditions, boundary conditions and gauge fixings that we use when constructing explicit models in~\secref{sec:sigmamodels}.
We start by describing the regularity conditions at simple zeroes of the twist function that ensure that we find relativistic models.
Having already briefly introduced boundary conditions for double poles and pairs of simple poles with equal and opposite residues in \secref{sec:general-theory}, we will then discuss these in more detail, including the associated defect Lie algebras, the surviving external gauge symmetries and $\Integer_N$-equivariant gauge fixings.
Finally, we present the general form of the actions for both the twisted $\eta$-models and $\lambda$-models.

We denote the set of poles of $\varphi_{\mathrm{trig}}^{(N)}$ by $\mathfrak{p}$ and the set of zeroes by $\mathfrak{z}$.
Recall that these poles and zeroes are away from $0$ and $\infty$.
Moreover, by construction, the sets $\mathfrak{p}$ and $\mathfrak{z}$ should be invariant under the $\Integer_N$ action~\eqref{eq:varsigma}.
Given that all the zeroes are simple, the meromorphic 1-form $\omega^{(N)}_{\mathrm{trig}}$~\eqref{eq:omegatrig} can then be written as
\begin{equation}\label{eq:trigomega}
\omega^{(N)}_{\mathrm{trig}} = c_0 \frac{\prod_{s\in\mathfrak{z}}(z-s)}{\prod_{r\in\mathfrak{p}}(z-r)^{m_r}} \frac{dz}{z} ~,
\end{equation}
where $m_r$ is the multiplicity of the pole $r\in\mathfrak{p}$ and $\sum_{s\in\mathfrak{z}} 1 = \sum_{r\in\mathfrak{p}} m_r$ by the Riemann-Roch theorem.
Our conventions for differentiation of meromorphic functions and integration over delta functions are
\begin{equation}\begin{gathered}
\partial_{\bar{z}} \left( \frac{1}{(z-t)^p} \right) = -2i\pi \frac{(-1)^{p-1}}{(p-1)!} \partial_{z}^{p-1} \delta^{(2)}(z-t) ~,
\\
\int_{\mathds{CP}^1} dz\wedge d\bar z \, f(z,\bar{z}) \, \partial_{z}^{p} \delta^{(2)}(z-t)= (-1)^{p} \partial_{z}^{p} f(z,\bar{z})\vert_{z=t} ~.
\end{gathered}\end{equation}

\subsection{Regularity conditions}

As discussed in~\secref{sec:general-theory}, the equation of motion $\omega^{(N)}_{\mathrm{trig}} \wedge \partial_{\bar{z}} \Lax = 0$ implies that $\Lax$ is meromorphic in $z$ and is allowed poles at the zeroes $\mathfrak{z}$ of the meromorphic 1-form $\omega^{(N)}_{\mathrm{trig}}$.
In general, a pole of order $p$ at point $z=s$ in the Lax connection $\Lax$ implies that the meromorphic 1-form $\omega^{(N)}_{\mathrm{trig}}$ must have a zero of order $p$ at $z=s$.
Given we assume that $\omega^{(N)}_{\mathrm{trig}}$ only has simple zeroes, the general form of the Lax connection is
\begin{equation}
\Lax = \tilde{V} + \sum_{s\in\mathfrak{z}} \frac{W^{(s)}}{z-s} ~,
\end{equation}
where $\tilde{V}$ and $W^{(s)}$ are $z$-independent 1-forms on $\Sigma$.

The zero-curvature equation for the Lax connection has two terms, $\omega^{(N)}_{\mathrm{trig}} \wedge d_\Sigma\Lax$ and $\frac12\omega^{(N)}_{\mathrm{trig}} \wedge[\Lax,\Lax]$, with their sum vanishing.
To ensure that the second term does not end up with double poles, overwhelming the zeroes in $\omega^{(N)}_{\mathrm{trig}}$, we split the zeroes of $\omega^{(N)}_{\mathrm{trig}}$ into two disjoint subsets of equal size, and demand that one set correspond to poles in $\Lax_+$ and the other to poles in $\Lax_-$.
This also ensures that the models we construct are relativistic \cite{Costello:2019tri, Delduc:2019whp}.
The general form for the Lax connection in light-cone coordinates is then given by
\begin{equation}\label{eq:lax:lc}
\Lax_\pm = \tilde{V}_\pm + \sum_{s\in\mathfrak{z}_\pm} \frac{W_\pm^{(s)}}{z-s} ~,
\end{equation}
where $\mathfrak{z}_+$ and $\mathfrak{z}_-$ are two disjoint subsets of the zeroes of $\omega^{(N)}_{\mathrm{trig}}$ such that $\mathfrak{z}_+ \cup \mathfrak{z}_- = \mathfrak{z}$.
In light-cone coordinates the zero-curvature equation for $\Lax_\pm$ is
\begin{equation}
F_{+-}(\Lax) = \partial_+ \Lax_- - \partial_- \Lax_+ + [\Lax_+,\Lax_-] = 0 ~.
\end{equation}

Regarding $\Integer_N$-equivariance, by construction the set of zeroes $\mathfrak{z}$ on the $N$-fold cover is invariant under the action of the $\Integer_N$ transformation~\eqref{eq:varsigma}.
The splitting of $\mathfrak{z}$ into $\mathfrak{z}_+$ and $\mathfrak{z}_-$ should respect this, with each subset invariant under the action~\eqref{eq:varsigma}.
As described above, to ensure this we specify the splitting on the original sheet, or in one sector of the $N$-fold cover, and extend to the rest of the space using $\Integer_N$-equivariance.

\subsection{Boundary conditions and external gauge symmetries}

Recalling that the meromorphic 1-form $\omega^{(N)}_{\mathrm{trig}}$~\eqref{eq:trigomega} has simple poles at $z=0$ and $z=\infty$, we can write it in the general form
\begin{equation}
\omega^{(N)}_{\mathrm{trig}} = \sum_{r\in\mathfrak{p}} \sum_{p_r=0}^{m_r-1} \frac{\ell_{r,p_r}}{(z-r)^{p_{r}+1}} dz + \ell_0\frac{dz}{z} = \ell_\infty \frac{d\zeta}{\zeta} + \mathcal{O}(\zeta^0) ~,
\end{equation}
where $\zeta = 1/z$.
The coefficients $\ell_{r,p_r}$, $\ell_0$ and $\ell_\infty$ are complex numbers called the levels of the theory.
As discussed in \secref{sec:general-theory}, in this paper we restrict to the case the sum of the residues at $z=0$ and $z=\infty$ vanishes, i.e.~$\ell_\infty + \ell_0 = 0$, which implies that
\begin{equation}
\sum_{r\in\mathfrak{p}} \ell_{r,0} = 0 ~,
\end{equation}
to ensure that the sum of all the residues vanishes.

The boundary term~\eqref{eq:boundaryterm} can then be written as
\begin{equation}\begin{split}\label{eq:btdoublepole}
\frac{1}{2i\pi}\int_{\CP^1\times \Sigma} d\omega^{(N)}_{\mathrm{trig}}\wedge\tr(A\delta A) & = \int_\Sigma d^2 \sigma \, \Big( \ell_0 (\tr(A\delta A)\vert_{z=0} - \tr(A\delta A)\vert_{z=\infty})
\\ & \qquad \qquad \qquad \qquad + \sum_{r\in\mathfrak{p}}\sum_{p_r=0}^{m_r-1} \frac{\ell_{r,p_r}}{p_r!} \partial_{z}^{p_r} \tr( A \delta A )\vert_{z=r} \Big) ~.
\end{split}
\end{equation}
From this expression it is possible to read off the defect Lie algebra, $\alg{d}$, and the bilinear form $\langle\!\langle,\rangle\!\rangle$.
We refer to \cite{Lacroix:2020flf} for a general analysis.
As already discussed, in this paper we consider twist functions that have real double poles with vanishing residue, pairs of real simple poles with equal and opposite residues, and pairs of complex conjugate simple poles with equal and opposite residues.
This means that the defect Lie algebra factorises into decoupled subalgebras.

\paragraph{Double poles with vanishing residues.}
Let us consider a double pole with vanishing residue at $z=r$.
The contribution to the boundary term~\eqref{eq:btdoublepole} is given by
\begin{equation}\label{eq:bcdp}
\int_{\Sigma} d^2\sigma \, \ell_{r,1} \partial_z \tr (A\delta A)\vert_{z=r} ~.
\end{equation}
Taking $z=r$ to be real, i.e.~$r\in \mathfrak{R}$, the defect Lie algebra is $\alg{g} \ltimes \alg{g}^{\mathrm{ab}}$ with Lie bracket
\begin{equation}
[[(X,X'),(Y,Y')]] = ([X,Y],[X,Y']+[X',Y]) ~,
\end{equation}
and bilinear form
\begin{equation}
\langle\!\langle(X,X'),(Y,Y')\rangle\!\rangle = \ell_{r,1} (\tr(XY') + \tr(X'Y)) ~,
\end{equation}
where, with a minor abuse of notation, $X,X',Y,Y' \in \alg{g}$ and $(X,X'),(Y,Y') \in \alg{g} \ltimes \alg{g}^{\mathrm{ab}}$.
Letting $\mathds{A} = (A,\partial_z A)\vert_{z=r} \in \alg{g} \ltimes \alg{g}^{\mathrm{ab}}$, we see that eq.~\eqref{eq:bcdp} can be written as
\begin{equation}\label{eq:311}
\int_\Sigma \langle\!\langle\mathds{A},\delta\mathds{A}\rangle\!\rangle ~,
\end{equation}
as described in \secref{sec:general-theory}.
At double poles we always consider Dirichlet boundary conditions
\begin{equation} \label{bc_undef_double}
A\vert_{z=r} = 0 ~,
\end{equation}
which ensures the vanishing of the boundary term~\eqref{eq:bcdp}.
This corresponds to taking $ \mathds{A}$ to be valued in the Lagrangian subalgebra $0\ltimes \alg{g}^{\mathrm{ab}}$, generated by elements of $\alg{d}$ of the form $(0,X')$.

The external gauge transformations~\eqref{eq:extgauge} that survive are those that preserve the boundary condition~\eqref{bc_undef_double}.
This implies that
\begin{equation}
(uAu^{-1} - du u^{-1}) \vert_{z=r} = -(du u^{-1})\vert_{z=r} = 0 ~,
\end{equation}
which means that $u$ is constant at the double pole $z=r$.
This eventually ends up as a global $\grp{G}$ symmetry of the 2d integrable sigma model.
Now considering how the surviving external gauge transformations act on the edge modes $\hat{g}\vert_{z=r}$ and $\partial_z \hat{g}\vert_{z=r}$, we have
\begin{equation}
\hat{g}\vert_{z=r} \to (u \hat{g})\vert_{z=r} ~, \qquad
\partial_z\hat{g}\vert_{z=r} \to (\partial_{z}u \hat{g} + u \partial_{z}\hat{g})\vert_{z=r} ~.
\end{equation}
While $u\vert_{z=r}$ is restricted to be constant, $\partial_{z}u\vert_{z=r}$ is not constrained.
This means that we can use the external gauge transformations to fix
\begin{equation}
\hat{g}\vert_{z=r} = g_{r} \in \grp{G} ~, \qquad \partial_{z}\hat{g}\vert_{z=r} = 0 ~.
\end{equation}

\paragraph{Pairs of simple poles with equal and opposite residues.}
We now consider a pair of simple poles at $z=r_1$ and $z=r_2$ with equal and opposite residues.
The contribution to the boundary term~\eqref{eq:btdoublepole} is given by
\begin{equation}\label{eq:bcsp}
\int_\Sigma d^2\sigma \, \ell_{r,0} (\tr(A\delta A)\vert_{z=r_1} - \tr(A\delta A)\vert_{z=r_2}) ~.
\end{equation}

If both $z=r_1$ and $z=r_2$ are real, i.e.~$r_1,r_2 \in \mathfrak{R}$, then the defect Lie algebra is $\alg{g} \oplus \alg{g}$ with Lie bracket
\begin{equation}
[[(X_1,X_2),(Y_1,Y_2)]] = ([X_1,Y_1],[X_2,Y_2]) ~,
\end{equation}
and bilinear form
\begin{equation}
\langle\!\langle (X_1,X_2) , (Y_1,Y_2) \rangle\!\rangle = \ell_{r,0} (\tr(X_1Y_1) - \tr(X_2Y_2)) ~,
\end{equation}
where $X_1,X_2,Y_1,Y_2\in\alg{g}$ and $(X_1,X_2),(Y_1,Y_2)\in \alg{g} \oplus \alg{g}$.
Letting $\mathds{A} = (A\vert_{z=r_1},A\vert_{z=r_2}) \in \alg{g}\oplus \alg{g}$, we again see that eq.~\eqref{eq:bcsp} can be written in the form~\eqref{eq:311}.

If $z=r_1$ and $z=r_2$ are complex conjugate, i.e.~$r_2 = \bar r_1$, then the defect Lie algebra is $\alg{g}^\Complex$ with Lie bracket
\begin{equation}
[[X+iX',Y+iY']] = ([X,Y] - [X',Y']) + i ([X,Y'] + [X',Y]) ~,
\end{equation}
and bilinear form
\begin{equation}
\langle\!\langle X+iX' , Y+i Y' \rangle\!\rangle = 2i\ell_{r,0} (\tr(XY') + \tr(X'Y)) ~,
\end{equation}
where $X,X',Y,Y'\in\alg{g}$ and $X+iX',Y+iY'\in \alg{g}^\Complex$.
To write eq.~\eqref{eq:bcsp} in the form~\eqref{eq:311}, we let $\mathds{A} = A\vert_{z=r_1} \in \alg{g}^\Complex$ and note that $\tau(\mathds{A}) = A\vert_{z=r_2}$.

We have previously introduced the $\eta$-type and $\lambda$-type boundary conditions in \secref{sec:general-theory} when discussing the boundary conditions at the ends of the cylinder, or the fixed points $z_{\mathrm{f.p.}} = 0,\infty$.
Here we give some more details for general pairs of simple poles, including an analysis of the surviving external gauge transformations and allowed gauge fixings.
In the following we introduce a parameter $c$ associated to the pair of simple poles, which is equal to $1$ if they are real and $i$ if they are complex conjugate.
Note that for the pair of simple poles at $z_{\mathrm{f.p.}} = 0,\infty$ the associated parameter is precisely the parameter $c_0$ introduced in \secref{sec:general-theory}.

\paragraph{$\eta$-type boundary conditions.}
The $\eta$-type boundary conditions are
\begin{equation}\label{etabc}
(\mathcal{R}+c)A\vert_{z=r_1}=(\mathcal{R}-c)A\vert_{z=r_2} ~,
\end{equation}
where the R-matrix $\mathcal{R}$ is an antisymmetric solution ot the modified classical Yang-Baxter equation on $\alg{g}$
\unskip\footnote{Note that the $\eta$-type boundary conditions only exist if an R-matrix exists for the real form $\alg{g}$, which need not be the case.
For the compact real form there is always a solution to the non-split modified classical Yang-Baxter equation, i.e.~with $c=i$, and no solution to the split modified classical Yang-Baxter equation, i.e.~with $c=1$.
On the other hand, for the split real form there is always a solution to the split modified classical Yang-Baxter equation.
For other real forms a more detailed analysis can be carried out.}
\begin{equation}\begin{gathered}\label{eq:mcybe}
[\mathcal{R}X, \mathcal{R}Y] - \mathcal{R}([\mathcal{R}X, Y] + [X, \mathcal{R}Y]) + c^2 [X, Y] = 0 ~,
\\ \tr(X\mathcal{R}Y) + \tr((\mathcal{R}X)Y) = 0 ~, \qquad X,Y \in \alg{g} ~.
\end{gathered}\end{equation}
Here the antisymmetry of $\mathcal{R}$ ensures that the boundary term~\eqref{eq:bcsp} vanishes, while the modified classical Yang-Baxter equation ensures that $\mathds{A}$ is valued in a Lagrangian subalgebra of $\alg{d}$.
In particular, using the vector space decompositions
\begin{equation}\begin{aligned}\label{eq:vsdecomp}
& c=1 : \qquad && \alg{d} = \alg{g} \oplus \alg{g} = \alg{g}_{\mathrm{diag}} \dot{+} \alg{g}_\mathcal{R} ~, &&
\\ &&&
\alg{g}_{\mathrm{diag}} = \operatorname{span}\{(X,X), \ X\in \alg{g}\} ~,\quad &&
\alg{g}_{\mathcal{R}} = \operatorname{span}\{((\mathcal{R}-1) X, (\mathcal{R}+1) X), \ X\in\alg{g}\} ~,
\\
& c=i: \qquad && \alg{d} = \alg{g}^\Complex = \alg{g}_{\mathrm{diag}} \dot{+} \alg{g}_{\mathcal{R}} ~, &&
\\ &&&
\alg{g}_{\mathrm{diag}} = \operatorname{span}\{X, \ X\in \alg{g}\} ~,\quad &&
\alg{g}_{\mathcal{R}} = \operatorname{span}\{(\mathcal{R}-i) X, \ X\in\alg{g}\} ~,
\end{aligned}\end{equation}
the boundary conditions~\eqref{etabc} correspond to taking $\mathds{A}$ to be valued in the Lagrangian subalgebra $\alg{g}_{\mathcal{R}}$ of $\alg{d}$.
Defining $\udsl = (u\vert_{z=r_1},u\vert_{z=r_2}) \in \grp{G} \times \grp{G}$ for $c=1$ or $\udsl = u\vert_{z=r_1} \in \grp{G}^\Complex$ for $c=i$, the surviving external gauge transformations, i.e.~those that preserve the boundary conditions~\eqref{etabc}, are parametrised by $\udsl \in \grp{G}_{\mathcal{R}}$, the Lie group corresponding to $\alg{g}_\mathcal{R}$.
Also defining $\gdsl = (\hat g\vert_{z=r_1},\hat g\vert_{z=r_2}) \in \grp{G} \times \grp{G}$ for $c=1$ or $\gdsl = \hat g\vert_{z=r_1} \in \grp{G}^\Complex$ for $c=i$, it follows that we can use the external gauge transformations to fix $\gdsl \in \grp{G}_{\mathrm{diag}}$, the Lie group corresponding to $\alg{g}_\mathrm{diag}$.
In summary, after fixing the external gauge transformations we find
\begin{equation}
\hat g\vert_{z=r_1} = \hat g\vert_{z=r_2} = g_r \in \grp{G} ~.
\end{equation}
Note that the same analysis holds for the simple poles at $z_{\mathrm{f.p.}} = 0,\infty$ after twisting, expect that we now end up with
\begin{equation}\label{eq:fpetagauge}
\hat g\vert_{z=0} = \hat g\vert_{z=\infty} = \tilde g \in \grp{G}_0 ~.
\end{equation}
This gauge fixing is compatible with $\Integer_N$-equivariance at the fixed points.

\paragraph{$\lambda$-type boundary conditions.}
The $\lambda$-type boundary conditions are
\begin{equation}\label{lambdabc}
A\vert_{z=r_1} = A\vert_{z=r_2} ~.
\end{equation}
These boundary conditions correspond to taking $\mathds{A}$ to be valued in the Lagrangian subalgebra $\alg{g}_{\mathrm{diag}}$ of $\alg{d}$ defined in eq.~\eqref{eq:vsdecomp}.
The surviving external gauge transformations are now parametrised by $\udsl \in \grp{G}_{\mathrm{diag}}$.
We only consider $\lambda$-type boundary conditions for $c=1$, in which case we can use the external gauge transformations to fix
\unskip\footnote{Note that this gauge fixing is not compatible with reality conditions for $c=i$ since these imply $\hat{g}\vert_{z=r_1} = \hat\tau (\hat{g}\vert_{z=r_2})$ where we recall that $r_2 = \bar r_1$.
For both $c=1$ and $c=i$, we could instead choose to fix $\gdsl \in \grp{G}_{\mathcal{R}}$ if an R-matrix satisfying~\eqref{eq:mcybe} and defining $\alg{g}_\mathcal{R}$ exists.}
\begin{equation}
\hat{g}\vert_{z=r_1} = g_r \in \grp{G} ~, \qquad
\hat{g}\vert_{z=r_2} = 1 ~.
\end{equation}
Again, the same analysis holds for the simple poles at $z_{\mathrm{f.p.}} = 0,\infty$ with $c_0 = 1$ after twisting, except that now we end up with
\begin{equation}\label{eq:fplambdagauge}
\hat{g}\vert_{z=0} = \tilde{g} \in \grp{G}_0 ~, \qquad
\hat{g}\vert_{z=\infty} = 1 ~,
\end{equation}
which is again compatible with $\Integer_N$-equivariance at the fixed points.

\subsection{Internal gauge symmetry}

Finally turning to the internal gauge symmetries, we recall from eq.~\eqref{eq:internal_gauge_symmetry} that after twisting these act as
\begin{equation}\label{eq:igs}
\hat g \to \hat g v ~, \qquad v \in \grp{G}_0 ~.
\end{equation}
where $v$ is $\CP^1$ independent.
The restriction of $v$ to $\grp{G}_0$ follows since it is restricted to be an element of the real form $\grp{G}$ along the real line $\mathfrak{R} \subset \CP^1$ and to be an element of $\grp{G}_0$ at the fixed points of $\varsigma$, $z= z_{\mathrm{f.p.}}$.

If we impose $\eta$-type boundary conditions at $ z_{\mathrm{f.p.}} = 0,\infty$ and the gauge fixing~\eqref{eq:fpetagauge} then the internal gauge transformations~\eqref{eq:igs} preserve this gauge choice and $\tilde g$ transforms as
\begin{equation}
\tilde g \to \tilde g v ~, \qquad \tilde g,v \in \grp{G}_0 ~.
\end{equation}
This means we can use the internal gauge symmetries to fix
\begin{equation} \label{eq:fpetagaugeint}
\tilde g = 1 ~.
\end{equation}

On the other hand, if we impose $\lambda$-type boundary conditions at $ z_{\mathrm{f.p.}} = 0,\infty$ and the gauge fixing~\eqref{eq:fplambdagauge} then the internal gauge transformations~\eqref{eq:igs} do not preserve this gauge choice.
Therefore, we introduce a compensating external gauge transformation
\begin{equation}
(\hat g\vert_{z=0},\hat g\vert_{z=\infty}) \to (v^{-1},v^{-1} )(\hat g\vert_{z=0},\hat g\vert_{z=\infty}) ~.
\end{equation}
This, combined with the internal gauge transformation~\eqref{eq:igs} now preserves the gauge choice $\hat g\vert_{z=\infty} = 1$ and $\tilde g$ transforms as
\begin{equation}\label{eq:gs}
\tilde g \to v^{-1} \tilde g v ~, \qquad \tilde g,v \in \grp{G}_0 ~.
\end{equation}
In the following we leave this unfixed and as a result the twisted $\lambda$-models will have a gauge symmetry.
In addition to acting on $\tilde g$ as in eq.~\eqref{eq:gs}, this gauge symmetry acts on fields $g_r$ associated to Dirichlet boundary conditions at double poles or $\eta$-type boundary conditions at pairs of simple poles as
\begin{equation}\label{eq:intgaugefixbc}
g_r \to g_r v ~,
\end{equation}
and on fields $g_r$ associated to $\lambda$-type boundary conditions at pairs of simple poles as
\begin{equation}\label{eq:intgaugefixbc2}
g_r \to v^{-1} g_r v ~.
\end{equation}

\subsection{Actions}

We conclude this section by giving the general form of the actions of the twisted trigonometric 2d integrable sigma models that follow from the construction outlined above.
The boundary conditions that we consider allow us to use the unifying action constructed from 4d Chern-Simons in~\cite{Delduc:2019whp}.

Defining the Wess-Zumino term
\begin{equation}\begin{split}
\Act_{\text{WZ}}(g) &= \frac{1}{3} \int_{[0,1]\times\Sigma} d^3\sigma \, \epsilon^{ijk} \tr\big( g^{-1}\partial_{i}g[g^{-1}\partial_{j}g, g^{-1}\partial_{k}g] \big) ~,
\end{split}\end{equation}
the general form of the $\eta$-models is given by
\begin{equation}\begin{split}\label{eq:action-etageneral}
\Act_\eta^{(N)}(\hat{g}) &= \sum_{r\in\alg{p}} \bigg(\int_\Sigma d^{2}\sigma \sum_{p_r=0}^{m_r-1} \ell_{r,p_r} \tr\big( (\partial_{z}^{p_r}\Lax_{+})\hat{g}^{-1}\partial_{-}\hat{g} - \hat{g}^{-1}\partial_{+}\hat{g}(\partial_{z}^{p_r}\Lax_{-}) \big)\vert_{z=r} + \ell_{r,0} \, \Act_{\text{WZ}}(\hat{g}\vert_{z=r}) \bigg)~.
\end{split}\end{equation}
Recall that the set of poles $\mathfrak{p}$ does not include the poles at $z_{\mathrm{f.p.}} = 0,\infty$.
There is no contribution from these poles to the $\eta$-model actions since, as a result of the $\eta$-type boundary conditions and the gauge fixings outlined above, we have $\hat{g}\vert_{z=0} = \hat{g}\vert_{z=\infty} = 1$, cf.~eqs.~\eqref{eq:fpetagauge,eq:fpetagaugeint}.

On the other hand, the general form of the $\lambda$-models is given by
\begin{equation}\begin{split}\label{eq:action-lambdageneral}
\Act_\lambda^{(N)}(\hat{g}) &= \sum_{r\in\alg{p}} \bigg(\int_\Sigma d^{2}\sigma \sum_{p_r=0}^{m_r-1} \ell_{r,p_r} \tr\big( (\partial_{z}^{p_r}\Lax_{+})\hat{g}^{-1}\partial_{-}\hat{g} - \hat{g}^{-1}\partial_{+}\hat{g}(\partial_{z}^{p_r}\Lax_{-}) \big)\vert_{z=r} + \ell_{r,0} \, \Act_{\text{WZ}}(\hat{g}\vert_{z=r}) \bigg)
\\ & \quad + \int_\Sigma d^2\sigma \, \ell_0 \tr\big(\Lax_{+}\tilde{g}^{-1}\partial_{-}\tilde{g} - \tilde{g}^{-1}\partial_{+}\tilde{g}\Lax_{-} \big)\vert_{z=0} + \ell_0 \, \Act_{\mathrm{WZ}}(\tilde{g}) ~.
\end{split}\end{equation}
In this case there is a contribution to the $\lambda$-model actions from the pole at $0$ since, as a consequence of the $\lambda$-type boundary conditions and the gauge fixings outlined above, we have $\hat{g}\vert_{z=0} = \tilde g$ and $\hat{g}\vert_{z=\infty} = 1$, cf.~eq.~\eqref{eq:fplambdagauge}.
As explained above, this action has a $\grp{G}_0$ gauge symmetry acting as in eqs.~\eqref{eq:gs,-,eq:intgaugefixbc2}.

Since we are describing the actions of the twisted models, we impose the $\Integer_N$-equivariance conditions
\begin{equation}
\hat{g}\vert_{z=\varsigma(r)} = \hat{\sigma}(\hat{g}\vert_{z=r}) ~, \qquad \tilde g \in \grp{G}_0 ~.
\qquad
\Lax\vert_{z=\varsigma(r)} = \sigma(\Lax\vert_{z=r}) ~,
\end{equation}
in both the actions~\eqref{eq:action-etageneral} and~\eqref{eq:action-lambdageneral}, which is possible since the set of zeroes $\mathfrak{z}$ and poles $\mathfrak{p}$ are both invariant under the $\Integer_N$ action~\eqref{eq:varsigma}.

\section{Construction of the \texorpdfstring{$\Integer_N$}{ZN}-twisted trigonometric sigma models}\label{sec:sigmamodels}

We now turn to the construction of specific $\Integer_N$-twisted trigonometric sigma models.
As summarised in~\secref{sec:general-theory} the models we will construct are:
\begin{itemize}
\item $\Integer_N$-twisted $\eta$-model,
\item $\Integer_N$-twisted $\lambda$-model,
\item YB-deformed $\Integer_2$-twisted $\eta$-model,
\item CC-deformed $\Integer_2$-twisted $\eta$-model,
\item YB-deformed $\Integer_2$-twisted $\lambda$-model,
\item CC-deformed $\Integer_2$-twisted $\lambda$-model.
\end{itemize}
For ease, we will refer to undeformed models, which are the $\Integer_N$-twisted $\eta$-model and $\lambda$-model, and deformed models, which are the YB-deformed and CC-deformed models.

\paragraph{Twist functions.}
For the undeformed models we start from the trigonometric 1-form with a single real double pole
\unskip\footnote{\label{foot:sl2c}The $\grp{SL}(2;\Complex)$ transformation $z \to \frac{z-c_0\eta}{z+c_0\eta}$ where $\eta = \frac{1}{c_0} \tanh \frac{N\alpha}{2}$ brings this $1$-form to the more familiar form $\omega \sim \frac{z^2-1}{z^2-c_0^2\eta^2} dz$ used to construct the $\eta$ deformation, i.e.~the inhomogeneous Yang-Baxter deformation, and $\lambda$ deformation, i.e.~the current-current deformation of the WZW model \cite{Delduc:2019whp}.}
\begin{equation}
\omega^{(1)}_{\mathrm{trig}} = -c_0\hay\frac{(\tilde{z}-e^{N\alpha})(\tilde{z}-e^{-N\alpha})}{(\tilde{z}-1)^{2}} \frac{d\tilde{z}}{\tilde{z}} ~.
\end{equation}
Here we have used the residual $\grp{SL}(2;\Complex)$ freedom~\eqref{eq:residsl2c} to place the double pole at $\tilde z=1$.
Note that $\tilde z=1$ is in $\mathfrak{R}$ for both $c=1$ and $c=i$, hence can be used for both choices.
The two simple zeroes are then only constrained to be real and such that the sum of the residues at $z=0$ and $z=\infty$ vanishes, i.e.~$\lim_{z\to\infty}\varphi_{\mathrm{trig}}(z) = \varphi_{\mathrm{trig}}(0)$.
This leaves a single free constant $\alpha$ parametrising the positions of the zeroes, with $\alpha \in \Real$ for $c_0 = 1$ and $\alpha \in i \Real$ for $c_0 = i$.
Finally, $\hay\in\Real$ parametrises the freedom in the overall normalisation of the 1-form and will eventually play the role of the sigma model coupling.
The twisted trigonometric 1-form on the $N$-fold cover then has $N$ double poles and is given by
\begin{equation}\label{undeftwist}
\omega^{(N)}_{\mathrm{trig}} = -c_0\hay\frac{(z^{N}-e^{N\alpha})(z^{N}-e^{-N\alpha})}{(z^{N}-1)^{2}} \frac{dz}{z} ~.
\end{equation}
Note that this generalises the $N=2$ twist function considered in~\cite{Fukushima:2020kta}.

For the deformed models, we split the double pole at $\tilde z=1$ into two simple poles
\begin{equation}
\omega^{(1)}_{\mathrm{trig}} = -c_0\hay\frac{(\tilde{z}-e^{N\alpha})(\tilde{z}-e^{-N\alpha})}{(\tilde{z}-e^{N\beta})(\tilde{z}-e^{-N\beta})} \frac{d\tilde{z}}{\tilde{z}} ~,
\end{equation}
with deformation parameter $\beta$.
Here we use the residual $\grp{SL}(2;\Complex)$ freedom~\eqref{eq:residsl2c} to set the product of the positions of the two simple poles equal to $\tilde z=1$.
Again this can be used for both $c_0 = 1$ and $c_0 = i$.
The same constraints as for the undeformed 1-form imply that the positions of the zeroes are parametrised in terms of a single parameter $\alpha$, with $\alpha \in \Real$ for $c_0 = 1$ and $\alpha \in i \Real$ for $c_0 = i$.
We consider setups where the simple poles are either real, $c=1$, or complex conjugate $c = i$, such that $\beta \in \Real$ for $c = c_0$ and $\beta \in i \Real$ for $c = \pm i c_0$.
\unskip\footnote{For $c_0 = 1$ the real line $\mathfrak{R} = \Real$.
For real simple poles at $e^{\pm N \beta}$ we have $c=1$ and $\beta \in \Real$, while for complex conjugate simple poles, $c=i$ and $\beta \in i \Real$.
On the other hand, for $c_0 = i$, the real line $\mathfrak{R} = \Unit$.
For real simple poles we again have $c=1$, but now $\beta \in i\Real$, while for complex conjugate simple poles, $c=i$ and $\beta \in \Real$.}
The reality conditions on parameters are summarised in \tabref{tab1}.
\begin{table}
\centering
\begin{tabular}{|c|c||c|c|}
\hline
$c_0 =$ & $c =$ & $\alpha \in$ & $\beta \in$ \\ \hline\hline
$i$ & $i$ & $i\Real$ & $\Real$ \\ \hline
$i$ & $1$ & $i\Real$ & $i\Real$ \\ \hline
$1$ & $i$ & $\Real$ & $i\Real$ \\ \hline
$1$ & $1$ & $\Real$ & $\Real$ \\ \hline
\end{tabular}
\caption{The relation between the values of $c_0$ and $c$ and the reality conditions for $\alpha$ and $\beta$.}\label{tab1}
\end{table}
The twisted trigonometric 1-form on the $N$-fold cover then has $2N$ simple poles and is given by
\begin{equation}\label{deftwist}
\omega^{(N)}_{\mathrm{trig}} = -c_0\hay \frac{(z^{N}-e^{N\alpha})(z^{N}-e^{-N\alpha})}{(z^{N}-e^{N\beta})(z^{N}-e^{-N\beta})} \frac{dz}{z} ~.
\end{equation}

\paragraph{Zeroes and poles.}
The meromorphic 1-form in both cases has a set of simple zeroes $\mathfrak{z}$, which we split into two disjoint subsets $\mathfrak{z}_\pm$.
The set of zeroes $\mathfrak{z}$ and its subsets $\mathfrak{z}_\pm$ are
\begin{equation}
\mathfrak{z} = \mathfrak{z}_+ \cup \mathfrak{z}_- ~, \qquad \mathfrak{z}_\pm = \left\{ e^{\mp\alpha+\frac{2i\pi a}{N}} \ : \ a\in\Integer_{N} \right\} ~,
\end{equation}
where the set $\Integer_N = \{0,\dots,N-1\}$.
Each pair of simple zeroes are separated by a distance $2 \sinh\alpha$ around the roots of unity $z=e^{\frac{2i\pi a}{N}}$ for $a\in\Integer_{N}$.
From eq.~\eqref{eq:lax:lc} it follows that a general ansatz for the Lax connection of our models is
\begin{equation}\label{eq:laxconnection:zn}
\Lax_{\pm}^{(N)}(z) = \tilde{V}_{\pm} + e^{-\frac{1\mp1}{2}N\alpha} \sum_{a\in\Integer_{N}} \frac{V_{\pm}^{(a)}z^{a}}{e^{\pm N\alpha}z^{N}-1} ~,
\end{equation}
where $\tilde{V}_{\pm}$ and $V_{\pm}^{(a)}$ are $\alg{g}^{\Complex}$-valued fields on $\Sigma$ that are independent of $z$.
To be compatible with the $\Integer_N$-equivariance condition~\eqref{eq:fieldtwisting} we expect to find that $\tilde{V}_\pm\in \alg{g}_0$ and $V_\pm^{(a)} \in \alg{g}_a$, the eigenspace of the $\Integer_N$ automorphism $\sigma$ with eigenvalue $e^{\frac{2i\pi a}{N}}$, after implementing the $\Integer_N$-twisting.

In both cases there are simple poles at $0$ and $\infty$, and we define the set $\mathfrak{p}_0 = \{0,\infty\}$.
In undeformed case we also have a set of double poles $\mathfrak{p}_2$
\begin{equation}\label{poles-undef}
\mathfrak{p} = \mathfrak{p}_{2} ~, \qquad \mathfrak{p}_{0} = \{0,\infty\} ~, \qquad \mathfrak{p}_{2} = \left\{ e^{\frac{2i\pi a}{N}} \ : \ a\in\Integer_{N} \right\} ~,
\end{equation}
and in the deformed case a set of simple poles $\mathfrak{p}_1$
\begin{equation}\label{poles-def}
\mathfrak{p} = \mathfrak{p}_1 ~, \qquad \mathfrak{p}_{0} = \{0,\infty\} ~, \qquad \mathfrak{p}_1 = \left\{ e^{\pm\beta+\frac{2i\pi a}{N}} \ : \ a\in\Integer_{N} \right\} ~.
\end{equation}
Again the simple poles in $\mathfrak{p}_1$ are separated by a distance $2\sinh\beta$ around the roots of unity $z=e^{\frac{2i\pi a}{N}}$ for $a \in \Integer_N$.

\paragraph{Boundary terms.}
The levels $\ell_{0}$ and $\ell_{\infty}$ associated with the simple poles at $0$ and $\infty$ are
\begin{equation}
\ell_{0} = -\ell_{\infty} = -c_0\hay ~,
\end{equation}
The levels associated with the $N$ double poles in the undeformed case are
\begin{equation}\label{eq:levels:undef}
\ell_{r,0} = 0 ~, \qquad \ell_{r,1} = -\frac{c_0\hay r}{N^{2}} (e^{N\alpha}-1)(e^{-N\alpha}-1) ~, \qquad r \in \left\{ e^{\frac{2i\pi a}{N}} \ : \ a\in\Integer_{N} \right\} ~,
\end{equation}
while those associated to the $2N$ simple poles in the deformed case are
\begin{equation}\label{eq:levelsdef}
\ell_{r_{\pm},0} = \mp \frac{c_0\hay}{N} \frac{(e^{N(\beta+\alpha)}-1)(e^{N(\beta-\alpha)}-1)}{e^{2N\beta}-1} ~, \qquad r_{\pm} \in \left\{ e^{\pm\beta+\frac{2i\pi a}{N}} \ : \ a\in\Integer_{N} \right\} ~.
\end{equation}
Therefore, in the undeformed case the total boundary term~\eqref{eq:btdoublepole} is
\begin{equation}
\begin{aligned}
-c_0 \hay \int_{\Sigma} d^{2}\sigma & \, \Big(\tr(A\delta A)\vert_{z=0} - \tr(A\delta A)\vert_{z=\infty} \\
& \qquad\qquad - \frac{4}{N^{2}} \sinh^{2}(\frac{N\alpha}{2}) \sum_{a\in\Integer_N} e^{\frac{2i\pi a}{N}} \partial_{z} \tr(A \delta A)\vert_{z=e^{\frac{2i\pi a}{N}}} \Big) ~, \label{eq:totaldefect:undef} \\
\end{aligned}
\end{equation}
while in the deformed case it is given by
\begin{equation}\label{eq:totaldefect:def}
\begin{aligned}
-c_0 \hay & \int_{\Sigma} d^{2}\sigma \, \Big( \tr(A\delta A)\vert_{z=0} - \tr(A\delta A)\vert_{z=\infty} \\
& \qquad \quad \ \ \,
- \frac{2}{N} \frac{\sinh\frac{N(\alpha+\beta)}{2} \sinh\frac{N(\alpha-\beta)}{2}}{\sinh(N\beta)} \sum_{a\in\Integer_N} \big( \tr(A\delta A)\vert_{z=e^{\beta+\frac{2i\pi a}{N}}} - \tr(A\delta A)\vert_{z=e^{-\beta+\frac{2i\pi a}{N}}} \big) \Big) ~.
\end{aligned}
\end{equation}

\paragraph{Fields and twisting.}
Following on from the discussions in \secsref{sec:general-theory}{sec:bcgs}, after fixing the external gauge symmetries, but before fixing the internal gauge symmetries and before the $\Integer_N$-twisting, the fields of the 2d integrable sigma models will be $N+1$ $\grp{G}$-valued fields.
We denote the field associated with the poles at $0$ and $\infty$ by $\tilde{g}$ and the remaining $N$ fields, associated to the $N$ $\Integer_N$-equivariant double poles or pairs of simple poles, by $g_a$, $a \in \Integer_N$.
We also define left-invariant Maurer–Cartan 1-forms associated with these group-valued fields as
\begin{equation}\label{j_in_terms_of_g}
\tilde{\jmath} = \tilde{g}^{-1}d\tilde{g} ~, \qquad j_{a} = g_{a}^{-1}dg_{a} ~, \qquad a \in \Integer_N ~.
\end{equation}
It is useful to also introduce the following rotated 1-forms
\begin{equation}\label{J_in_terms_of_j}
J^{(a)} = \frac{1}{N} \sum_{b\in\Integer_{N}} e^{-\frac{2i\pi ab}{N}} j_{b} ~, \qquad a\in\Integer_{N} ~.
\end{equation}

To implement the $\Integer_N$-twisting, we first recall that
\begin{equation}
\tilde g \in \grp{G}_0 ~, \qquad \tilde\jmath \in \alg{g}_0 ~.
\end{equation}
For the remaining fields we have
\unskip\footnote{While $\sigma^a(\alg{g})$ is automorphic to $\alg{g}$ by definition, they are only identical subalgebras of $\alg{g}^\Complex$ when $[\tau,\sigma] = 0$, which is the condition we demand for $c_0 = i$~\eqref{eq:realc0i}.
For $c_0 = 1$ the subalgebras are not identical with $\tau\sigma = \sigma^{-1} \tau$.}
\begin{equation}\label{eq:gtwisting}
g_a = \hat\sigma^a (g_0) \in \hat\sigma^a(\grp{G}) ~, \qquad j_a = \sigma^a (j_0) \in \sigma^a(\alg{g}) ~.
\end{equation}
This leaves one $\grp{G}_0$-valued field, $\tilde g$, and one $\grp{G}$-valued field, $g_0$, which, accounting for the $\grp{G}_0$ internal gauge symmetries that have not yet been fixed, implies that the $\Integer_N$-twisted trigonometric sigma models all have $\dim \grp{G}$ degrees of freedom.

Substituting eq.~\eqref{eq:gtwisting} into eq.~\eqref{J_in_terms_of_j} we find
\unskip\footnote{$P_a(\alg{g})$ is only a subspace of $\alg{g}$ if $[\tau,P_a] = 0$, which is the condition we demand for $c_0 = 1$~\eqref{eq:realc01}.
For $c_0 = i$ we have $P_a(\alg{g}) \subset \alg{g}^\Complex$ with $\tau P_a = P_{N-a}\tau$.}
\begin{equation}\label{eq:projjj}
J^{(a)} = \frac{1}{N} \sum_{b\in\Integer_{N}} e^{-\frac{2i\pi ab}{N}} \sigma^b(j_0) = P_a j_0 \in P_a(\alg{g}) ~, \qquad a\in\Integer_N ~,
\end{equation}
where $P_a$ is the projector onto the eigenspace $\alg{g}_a$ of the $\Integer_N$ automorphism $\sigma$ with eigenvalue $e^{\frac{2i\pi a}{N}}$ defined in eq.~\eqref{eq:pa}.
From now on, we take $J^{(a)}$ to denote the projection of the Maurer-Cartan 1-form $j_0$ as defined in eq.~\eqref{eq:projjj} and drop the index $0$ on both the 1-form $j_0$ and the field $g_0$.

\subsection{\texorpdfstring{$\Integer_{N}$}{ZN}-twisted \texorpdfstring{$\eta$}{η}-model and \texorpdfstring{$\lambda$}{λ}-model}

Let us now construct the $\Integer_N$-twisted $\eta$-model and $\lambda$-model starting from the twisted trigonometric 1-form~\eqref{undeftwist}.
As outlined in \secsref{sec:general-theory}{sec:bcgs} we impose Dirichlet boundary conditions at the double pole at $z = 1$ and use the external gauge symmetries to set $\partial_z \hat g\vert_{z = 1} = 0$.
The boundary conditions and the fixing of the external gauge symmetries at the remaining double poles in $\mathfrak{p}_2$ are then fixed by $\Integer_N$-equivariance
\begin{equation}\label{eq:undef:bc:A:dp}
A\vert_{z=e^{\frac{2i\pi a}{N}}} = 0 ~, \qquad \partial_z \hat g\vert_{z=e^{\frac{2i\pi a}{N}}} = 0 ~, \qquad a\in\Integer_N ~.
\end{equation}
For the $\eta$-model we impose the $\eta$-type boundary conditions~\eqref{eq:bcetac0} at $z=z_{\mathrm{f.p.}}$, while for the $\lambda$-model we impose the $\lambda$-type boundary conditions~\eqref{eq:bclambdac0}.
We also recall that for the $\eta$-type boundary conditions at $z=z_{\mathrm{f.p.}}$ we use the external and internal gauge symmetries to set $\hat{g}\vert_{z=0} = \hat{g}\vert_{z=1} = 1$, while for the $\lambda$-type boundary conditions we use the external gauge symmetries to set $\hat{g}\vert_{z=0} = \tilde g \in \grp{G}_0$ and $\hat{g}\vert_{z=1} = 1$, with the internal gauge symmetries remaining unfixed.
To summarise, we have the following field content
\begin{equation}
\begin{aligned}
\textrm{$\Integer_{N}$-twisted $\eta$-model:} \quad &&& \hat{g}\vert_{z=0} = 1~, \qquad && \hspace{-60pt} \hat{g}\vert_{z=\infty} = 1~,
\\ &&& \hat{g}\vert_{z=e^{\frac{2i\pi a}{N}}} = \hat\sigma^a (g) \in \hat\sigma^a(\grp{G}) ~, \qquad && a \in \Integer_{N} ~, \\
\textrm{$\Integer_{N}$-twisted $\lambda$-model:} \quad &&& \hat{g}\vert_{z=0} = \tilde{g} \in \grp{G}_0 ~, \qquad && \hspace{-60pt} \hat{g}\vert_{z=\infty} = 1 ~,
\\ &&& \hat{g}\vert_{z=e^{\frac{2i\pi a}{N}}} = \hat\sigma^a(g) \in \hat\sigma^a(\grp{G}) ~, \qquad && a \in \Integer_{N} ~.
\end{aligned}
\end{equation}

\paragraph{Lax connections.}
Substituting~\eqref{eq:Laxparam} into the boundary conditions and using the ansatz for the Lax connection~\eqref{eq:laxconnection:zn}, we find the following equations for the undetermined fields in the Lax connection
\begin{equation}\label{etalamp0}
\begin{aligned}
&\textrm{$\eta$-type at $\mathfrak{p}_{0}$:} \qquad && (\tilde{\mathcal{R}}+c_0) \left(\tilde{V}_{\pm}-e^{-\frac{1\mp1}{2}N\alpha}V_{\pm}^{(0)}\right) = (\tilde{\mathcal{R}}-c_0) \tilde{V}_{\pm} ~, \\
&\text{$\lambda$-type at $\mathfrak{p}_{0}$:} \qquad && \Ad_{\tilde{g}}\left(\tilde{V}_{\pm} -e^{-\frac{1\mp1}{2}N\alpha}V_{\pm}^{(0)}-\tilde{\jmath}_{\pm}\right) = \tilde{V}_{\pm} ~,
\end{aligned}
\end{equation}
\begin{equation}
\text{Dirichlet at $\mathfrak{p}_{2}$:} \qquad\quad \tilde{V}_{\pm} \pm \sum_{b\in\Integer_{N}} \frac{V_{\pm}^{(b)}e^{\frac{2i\pi ab}{N}}}{e^{N\alpha}-1} = \sigma^a(j_{\pm})~, \quad a \in \Integer_{N} ~,
\end{equation}

Solving these equations for the $\Integer_N$-twisted $\eta$-model gives
\begin{equation} \begin{aligned}
V_{\pm}^{(0)} &= \pm\frac{2e^{\frac{1\mp1}{2}N\alpha} \tanh\left(\frac{N\alpha}{2}\right)}{1\pm\frac{1}{c_0}\tanh\left(\frac{N\alpha}{2}\right)\tilde{\mathcal{R}}} J_{\pm}^{(0)} = \pm 2c_0\eta\frac{1+c_0\eta}{1\pm c_0\eta} \frac{1}{1\pm\eta\tilde{\mathcal{R}}} J_{\pm}^{(0)} ~, \\
V_{\pm}^{(a)} &= \pm(e^{N\alpha}-1)J_{\pm}^{(a)} = \pm\frac{2c_0\eta}{1-c_0\eta} J_{\pm}^{(a)} ~, \qquad a \in \Integer_N\backslash\{0\} ~, \\
\tilde{V}_{\pm} &= \pm\frac{1}{c_0}\frac{\tanh\left(\frac{N\alpha}{2}\right) (c_0+\tilde{\mathcal{R}})}{1\pm\frac{1}{c_0}\tanh\left(\frac{N\alpha}{2}\right)\tilde{\mathcal{R}}} J_{\pm}^{(0)} = \pm\frac{\eta(c_0+\tilde{\mathcal{R}})}{1\pm\eta\tilde{\mathcal{R}}} J_{\pm}^{(0)} ~,
\end{aligned}\end{equation}
where
\begin{equation}
\eta = \frac{1}{c_0} \tanh\Big(\frac{N\alpha}{2}\Big) ~.
\end{equation}
Substituting these expressions into the ansatz~\eqref{eq:laxconnection:zn}, we find that the Lax connection for $\Integer_{N}$-twisted $\eta$-model is
\begin{equation}\label{eq:laxetatwist}
\Lax_{\eta\pm}^{(N)} = \Big(c_0+\frac{2c_0}{\frac{1\pm c_0\eta}{1\mp c_0\eta}z^{N}-1} + \tilde{\mathcal{R}} \Big) \frac{\pm\eta}{1\pm\eta\tilde{\mathcal{R}}} J_{\pm}^{(0)}
\pm \frac{2c_0\eta}{1\mp c_0\eta} \frac{1}{\frac{1\pm c_0\eta}{1\mp c_0\eta}z^{N}-1} \sum_{a\in\Integer_N\backslash\{0\}} z^{a} J_{\pm}^{(a)} ~.
\end{equation}

On the other hand, solving the equations for the $\Integer_{N}$-twisted $\lambda$-model gives
\begin{equation}
\begin{aligned}
V_{\pm}^{(0)} &= \pm \frac{e^{N\alpha}-1}{e^{\pm N\alpha}-\Ad_{\tilde{g}}^{-1}} \big( (1-\Ad_{\tilde{g}}^{-1})J_{\pm}^{(0)} - \tilde{\jmath}_{\pm} \big) = \pm \frac{\lambda-1}{\lambda^{\pm1}-\Ad_{\tilde{g}}^{-1}} \big( (1-\Ad_{\tilde{g}}^{-1})J_{\pm}^{(0)} - \tilde{\jmath}_{\pm} \big) ~, \\
V_{\pm}^{(a)} &= \pm(e^{N\alpha}-1)J_{\pm}^{(a)} = \pm(\lambda-1)J_{\pm}^{(a)} ~, \qquad a \in \Integer_N\backslash\{0\} ~, \\
\tilde{V}_{\pm} &= \frac{1}{e^{\pm N\alpha}-\Ad_{\tilde{g}}^{-1}} \big((e^{\pm N\alpha}-1)J_{\pm}^{(0)}+\tilde{\jmath}_{\pm}\big) = \frac{1}{\lambda^{\pm1}-\Ad_{\tilde{g}}^{-1}} \big((\lambda^{\pm1}-1)J_{\pm}^{(0)}+\tilde{\jmath}_{\pm}\big) ~.
\end{aligned}
\end{equation}
where
\begin{equation}
\lambda = e^{N\alpha} ~.
\end{equation}
As for the $\eta$-model, substituting these expressions into the ansatz~\eqref{eq:laxconnection:zn}, we find that the Lax connection for $\Integer_{N}$-twisted $\lambda$-model is
\begin{equation}\begin{split}\label{eq:laxlambdatwist}
\Lax_{\lambda\pm}^{(N)} & = \frac{z^{N}-1}{z^{N}-\lambda^{\mp1}} \frac{1}{\lambda^{\pm1}-\Ad_{\tilde{g}}^{-1}} \tilde{\jmath}_{\pm} - \frac{\lambda^{\mp1}-1}{z^{N}-\lambda^{\mp1}} \Big( 1 + \lambda^{\pm1}\frac{z^{N}-1}{\lambda^{\pm1}-\Ad_{\tilde{g}}^{-1}} \Big) J_{\pm}^{(0)}
\\ & \quad
- (\lambda^{\mp1}-1) \sum_{a\in\Integer_N\backslash\{0\}} \frac{z^{a}}{z^{N}-\lambda^{\mp1}} J_{\pm}^{(a)} ~.
\end{split}\end{equation}

\paragraph{Actions.}
To find the action of the $\Integer_N$-twisted $\eta$-model we substitute the Lax connection that we have constructed in eq.~\eqref{eq:laxetatwist} and the levels~\eqref{eq:levels:undef} into the general form of the action for the twisted $\eta$-models~\eqref{eq:action-etageneral} to give
\begin{equation}\label{eq:actionetatwist}
\Act_{\eta}^{(N)}(g) = -\frac{4c_0^2\hay\eta}{1-c_0^{2}\eta^{2}} \int d^{2}\sigma \, \tr\Big( J_{+}^{(0)}\frac{1-c_0^{2}\eta^{2}}{1-\eta\tilde{\mathcal{R}}}J_{-}^{(0)} + \sum_{a\in\Integer_N\backslash\{0\}} \Big(1+c_0\eta \Big(1-\frac{2a}{N}\Big) \Big) J_{+}^{(a)}J_{-}^{(N-a)} \Big) ~.
\end{equation}
Note that there is no Wess-Zumino term in the $\Integer_{N}$-twisted $\eta$-model since the residues at the double poles vanish, i.e.~$\ell_{r,0} = 0$ for $r=e^{\frac{2i\pi a}{N}}$ and we have gauge fixed $\hat{g}\vert_{0} = \hat{g}\vert_{\infty}=1$.
Note that rescaling $\hay \to \eta^{-1} \hay$ and taking $\eta \to 0$, we find that the dependence on $N$ drops out and we recover the principal chiral model.
\begin{equation}
\Act_{\mathrm{PCM}} = -4c_0^2 \hay \int d^2\sigma \, \tr\big(j_+ j_-\big) ~.
\end{equation}
This may be expected since the twisting is closely tied to the trigonometric nature of the model.
Let us briefly comment that for $c_0 = 1$ the action~\eqref{eq:actionetatwist} is real if $\tau(J_\pm^{(a)}) = J_\pm^{(a)}$, which corresponds to the condition~\eqref{eq:realc01}, while for $c_0 = i$ we require that $\tau(J_\pm^{(a)}) = \tau(J_\pm^{(N-a)})$, corresponding to eq.~\eqref{eq:realc0i}.

The action of the $\Integer_N$-twisted $\lambda$-model is similarly found by substituting the Lax connection~\eqref{eq:laxlambdatwist} and the levels~\eqref{eq:levels:undef} into the general form of the action for the twisted $\lambda$-models~\eqref{eq:action-lambdageneral} to give
\begin{equation}\begin{split}\label{eq:actionlambdatwist}
\Act_{\lambda}^{(N)}(g,\tilde g) &= -2\hay\int d^{2}\sigma \, \bigg( \tr\Big(\big( (1-\Ad_{\tilde{g}}^{-1})J_{+}^{(0)} - \tilde\jmath_{+} \big) \frac{1}{1-\lambda^{-1}\Ad_{\tilde{g}}} \big((1-\Ad_{\tilde{g}}^{-1}) J_{-}^{(0)} - \tilde\jmath_{-} \big)\Big)
\\ & \qquad \qquad \qquad \quad + \tr\Big( \sum_{a\in\Integer_N\backslash\{0\}} (\lambda-1) \big(1+\frac{a}{N}(\lambda^{-1}-1)\big) J_{+}^{(a)}J_{-}^{(N-a)} \Big) \bigg)
\\ & \quad + 2\hay \int d^{2}\sigma \, \tr\Big( \frac12 \tilde\jmath_{+}\tilde\jmath_{-} - J_{+}^{(0)}\tilde\jmath_{-} + \tilde\jmath_{+}\Ad_{\tilde{g}}^{-1}J_{-}^{(0)} + J_{+}^{(0)}(1-\Ad_{\tilde{g}}^{-1})J_{-}^{(0)}\Big) - \hay \Act_{\mathrm{WZ}}(\tilde g) ~,
\end{split}\end{equation}
where we recall that our gauge fixings for the $\lambda$-type boundary conditions require $c_0 = 1$, hence for reality we have that $\tau(J_\pm^{(a)}) = J_\pm^{(a)}$.
In this case the Wess-Zumino term survives since we have only gauge fixed $\hat{g}\vert_{\infty}=1$, while $\hat{g}\vert_{0}=\tilde g$.
Correspondingly, this action is invariant under the internal gauge symmetries, which from eqs.~\eqref{eq:gs,eq:intgaugefixbc} act as
\unskip\footnote{This is manifest since it is straightforward to check that $\tilde\jmath \to \Ad_v^{-1}\big( \tilde \jmath + (1-\Ad_{\tilde g}^{-1}) dv v^{-1}\big)$, $J^{(0)} \to \Ad_{v}^{-1} \big(J^{(0)} + d v v^{-1}\big)$, and $J^{(a)} \to \Ad_v^{-1} J^{(a)}$ for $a \in \Integer_N\backslash\{0\}$, which means the first two lines of the action~\eqref{eq:actionlambdatwist} are invariant.
For the final line, we note that this take the form of a gauged WZW action~\cite{Witten:1991mm,Figueroa-OFarrill:1994vwl} with $J^{(0)}$ playing the role of the gauge field.}
\begin{equation}
g\to g v ~, \qquad \tilde g \to v^{-1} \tilde g v ~, \qquad v \in \grp{G}_0 ~.
\end{equation}
Taking $\lambda \to 1$ in the $\Integer_N$-twisted $\lambda$-model, which corresponds to $\eta \to 0$, requires simultaneously taking $\tilde g \to 1$ in a non-abelian dual type limit~\cite{Sfetsos:2013wia,Borsato:2024alk} and gives the non-abelian dual with respect to $\grp{G}_0$ of the principal chiral model.

\paragraph{Untwisted and $N=2$ models.} The actions of the $\Integer_N$-twisted $\eta$-model~\eqref{eq:actionetatwist} and $\lambda$-model~\eqref{eq:actionlambdatwist} are known for $N=1$ and $N=2$.
For $N=1$ the untwisted $\eta$-model action is
\begin{equation}
\Act_{\eta}^{(1)}(g) = -4c_0^2\hay\eta \int d^{2}\sigma \, \tr\Big( j_{+}\frac{1}{1-\eta\tilde{\mathcal{R}}}j_{-} \Big) ~,
\end{equation}
which is the action of the inhomogeneous Yang-Baxter deformation of the principal chiral model~\cite{Klimcik:2002zj}.
The Lax connection that follows from eq.~\eqref{eq:laxetatwist} is
\unskip\footnote{Implementing the $\grp{SL}(2;\Complex)$ transformation from \foottref{foot:sl2c}, this Lax connection becomes $\Lax_{\eta\pm}^{(1)} = j_\pm - \frac{1 - c_0^2 \eta^2}{(1\mp z)(1\pm \eta \tilde{\mathcal{R})}} j_\pm$, which is related to the usual Lax connection of the inhomogeneous Yang-Baxter deformation of the principal chiral model built from a flat conserved current, $\tilde{\Lax}_{\eta\pm}^{(1)} = - \frac{1 - c_0^2 \eta^2}{(1\mp z)(1\pm \eta \tilde{\mathcal{R}}_{g^{-1}})} \partial_\pm g g^{-1}$ where $\tilde{\mathcal{R}}_{g^{-1}} = \Ad_g \tilde{\mathcal{R}} \Ad_{g}^{-1}$, by a gauge transformation $\tilde{\Lax}_{\eta\pm}^{(1)} = \Ad_g (\Lax_{\eta\pm}^{(1)} - j_\pm)$.
Correspondingly, to recover this form of the Lax connection directly from 4d Chern-Simons, we use the alternative gauge fixing $\hat{g}\vert_{z=0} = \hat{g}\vert_{z=\infty} = g^{-1}$ and $\hat{g}\vert_{z=1} = 1$.}
\begin{equation}
\Lax_{\eta\pm}^{(1)} = \Big(c_0 + \frac{2c_0}{\frac{1\pm c_0 \eta}{1\mp c_0 \eta} z -1} + \tilde{\mathcal{R}} \Big) \frac{\pm \eta}{1\pm \eta \tilde{\mathcal{R}}} j_\pm
= \frac{\tilde{\mathcal{R}} - i c_0 \cot\frac{w\pm \nu}{2}}{\tilde{\mathcal{R}} \mp i c_0 \cot\frac\nu2 }j_\pm ~,
\end{equation}
where the second form is written in terms of the trigonometric coordinate $w$~\eqref{eq:maptrigplane} and $\eta = \frac{1}{c_0}\tanh\frac\alpha2 = \frac{i}{c_0} \tan \frac\nu2$.
For $N=2$ the $\Integer_2$-twisted $\eta$-model action was first constructed in~\cite{Fukushima:2020kta} and is given by
\begin{equation}
\Act_{\eta}^{(2)}(g) = -\frac{4c_0^2\hay\eta}{1-c_0^{2}\eta^{2}} \int d^{2}\sigma \, \tr\Big(J_{+}^{(0)} \frac{1-c_0^{2}\eta^{2}}{1-\eta\tilde{\mathcal{R}}} J_{-}^{(0)} + J_{+}^{(1)} J_{-}^{(1)} \Big) ~.
\end{equation}
In this case the Lax connection is
\unskip\footnote{Note that for the $\grp{SU}(2)$ models discussed in \secref{sec:examples} we have $P_0 \mathcal{\tilde{R}} = \mathcal{\tilde{R}} P_0 = 0$.
Therefore, the dependence on $\mathcal{\tilde{R}}$ drops out and we recover the familiar form of the Lax connection for the trigonometric deformation of the $\grp{SU}(2)$ principal chiral model~\cite{Cherednik:1981df}.}
\begin{equation}\begin{split}
\Lax_{\eta\pm}^{(2)} & = \Big(c_0 + \frac{2c_0}{\frac{1\pm c_0 \eta}{1\mp c_0 \eta} z^2 -1} + \tilde{\mathcal{R}} \Big) \frac{\pm \eta}{1\pm \eta \tilde{\mathcal{R}}} J^{(0)}_\pm \pm \frac{2c_0\eta}{1\mp c_0\eta} \frac{z}{\frac{1\pm c_0 \eta}{1\mp c_0 \eta} z^2 -1} J_\pm^{(1)}
\\ & = \frac{\tilde{\mathcal{R}} - i c_0 \cot (w\pm \nu)}{\tilde{\mathcal{R}} \mp i c_0 \cot\nu }J^{(0)}_\pm \pm \frac{\sin\nu}{\sin(w\pm \nu)} J^{(1)}_\pm ~.
\end{split}\end{equation}
For $c_0 = i$ we have $\alpha \in i\Real$, hence $\nu \in \Real$ and the Lax connections for both $N=1$ and $N=2$ are real for $w \in \Real$.
On the other hand, for $c_0 = 1$ we have $\alpha \in \Real$, the Lax connections are real for $w \in i \Real$, and are more naturally written in terms of hyperbolic functions.

For $N=1$ the untwisted $\lambda$-model action is
\begin{equation}
\Act_{\lambda}^{(1)}(\tilde g) = -2\hay \int d^{2}\sigma \, \tr\Big(\tilde\jmath_{+} \frac{1}{1-\lambda^{-1}\Ad_{\tilde{g}}} \tilde\jmath_{-} \Big) + \hay \int d^{2}\sigma \, \tr\big(\tilde\jmath_{+}\tilde\jmath_{-}\big) - \hay \Act_{\mathrm{WZ}}(\tilde g) ~,
\end{equation}
where we have set $g = 1$ using the internal gauge symmetries since for $N=1$ we have $\grp{G}_0 = \grp{G}$.
This also means that the field $\tilde g \in \grp{G}$.
This is the action of the current-current deformation of the WZW model~\cite{Sfetsos:2013wia}.
For $N=2$ the $\Integer_2$-twisted $\lambda$-model action was first constructed in~\cite{Borsato:2024alk} and is given by
\begin{equation}\begin{split}
\Act_{\lambda}^{(2)}(g,\tilde g) &= -2\hay\int d^{2}\sigma \, \bigg( \tr\Big(\big( (1-\Ad_{\tilde{g}}^{-1})J_{+}^{(0)} - \tilde\jmath_{+} \big) \frac{1}{1-\lambda^{-1}\Ad_{\tilde{g}}} \big((1-\Ad_{\tilde{g}}^{-1}) J_{-}^{(0)} - \tilde\jmath_{-} \big)\Big)
\\ & \qquad \qquad \qquad \quad + \tr\Big(\frac{1}{2}(\lambda-1)(\lambda^{-1}+1) J_{+}^{(1)}J_{-}^{(1)} \Big) \bigg)
\\ & \quad + 2\hay \int d^{2}\sigma \, \tr\Big( \frac12 \tilde\jmath_{+}\tilde\jmath_{-} - J_{+}^{(0)}\tilde\jmath_{-} + \tilde\jmath_{+}\Ad_{\tilde{g}}^{-1}J_{-}^{(0)} + J_{+}^{(0)}(1-\Ad_{\tilde{g}}^{-1})J_{-}^{(0)}\Big) - \hay \Act_{\mathrm{WZ}}(\tilde g) ~.
\end{split}\end{equation}
The Lax connections following from \eqref{eq:laxlambdatwist} are
\unskip\footnote{Again implementing the $\grp{SL}(2;\Complex)$ transformation from \foottref{foot:sl2c}, which in terms of $\lambda$ and for $c_0 = 1$ is given by $z \to \frac{(1+\lambda)z + (1-\lambda)}{(1+\lambda)z - (1-\lambda)}$, the Lax connection for the untwisted $\lambda$-model in the gauge $g=1$ becomes the usual one for the current-current deformation of the WZW model built from a flat conserved current, $\Lax_{\lambda\pm}^{(1)} = \frac{2}{1+\lambda^{\pm1}}\frac{1}{1\mp z} \frac{1}{1-\lambda^{\mp 1}\Ad_{\tilde g}^{-1}}\tilde\jmath_\pm$~\cite{Sfetsos:2013wia}.}
\begin{equation}
\begin{split}
\Lax_{\lambda\pm}^{(1)} & = \frac{z-1}{z-\lambda^{\mp1}} \frac{1}{\lambda^{\pm1}-\Ad_{\tilde{g}}^{-1}} \tilde{\jmath}_{\pm}
=\frac{\sinh\frac{iw}{2}\csch\frac{iw\pm\alpha}{2}}{e^{\pm\frac{\alpha}{2}} - e^{\mp\frac{\alpha}{2}} \Ad_{\tilde g}^{-1}}\tilde{\jmath}_{\pm} ~,
\\
\Lax_{\lambda\pm}^{(2)} & = \frac{z^2-1}{z^2-\lambda^{\mp1}} \frac{1}{\lambda^{\pm1}-\Ad_{\tilde{g}}^{-1}} \tilde{\jmath}_{\pm} - \frac{\lambda^{\mp1}-1}{z^2-\lambda^{\mp1}} \Big(\big( 1 + \lambda^{\pm1}\frac{z^2-1}{\lambda^{\pm1}-\Ad_{\tilde{g}}^{-1}} \big) J_{\pm}^{(0)} + z J_{\pm}^{(1)}\Big) ~,
\\ &
=\frac{\sinh iw \csch(iw\pm\alpha)}{e^{\pm\alpha} - e^{\mp\alpha} \Ad_{\tilde g}^{-1}}\tilde{\jmath}_{\pm}
\pm \frac{\sinh\alpha}{\sinh(iw \pm \alpha)} \Big(\frac{e^{iw\pm\alpha} - e^{-iw\mp\alpha} \Ad_{\tilde g}^{-1}}{e^{\pm\alpha} - e^{\mp\alpha} \Ad_{\tilde g}^{-1}} J_\pm^{(0)} + J_\pm^{(1)} \Big)~,
\end{split}
\end{equation}
where for $N=1$ we have again gauge fixed $g=1$.
Since $c_0 = 1$ for the twisted $\lambda$-models, $\alpha \in \Real$ and we have written the Lax connections in terms of hyperbolic functions, explicitly showing that they are real for $w \in i\Real$.

\subsection{Deformed \texorpdfstring{$\Integer_{N}$}{ZN}-twisted \texorpdfstring{$\eta$}{η}-models and \texorpdfstring{$\lambda$}{λ}-models}\label{sec:double-deformation}

To construct the deformed $\Integer_N$-twisted $\eta$-models and $\lambda$-models we consider the twisted trigonometric 1-form~\eqref{deftwist} with either $\eta$-type or $\lambda$-type boundary conditions at the simple poles $z=\{e^{\beta},e^{-\beta}\}$, and with the boundary conditions and gauge fixings at the remaining simple poles in $\mathfrak{p}_1$ fixed by $\Integer_N$-equivariance
\begin{equation}\label{eq:def:bc:A}
\begin{aligned}
& \text{$\eta$-type at $z \in \mathfrak{p}_1$} \qquad && (\mathcal{R}_{a}+c)A_{\pm}\vert_{z=e^{\beta+\frac{2i\pi a}{N}}} = (\mathcal{R}_{a}-c)A_{\pm}\vert_{z=e^{-\beta+\frac{2i\pi a}{N}}} ~, && \qquad a\in \Integer_N ~,
\\
& \text{$\lambda$-type at $z\in\mathfrak{p}_1$} \qquad && A_{\pm}\vert_{z=e^{\beta+\frac{2i\pi a}{N}}} = A_{\pm}\vert_{z=e^{-\beta+\frac{2i\pi a}{N}}} ~, && \qquad a\in\Integer_{N} ~.
\end{aligned}
\end{equation}
Note that $\Integer_N$-equivariance implies that
\begin{equation}
\mathcal{R}_a = \sigma^a \mathcal{R}_0 \sigma^{-a} ~, \qquad a \in \Integer_N ~,
\end{equation}
and from now on we drop the index $0$ on $\mathcal{R}_0$.
To summarise, we consider either $\eta$-type~\eqref{eq:bcetac0} or $\lambda$-type~\eqref{eq:bclambdac0} boundary conditions at the simple poles in $\mathfrak{p}_0$, and either $\eta$-type or $\lambda$-type boundary conditions at the simple poles in $\mathfrak{p}_1$.
Therefore, we construct four deformed models:
\begin{itemize}
\item The YB deformed $\Integer_{N}$-twisted $\eta$-model (YB-$\eta$): $\eta$-type boundary conditions at $z=\{0,\infty\}$ and $\eta$-type boundary conditions at $z=\{e^{\beta+\frac{2i\pi a}{N}},e^{-\beta+\frac{2i\pi a}{N}}\}$ for $a \in \Integer_N$.
\item The CC-deformed $\Integer_{N}$-twisted $\eta$-model (CC-$\eta$): $\eta$-type boundary conditions at $z=\{0,\infty\}$ and $\lambda$-type boundary conditions at $z=\{e^{\beta+\frac{2i\pi a}{N}},e^{-\beta+\frac{2i\pi a}{N}}\}$ for $a \in \Integer_N$.
\item The YB-deformed $\Integer_{N}$-twisted $\lambda$-model (YB-$\lambda$): $\lambda$-type boundary conditions at $z=\{0,\infty\}$ and $\eta$-type boundary conditions at $z=\{e^{\beta+\frac{2i\pi a}{N}},e^{-\beta+\frac{2i\pi a}{N}}\}$ for $a \in \Integer_N$.
\item The CC-deformed $\Integer_{N}$-twisted $\lambda$-model (CC-$\lambda$): $\lambda$-type boundary conditions at $z=\{0,\infty\}$ and $\lambda$-type boundary conditions at $z=\{e^{\beta+\frac{2i\pi a}{N}},e^{-\beta+\frac{2i\pi a}{N}}\}$ for $a \in \Integer_N$.
\end{itemize}
For the moment we continue with general $N$, but to solve the boundary conditions and construct the Lax connections we will restrict to $N=2$.
Following the analysis in \secref{sec:bcgs} we fix all the external gauge symmetries, and the internal gauge symmetries for the $\eta$-models, to give the following field content
\begin{equation}
\begin{aligned}
& \text{YB-$\eta$:} \qquad & \hat{g}\vert_{z=0} &= 1 ~, \qquad & \hat{g}\vert_{z=\infty} &= 1 ~, \qquad & \hat{g}\vert_{z=e^{\beta+\frac{2i\pi a}{N}}} &= \hat{\sigma}^a(g) ~, \qquad & \hat{g}\vert_{z=e^{-\beta+\frac{2i\pi a}{N}}} &= \hat{\sigma}^a(g) ~, \\
& \text{CC-$\eta$:} & \hat{g}\vert_{z=0} &= 1~, & \hat{g}\vert_{z=\infty} &= 1~, & \hat{g}\vert_{z=e^{\beta+\frac{2i\pi a}{N}}} &= \hat{\sigma}^a(g)~, & \hat{g}\vert_{z=e^{-\beta+\frac{2i\pi a}{N}}} &= 1~, \\
& \text{YB-$\lambda$:} & \hat{g}\vert_{z=0} &= \tilde{g}~, & \hat{g}\vert_{z=\infty} &= 1~, & \hat{g}\vert_{z=e^{\beta+\frac{2i\pi a}{N}}} &= \hat{\sigma}^a(g)~, & \hat{g}\vert_{z=e^{-\beta+\frac{2i\pi a}{N}}} &= \hat{\sigma}^a(g)~, \\
& \text{CC-$\lambda$:} & \hat{g}\vert_{z=0} &= \tilde{g}~, & \hat{g}\vert_{z=\infty} &= 1~, & \hat{g}\vert_{z=e^{\beta+\frac{2i\pi a}{N}}} &= \hat{\sigma}^a(g)~, & \hat{g}\vert_{z=e^{-\beta+\frac{2i\pi a}{N}}} &= 1 ~,
\end{aligned}
\end{equation}
with $a\in\Integer_N$.

\paragraph{Lax connections.}
As for the undeformed models, we substitute~\eqref{eq:Laxparam} into the boundary conditions and use the ansatz for the Lax connection~\eqref{eq:laxconnection:zn}.
From the simple poles in $\mathfrak{p}_0$, we again find the two equations in eq.~\eqref{etalamp0}, while the simple poles in $\mathfrak{p}_1$ lead to the following equations
\unskip\footnote{Note that here we have used that $\Ad_{\hat\sigma^a(g)} = \sigma^a \Ad_g \sigma^{-a}$ and $\mathcal{R}_a = \sigma^a \mathcal{R}\sigma^{-a}$.}
\begin{equation}\label{eq:etalamp1}
\begin{aligned}
\textrm{$\eta$-type at $\mathfrak{p}_1$:} \qquad & (\mathcal{R}_{g,a} +c) \Big(\tilde{V}_{\pm} + e^{-\frac{1\mp1}{2}N\alpha} \sum_{b\in\Integer_{N}} \frac{V_{\pm}^{(b)}e^{b\beta+\frac{2i\pi ab}{N}}}{e^{N(\pm\alpha+\beta)}-1} - \sigma^a(j_{\pm})\Big) \\
& \qquad\qquad = (\mathcal{R}_{g,a} -c) \Big(\tilde{V}_{\pm} + e^{-\frac{1\mp1}{2}N\alpha} \sum_{b\in\Integer_{N}} \frac{V_{\pm}^{(b)}e^{-b\beta+\frac{2i\pi ab}{N}}}{e^{N(\pm\alpha-\beta)}-1} - \sigma^a(j_{\pm})\Big) ~, \\
\textrm{$\lambda$-type at $\mathfrak{p}_1$:} \qquad & \Ad_{g,a} \Big(\tilde{V}_{\pm} + e^{-\frac{1\mp1}{2}N\alpha} \sum_{b\in\Integer_{N}} \frac{V_{\pm}^{(b)}e^{b\beta+\frac{2i\pi ab}{N}}}{e^{N(\pm\alpha+\beta)}-1} - \sigma^a(j_{\pm})\Big) \\
& \qquad\qquad\qquad\qquad\qquad\qquad = \tilde{V}_{\pm} + e^{-\frac{1\mp1}{2}N\alpha} \sum_{b\in\Integer_{N}} \frac{V_{\pm}^{(b)}e^{-b\beta+\frac{2i\pi ab}{N}}}{e^{N(\pm\alpha-\beta)}-1} ~,
\end{aligned}
\end{equation}
for $a\in \Integer_N$ and we have defined the dressed adjoint action and dressed R-matrices
\begin{equation}
\Ad_{g,a} = \sigma^a \Ad_g \sigma^{-a} ~, \qquad
\mathcal{R}_{g,a} = \sigma^a \Ad_g^{-1} \mathcal{R} \Ad_g \sigma^{-a} ~.
\end{equation}

To proceed, we solve the equations~\eqref{etalamp0} for $\tilde{V}_\pm$ for the $\eta$-models and for $V_\pm^{(0)}$ for the $\lambda$-models, and substitute these expressions into the equations~\eqref{eq:etalamp1}.
For the $\eta$-models we find equations of the following general form
\begin{equation}\label{bc_general_deformed_form_eta}
\rho_{\pm} \mathcal{U}_{\pm,a} V_{\pm}^{(0)} + \sum_{b\in\Integer_N\backslash\{0\}} \xi_{\pm,a}^{(b)} \mathcal{V}_{\pm,a}^{(b)} V_{\pm}^{(b)} = \sigma^a(j_{\pm}) ~, \qquad a\in\Integer_{N} ~,
\end{equation}
while for the $\lambda$-models they take the form
\begin{equation}\label{bc_general_deformed_form_lambda}
\rho_{\pm} \mathcal{U}_{\pm,a} \tilde V_{\pm} + \sum_{b\in\Integer_N\backslash\{0\}} \xi_{\pm,a}^{(b)} \mathcal{V}_{\pm,a}^{(b)} V_{\pm}^{(b)} = \sigma^a(j_{\pm}) + e^{\frac{1\mp1}{2}N\alpha} \xi_{\pm,a}^{(0)} \mathcal{V}_{\pm,a}^{(0)} \tilde{\jmath}_{\pm} ~, \qquad a\in\Integer_{N} ~.
\end{equation}
The operators $\mathcal{U}_{\pm,a}$ and $\mathcal{V}_{\pm,a}^{(b)}$ for the deformed models are
\begin{equation}\label{eq:operators}
\begin{aligned}
& \text{YB-$\eta$:} & \qquad &
\mathcal{U}_{\pm,a} = 1 + \nu_{1\pm} \tilde{\mathcal{R}} + \nu_{2\pm} \mathcal{R}_{g,a} ~, & \qquad &
\mathcal{V}_{\pm,a}^{(b)} = 1 + \gamma_{\pm}^{(b)} \mathcal{R}_{g,a} ~, \\
& \text{CC-$\eta$:} &&
\mathcal{U}_{\pm,a} = 1 + \nu_{1\pm} (1-\Ad_{g,a}^{-1}) \tilde{\mathcal{R}} + \nu_{2\pm} \Ad_{g,a}^{-1} ~, &&
\mathcal{V}_{\pm,a}^{(b)} = 1 + \gamma_{\pm}^{(b)} \Ad_{g,a}^{-1} ~, \\
& \text{YB-$\lambda$:} &&
\mathcal{U}_{\pm,a} = 1 + \nu_{1\pm} \Ad_{\tilde{g}}^{-1} + \nu_{2\pm} \mathcal{R}_{g,a} (1-\Ad_{\tilde{g}}^{-1}) ~, &&
\mathcal{V}_{\pm,a}^{(b)} = 1 + \gamma_{\pm}^{(b)} \mathcal{R}_{g,a} ~, \\
& \text{CC-$\lambda$:} &&
\mathcal{U}_{\pm,a} = 1 + \nu_{1\pm} \Ad_{\tilde{g}}^{-1} + \nu_{2\pm} \Ad_{g,a}^{-1} + \nu_{3\pm} \Ad_{g,a}^{-1}\Ad_{\tilde{g}}^{-1} ~, &&
\mathcal{V}_{\pm,a}^{(b)} = 1 + \gamma_{\pm}^{(b)} \Ad_{g,a}^{-1}~, \\
\end{aligned}
\end{equation}
and the coefficients $\rho_\pm$, $\xi_{\pm,a}^{(b)}$, $\gamma_\pm^{(b)}$ and $\nu_{i\pm}$ are given in \tabref{tab:coefficients}.
\begin{table}[t]
\centering
\begin{tabular}{|l||ll|}
\hline
YB-$\eta$
&
\multicolumn{2}{l|}{$\rho_{\pm} = \pm \frac{1}{2} e^{-\frac{1\mp1}{2}N\alpha} \frac{e^{2N\alpha}-1}{(e^{N(\alpha+\beta)}-1)(e^{N(\alpha-\beta)}-1)}$}
\\
&
\multicolumn{2}{l|}{$\xi_{\pm,a}^{(b)} = \pm \frac{1}{2} e^{\frac{2i\pi ab}{N}} \frac{e^{-(\frac{1\mp1}{2}N\pm b)\beta}(e^{N(\alpha+\beta)}-1) + e^{(\frac{1\mp1}{2}N\pm b)\beta}(e^{N(\alpha-\beta)}-1)}{(e^{N(\alpha+\beta)}-1)(e^{N(\alpha-\beta)}-1)}$}
\\
&
\multicolumn{2}{l|}{$\gamma_{\pm}^{(b)} = \mp \frac{1}{c} \frac{e^{-(\frac{1\mp1}{2}N\pm b)\beta}(e^{N(\alpha+\beta)}-1) - e^{(\frac{1\mp1}{2}N\pm b)\beta}(e^{N(\alpha-\beta)}-1)}{e^{-(\frac{1\mp1}{2}N\pm b)\beta}(e^{N(\alpha+\beta)}-1) + e^{(\frac{1\mp1}{2}N\pm b)\beta}(e^{N(\alpha-\beta)}-1)}$}
\\
&
$\nu_{1\pm} = \pm \frac{1}{c_0} \frac{(e^{N(\alpha+\beta)}-1)(e^{N(\alpha-\beta)}-1)}{e^{2N\alpha}-1}$
\qquad\qquad\qquad
&
$\nu_{2\pm} = \mp \frac{1}{c} e^{N(\alpha-\beta)} \frac{e^{2N\beta}-1}{e^{2N\alpha}-1}$
\\\hline
CC-$\eta$
&
$\rho_{\pm} = \pm \frac{e^{-\frac{1\mp1}{2}N\alpha}}{2} \frac{e^{N(\alpha\pm\beta)}+1}{e^{N(\alpha\pm\beta)}-1}$
&
$\nu_{1\pm} = \pm \frac{1}{c_0} \frac{e^{N(\alpha\pm\beta)}-1}{e^{N(\alpha\pm\beta)}+1}$
\\
&
$\xi_{\pm,a}^{(b)} = \pm \frac{e^{-(\frac{1\mp1}{2}N-b)\beta+\frac{2i\pi ab}{N}}}{e^{N(\alpha\pm\beta)}-1}$
&
$\nu_{2\pm} = - \frac{(e^{N(\alpha\pm\beta)}-1) (e^{N(\alpha\mp\beta)}+1)}{(e^{N(\alpha\pm\beta)}+1) (e^{N(\alpha\mp\beta)}-1)}$
\\
&
$\gamma_{\pm}^{(b)} = -e^{2(\frac{1\mp1}{2}N-b)\beta} \frac{e^{N(\alpha\pm\beta)}-1}{e^{N(\alpha\mp\beta)}-1}$
&
\\\hline
YB-$\lambda$
&
\multicolumn{2}{l|}{$\rho_{\pm} = \frac{1}{2} \frac{1+e^{2N\beta}-2e^{N(\beta\pm\alpha)}}{(e^{N(\beta+\alpha)}-1)(e^{N(\beta-\alpha)}-1)}$}
\\
&
\multicolumn{2}{l|}{$\xi_{\pm,a}^{(b)} = \pm \frac{1}{2} e^{\frac{2i\pi ab}{N}} \frac{e^{-(\frac{1\mp1}{2}N\pm b)\beta}(e^{N(\alpha+\beta)}-1) + e^{(\frac{1\mp1}{2}N\pm b)\beta}(e^{N(\alpha-\beta)}-1)}{(e^{N(\alpha+\beta)}-1)(e^{N(\alpha-\beta)}-1)}$}
\\
&
\multicolumn{2}{l|}{$\gamma_{\pm}^{(b)} = \mp \frac{1}{c} \frac{e^{-(\frac{1\mp1}{2}N\pm b)\beta}(e^{N(\alpha+\beta)}-1) - e^{(\frac{1\mp1}{2}N\pm b)\beta}(e^{N(\alpha-\beta)}-1)}{e^{-(\frac{1\mp1}{2}N\pm b)\beta}(e^{N(\alpha+\beta)}-1) + e^{(\frac{1\mp1}{2}N\pm b)\beta}(e^{N(\alpha-\beta)}-1)}$}
\\
&
$\nu_{1\pm} = \frac{1+e^{2N\beta}-2e^{N(\beta\mp\alpha)}}{1+e^{2N\beta}-2e^{N(\beta\pm\alpha)}}$
&
$\nu_{2\pm} = \frac{1}{c} \frac{e^{2N\beta}-1}{1+e^{2N\beta}-2e^{N(\beta\pm\alpha)}}$
\\\hline
CC-$\lambda$
&
$\rho_{\pm} = -\frac{1}{e^{-N(\beta\pm\alpha)}-1}$
&
$\nu_{1\pm} = -e^{-N(\beta\pm\alpha)}$
\\
&
$\xi_{\pm,a}^{(b)} = \pm \frac{e^{-(\frac{1\mp1}{2}N-b)\beta+\frac{2i\pi ab}{N}}}{e^{N(\alpha\pm\beta)}-1}$
&
$\nu_{2\pm} = -\frac{e^{-N(\beta\pm\alpha)}-1}{e^{N(\beta\mp\alpha)}-1}$
\\
&
$\gamma_{\pm}^{(b)} = -e^{2(\frac{1\mp1}{2}N-b)\beta} \frac{e^{N(\alpha\pm\beta)}-1}{e^{N(\alpha\mp\beta)}-1}$
&
$\nu_{3\pm} = -\frac{e^{-N(\beta\pm\alpha)}-1}{e^{-N(\beta\mp\alpha)}-1}$
\\\hline
\end{tabular}
\caption{The coefficients appearing in the operators~(\ref{eq:operators}) and the equations~(\ref{bc_general_deformed_form_eta}) and~(\ref{bc_general_deformed_form_lambda}) following from the boundary conditions for the deformed $\Integer_N$-twisted $\eta$-models and $\lambda$-models.}\label{tab:coefficients}
\end{table}

In order to solve eqs.~\eqref{bc_general_deformed_form_eta,bc_general_deformed_form_lambda} we restrict to $N=2$.
In this case there are two equations for $V_\pm^{(0)}$ and $V_\pm^{(1)}$ for the $\eta$-models, and for $\tilde V_\pm$ and $V_\pm^{(1)}$ for the $\lambda$-models.
The operators acting on these are non-commutative, hence we use the general formula for the inversion of a $2\times2$ matrix whose entries are invertible non-commutative operators
\begin{equation}
\begin{pmatrix}
M_{11} & M_{12} \\
M_{21} & M_{22}
\end{pmatrix}^{-1} =
\begin{pmatrix}
(M_{12}^{-1}M_{11}-M_{22}^{-1}M_{21})^{-1}M_{12}^{-1} & -(M_{12}^{-1}M_{11}-M_{22}^{-1}M_{21})^{-1}M_{22}^{-1} \\
(M_{11}^{-1}M_{12}-M_{21}^{-1}M_{22})^{-1}M_{11}^{-1} & -(M_{11}^{-1}M_{12}-M_{21}^{-1}M_{22})^{-1}M_{21}^{-1}
\end{pmatrix} ~.
\end{equation}
Recalling the projections of the Maurer-Cartan 1-form $j$~\eqref{eq:projjj}
\begin{equation}\label{rotated_currents_2}
J^{(0)} = \frac{1}{2} (j+\sigma(j)) \in \alg{g}_0 ~, \qquad J^{(1)} = \frac{1}{2} (j-\sigma(j)) \in \alg{g}_1 ~,
\end{equation}
where $\alg{g}_0$ and $\alg{g}_1$ are the eigenspaces of the $\Integer_2$ automorphism $\sigma$ with eigenvalues $+1$ and $-1$ respectively, the general form of the solution for the $\eta$-models is
\begin{equation}\begin{gathered}
V^{(0)}_\pm = \mathcal{O}^{\mathcal{M}}_\pm \left( \mathcal{M}_\pm J_\pm^{(0)} + \tilde{\mathcal{M}}_\pm J_\pm^{(1)} \right) ~, \qquad
V^{(1)}_\pm = \mathcal{O}^{\mathcal{N}}_\pm \left( \tilde{\mathcal{N}}_\pm J_\pm^{(0)} + \mathcal{N}_\pm J_\pm^{(1)} \right) ~.
\\
\tilde{V}_{\pm} = \frac{e^{-(1\mp1)\alpha}}{2c_0} (\tilde{\mathcal{R}}+c_0) \mathcal{O}^{\mathcal{M}}_{\pm} \left( \mathcal{M}_{\pm} J_{\pm}^{(0)} + \tilde{\mathcal{M}}_{\pm} J_{\pm}^{(1)} \right) ~,
\end{gathered}\end{equation}
while for the $\lambda$-models it is
\begin{equation}\begin{gathered}
V_{\pm}^{(0)} = e^{(1\mp1)\alpha} (1-\Ad_{\tilde{g}}^{-1}) \mathcal{O}^{\mathcal{M}}_{\pm} \left( \mathcal{M}_{\pm} J_{\pm}^{(0)} + \tilde{\mathcal{M}}_{\pm} J_{\pm}^{(1)} + \mathcal{P}_{\pm} \tilde{\jmath} \right) - e^{(1\mp 1)\alpha} \tilde{\jmath}_{\pm} ~,
\\
V^{(1)}_\pm = \mathcal{O}^{\mathcal{N}}_\pm \left( \tilde{\mathcal{N}}_\pm J_\pm^{(0)} + \mathcal{N}_\pm J_\pm^{(1)} + \tilde{\mathcal{P}}_\pm \tilde{\jmath}_\pm \right) ~,
\\
\tilde V_\pm = \mathcal{O}^{\mathcal{M}}_\pm \left( \mathcal{M}_\pm J_\pm^{(0)} + \tilde{\mathcal{M}}_\pm J_\pm^{(1)} + \mathcal{P}_\pm \tilde{\jmath}_\pm \right) ~,
\end{gathered}\end{equation}
where
\begin{equation}\label{eq:mnpops}
\begin{split}
&
\begin{aligned}
& \mathcal{M}_\pm = (\xi^{(1)}_{\pm,0}\mathcal{V}_{\pm,0}^{(1)} )^{-1} - (\xi_{\pm,1}^{(1)}\mathcal{V}_{\pm,1}^{(1)} )^{-1} ~, \qquad &
& \tilde{\mathcal{M}}_\pm = (\xi^{(1)}_{\pm,0}\mathcal{V}_{\pm,0}^{(1)} )^{-1} + (\xi_{\pm,1}^{(1)}\mathcal{V}_{\pm,1}^{(1)} )^{-1} ~,
\\
& \mathcal{N}_\pm = (\rho_\pm \mathcal{U}_{\pm,0})^{-1} + (\rho_\pm \mathcal{U}_{\pm,1})^{-1} ~, \qquad &
& \tilde{\mathcal{N}}_\pm = (\rho_\pm \mathcal{U}_{\pm,0})^{-1} - (\rho_\pm \mathcal{U}_{\pm,1})^{-1} ~,
\end{aligned}
\\ &
\mathcal{P}_\pm = (\xi^{(1)}_{\pm,0}\mathcal{V}_{\pm,0}^{(1)} )^{-1}(e^{(1\mp 1)\alpha} \xi_{\pm,0}^{(0)}\mathcal{V}_{\pm,0}^{(0)}) - (\xi_{\pm,1}^{(1)}\mathcal{V}_{\pm,1}^{(1)} )^{-1}( e^{(1\mp 1)\alpha} \xi_{\pm,1}^{(0)}\mathcal{V}_{\pm,1}^{(0)}) ~,
\\ &
\tilde{\mathcal{P}}_\pm = (\rho_\pm \mathcal{U}_{\pm,0})^{-1}(e^{(1\mp 1)\alpha} \xi_{\pm,0}^{(0)}\mathcal{V}_{\pm,0}^{(0)}) - (\rho_\pm \mathcal{U}_{\pm,1})^{-1}( e^{(1\mp 1)\alpha} \xi_{\pm,1}^{(0)}\mathcal{V}_{\pm,1}^{(0)}) ~,
\end{split}
\end{equation}
and
\begin{equation}\label{eq:oos}
\begin{aligned}
\mathcal{O}^{\mathcal{M}}_{\pm} &= \left( (\mathcal{M}_{\pm} + \tilde{\mathcal{M}}_{\pm})(\mathcal{N}_{\pm} + \tilde{\mathcal{N}}_{\pm})^{-1}+ (\mathcal{M}_{\pm} - \tilde{\mathcal{M}}_{\pm})(\mathcal{N}_{\pm} - \tilde{\mathcal{N}}_{\pm})^{-1} \right)^{-1} ~, \\
\mathcal{O}^{\mathcal{N}}_{\pm} &= \left( (\mathcal{N}_{\pm} + \tilde{\mathcal{N}}_{\pm})(\mathcal{M}_{\pm} + \tilde{\mathcal{M}}_{\pm})^{-1}+ (\mathcal{N}_{\pm} - \tilde{\mathcal{N}}_{\pm})(\mathcal{M}_{\pm} - \tilde{\mathcal{M}}_{\pm})^{-1} \right)^{-1} ~.
\end{aligned}
\end{equation}
Defining the parameters
\begin{equation}
\lambda = e^{2\alpha} ~, \qquad \chi = e^{2\beta} ~,
\end{equation}
the expressions for the operators $\mathcal{M}_\pm$, $\tilde{\mathcal{M}}_\pm$, $\mathcal{N}_\pm$, $\tilde{\mathcal{N}}_\pm$, $\mathcal{P}_\pm$ and $\tilde{\mathcal{P}}_\pm$ for the YB-$\eta$, CC-$\eta$, YB-$\lambda$ and CC-$\lambda$ models are given in \tabref{tab:z2}.
\begin{table}
\centering
\begin{tabular}{|l||lll|}
\hline
YB-$\eta$
&
$\mathcal{M}_{\pm} = \frac{1}{\xi_{\pm}} \big( \frac{1}{1 + \gamma_{\pm} \mathcal{R}_{g,0}} + \frac{1}{1 + \gamma_{\pm} \mathcal{R}_{g,1}} \big)$
&
\multicolumn{2}{|l|}{$\rho_{\pm} = \pm \frac{1}{2} \lambda^{-\frac{1\mp1}{2}} \frac{\lambda^{2}-1}{(\lambda\chi-1)(\lambda\chi^{-1}-1)}$}
\\
&
$\tilde{\mathcal{M}}_{\pm} = \frac{1}{\xi_{\pm}} \big( \frac{1}{1 + \gamma_{\pm} \mathcal{R}_{g,0}} - \frac{1}{1 + \gamma_{\pm} \mathcal{R}_{g,1}} \big)$
&
\multicolumn{2}{|l|}{$\xi_{\pm} = \pm \frac{1}{2} \frac{\chi^{-\frac{1}{2}}(\lambda\chi-1) + \chi^{\frac{1}{2}}(\lambda\chi^{-1}-1)}{(\lambda\chi-1)(\lambda\chi^{-1}-1)}$}
\\
&
$\mathcal{N}_{\pm} = \frac{1}{\rho_{\pm}} \big( \frac{1}{1 + \nu_{1\pm} \tilde{\mathcal{R}} + \nu_{2\pm} \mathcal{R}_{g,0}} + \frac{1}{1 + \nu_{1\pm} \tilde{\mathcal{R}} + \nu_{2\pm} \mathcal{R}_{g,1}} \big)$
&
\multicolumn{2}{|l|}{$\gamma_{\pm} = \mp \frac{1}{c} \frac{\chi^{-\frac{1}{2}}(\lambda\chi-1) - \chi^{\frac{1}{2}}(\lambda\chi^{-1}-1)}{\chi^{-\frac{1}{2}}(\lambda\chi-1) + \chi^{\frac{1}{2}}(\lambda\chi^{-1}-1)}$}
\\
&
$\tilde{\mathcal{N}}_{\pm} = \frac{1}{\rho_{\pm}} \big( \frac{1}{1 + \nu_{1\pm} \tilde{\mathcal{R}} + \nu_{2\pm} \mathcal{R}_{g,0}} - \frac{1}{1 + \nu_{1\pm} \tilde{\mathcal{R}} + \nu_{2\pm} \mathcal{R}_{g,1}} \big)$
&
\multicolumn{2}{|l|}{$\nu_{1\pm} = \pm \frac{1}{c_0} \frac{(\lambda\chi-1)(\lambda\chi^{-1}-1)}{\lambda^{2}-1}$}
\\
& &
\multicolumn{2}{|l|}{$\nu_{2\pm} = \mp \frac{1}{c} \lambda\chi^{-1} \frac{\chi^{2}-1}{\lambda^{2}-1}$}
\\\hline
CC-$\eta$
&
$\mathcal{M}_{\pm} = \frac{1}{\xi_{\pm}} \big( \frac{1}{1 + \gamma_{\pm} \Ad_{g,0}^{-1}} + \frac{1}{1 + \gamma_{\pm} \Ad_{g,1}^{-1}} \big)$
&
\multicolumn{2}{|l|}{$\rho_{\pm} = \pm \frac{1}{2} \lambda^{-\frac{1\mp1}{2}} \frac{\lambda\chi^{\pm1}+1}{\lambda\chi^{\pm1}-1}$}
\\
&
$\tilde{\mathcal{M}}_{\pm} = \frac{1}{\xi_{\pm}} \big( \frac{1}{1 + \gamma_{\pm} \Ad_{g,0}^{-1}} - \frac{1}{1 + \gamma_{\pm} \Ad_{g,1}^{-1}} \big)$
&
\multicolumn{2}{|l|}{$\xi_{\pm} = \pm \chi^{\pm\frac{1}{2}} \frac{1}{\lambda\chi^{\pm1}-1}$}
\\
&
$\mathcal{N}_{\pm} = \frac{1}{\rho_{\pm}} \big( \frac{1}{1 + \nu_{1\pm} (1-\Ad_{g,0}^{-1})\tilde{\mathcal{R}} + \nu_{2\pm} \Ad_{g,0}^{-1}}$
&
\multicolumn{2}{|l|}{$\gamma_{\pm} = - \chi^{\mp1} \frac{\lambda\chi^{\pm1}-1}{\lambda\chi^{\mp1}-1}$}
\\
&
$\qquad\qquad\qquad + \frac{1}{1 + \nu_{1\pm} (1-\Ad_{g,1}^{-1})\tilde{\mathcal{R}} + \nu_{2\pm} \Ad_{g,1}^{-1}} \big)$
&
\multicolumn{2}{|l|}{$\nu_{1\pm} = \pm \frac{1}{c_0} \frac{\lambda\chi^{\pm1}-1}{\lambda\chi^{\pm1}+1}$}
\\
&
$\tilde{\mathcal{N}}_{\pm} = \frac{1}{\rho_{\pm}} \big( \frac{1}{1 + \nu_{1\pm} (1-\Ad_{g,0}^{-1})\tilde{\mathcal{R}} + \nu_{2\pm} \Ad_{g,0}^{-1}}$
&
\multicolumn{2}{|l|}{$\nu_{2\pm} = - \frac{(\lambda\chi^{\pm1}-1) (\lambda\chi^{\mp1}+1)}{(\lambda\chi^{\pm1}+1) (\lambda\chi^{\mp1}-1)}$}
\\
&
$\qquad\qquad\qquad - \frac{1}{1 + \nu_{1\pm} (1-\Ad_{g,1}^{-1})\tilde{\mathcal{R}} + \nu_{2\pm} \Ad_{g,1}^{-1}} \big)$
&
\multicolumn{2}{|l|}{}
\\\hline
YB-$\lambda$
&
$\mathcal{M}_{\pm} = \frac{1}{\xi_{\pm}} \big( \frac{1}{1 + \gamma_{\pm} \mathcal{R}_{g,0}} + \frac{1}{1 + \gamma_{\pm} \mathcal{R}_{g,1}} \big)$
&
\multicolumn{2}{|l|}{$\rho_{\pm} = \frac{1}{2} \frac{1+\chi^{2}-2\lambda^{\pm1}\chi}{(\lambda\chi-1)(\lambda^{-1}\chi-1)}$}
\\
&
$\tilde{\mathcal{M}}_{\pm} = \frac{1}{\xi_{\pm}} \big( \frac{1}{1 + \gamma_{\pm} \mathcal{R}_{g,0}} - \frac{1}{1 + \gamma_{\pm} \mathcal{R}_{g,1}} \big)$
&
\multicolumn{2}{|l|}{$\xi_{\pm} = \pm \frac{1}{2} \frac{\chi^{-\frac{1}{2}}(\lambda\chi-1) + \chi^{\frac{1}{2}}(\lambda\chi^{-1}-1)}{(\lambda\chi-1)(\lambda\chi^{-1}-1)}$}
\\
&
$\mathcal{N}_{\pm} = \frac{1}{\rho_{\pm}} \big( \frac{1}{1 + \nu_{1\pm} \Ad_{\tilde{g}}^{-1} + \nu_{2\pm} \mathcal{R}_{g,0} (1-\Ad_{\tilde{g}}^{-1})}$
&
\multicolumn{2}{|l|}{$\bar{\xi}_{\pm} = \pm \frac{1}{2} \frac{\chi^{-\frac{1\mp1}{2}}(\lambda\chi-1) + \chi^{\frac{1\mp1}{2}}(\lambda\chi^{-1}-1)}{(\lambda\chi-1)(\lambda\chi^{-1}-1)}$}
\\
&
$\qquad\qquad\qquad + \frac{1}{1 + \nu_{1\pm} \Ad_{\tilde{g}}^{-1} + \nu_{2\pm} \mathcal{R}_{g,1} (1-\Ad_{\tilde{g}}^{-1})} \big)$
&
\multicolumn{2}{|l|}{$\gamma_{\pm} = \mp \frac{1}{c} \frac{\chi^{-\frac{1}{2}}(\lambda\chi-1) - \chi^{\frac{1}{2}}(\lambda\chi^{-1}-1)}{\chi^{-\frac{1}{2}}(\lambda\chi-1) + \chi^{\frac{1}{2}}(\lambda\chi^{-1}-1)}$}
\\
&
$\tilde{\mathcal{N}}_{\pm} = \frac{1}{\rho_{\pm}} \big( \frac{1}{1 + \nu_{1\pm} \Ad_{\tilde{g}}^{-1} + \nu_{2\pm} \mathcal{R}_{g,0} (1-\Ad_{\tilde{g}}^{-1})}$
&
\multicolumn{2}{|l|}{$\bar{\gamma}_{\pm} = \mp \frac{1}{c} \frac{\chi^{-\frac{1\mp1}{2}}(\lambda\chi-1) - \chi^{\frac{1\mp1}{2}}(\lambda\chi^{-1}-1)}{\chi^{-\frac{1\mp1}{2}}(\lambda\chi-1) + \chi^{\frac{1\mp1}{2}}(\lambda\chi^{-1}-1)}$}
\\
&
$\qquad\qquad\qquad - \frac{1}{1 + \nu_{1\pm} \Ad_{\tilde{g}}^{-1} + \nu_{2\pm} \mathcal{R}_{g,1} (1-\Ad_{\tilde{g}}^{-1})} \big)$
&
\multicolumn{2}{|l|}{$\nu_{1\pm} = \frac{1+\chi^{2}-2\lambda^{\mp1}\chi}{1+\chi^{2}-2\lambda^{\pm1}\chi}$}
\\
&
$\mathcal{P}_{\pm} = \frac{\lambda^{\frac{1\mp1}{2}}\bar{\xi}_{\pm}}{\xi_{\pm}} \big( \frac{1 + \bar{\gamma}_{\pm} \mathcal{R}_{g,0}}{1 + \gamma_{\pm} \mathcal{R}_{g,0}} + \frac{1 + \bar{\gamma}_{\pm} \mathcal{R}_{g,1}}{1 + \gamma_{\pm} \mathcal{R}_{g,1}} \big)$
&
\multicolumn{2}{|l|}{$\nu_{2\pm} = \frac{1}{c} \frac{\chi^{2}-1}{1+\chi^{2}-2\lambda^{\pm1}\chi}$}
\\
&
$\tilde{\mathcal{P}}_{\pm} = \frac{\lambda^{\frac{1\mp1}{2}}\bar{\xi}_{\pm}}{\rho_{\pm}} \big( \frac{1}{1 + \nu_{1\pm} \Ad_{\tilde{g}}^{-1} + \nu_{2\pm} \mathcal{R}_{g,0} (1-\Ad_{\tilde{g}}^{-1})} {\scriptstyle (1 + \bar{\gamma}_{\pm} \mathcal{R}_{g,0})}$
&
\multicolumn{2}{|l|}{}
\\
&
$\qquad\qquad\quad \ - \frac{1}{1 + \nu_{1\pm} \Ad_{\tilde{g}}^{-1} + \nu_{2\pm} \mathcal{R}_{g,1} (1-\Ad_{\tilde{g}}^{-1})} {\scriptstyle(1 + \bar{\gamma}_{\pm} \mathcal{R}_{g,1})} \big)$
&
\multicolumn{2}{|l|}{}
\\\hline
CC-$\lambda$
&
\multicolumn{2}{l|}{$\mathcal{M}_{\pm} = \frac{1}{\xi_{\pm}} \big( \frac{1}{1 + \gamma_{\pm} \Ad_{g,0}^{-1}} + \frac{1}{1 + \gamma_{\pm} \Ad_{g,1}^{-1}} \big)$}
&
$\rho_{\pm} = -\frac{1}{\lambda^{\mp1}\chi^{-1}-1}$
\\
&
\multicolumn{2}{l|}{$\tilde{\mathcal{M}}_{\pm} = \frac{1}{\xi_{\pm}} \big( \frac{1}{1 + \gamma_{\pm} \Ad_{g,0}^{-1}} - \frac{1}{1 + \gamma_{\pm} \Ad_{g,1}^{-1}} \big)$}
&
$\xi_{\pm} = \pm \frac{\chi^{\pm\frac{1}{2}}}{\lambda\chi^{\pm1}-1}$
\\
&
\multicolumn{2}{l|}{$\mathcal{N}_{\pm} = \frac{1}{\rho_{\pm}} \big( \frac{1}{1 + \nu_{1\pm} \Ad_{\tilde{g}}^{-1} + \nu_{2\pm} \Ad_{g,0}^{-1} + \nu_{3\pm} \Ad_{g,0}^{-1}\Ad_{\tilde{g}}^{-1}}$}
&
$\bar{\xi}_{\pm} = \pm \frac{\chi^{-\frac{1\mp1}{2}}}{\lambda\chi^{\pm1}-1}$
\\
&
\multicolumn{2}{l|}{$\qquad\qquad\qquad + \frac{1}{1 + \nu_{1\pm} \Ad_{\tilde{g}}^{-1} + \nu_{2\pm} \Ad_{g,1}^{-1} + \nu_{3\pm} \Ad_{g,1}^{-1}\Ad_{\tilde{g}}^{-1}} \big)$}
&
$\gamma_{\pm} = - \frac{\lambda\chi^{\pm1}-1}{\lambda-\chi^{\pm1}}$
\\
&
\multicolumn{2}{l|}{$\tilde{\mathcal{N}}_{\pm} = \frac{1}{\rho_{\pm}} \big( \frac{1}{1 + \nu_{1\pm} \Ad_{\tilde{g}}^{-1} + \nu_{2\pm} \Ad_{g,0}^{-1} + \nu_{3\pm} \Ad_{g,0}^{-1}\Ad_{\tilde{g}}^{-1}}$}
&
$\bar{\gamma}_{\pm} = -\chi \frac{\lambda\chi^{\pm1}-1}{\lambda-\chi^{\pm1}}$
\\
&
\multicolumn{2}{l|}{$\qquad\qquad\qquad - \frac{1}{1 + \nu_{1\pm} \Ad_{\tilde{g}}^{-1} + \nu_{2\pm} \Ad_{g,1}^{-1} + \nu_{3\pm} \Ad_{g,1}^{-1}\Ad_{\tilde{g}}^{-1}} \big)$}
&
$\nu_{1\pm} = -\lambda^{\mp1}\chi^{-1}$
\\
&
\multicolumn{2}{l|}{$\mathcal{P}_{\pm} = \frac{\lambda^{\frac{1\mp1}{2}}\bar{\xi}_{\pm}}{\xi_{\pm}} \big( \frac{1 + \bar{\gamma}_{\pm} \Ad_{g,0}^{-1}}{1 + \gamma_{\pm} \Ad_{g,0}^{-1}} + \frac{1 + \bar{\gamma}_{\pm} \Ad_{g,1}^{-1}}{1 + \gamma_{\pm} \Ad_{g,1}^{-1}} \big)$}
&
$\nu_{2\pm} = -\frac{\lambda^{\pm1}-\chi^{-1}}{\lambda^{\pm1}-\chi}$
\\
&
\multicolumn{2}{l|}{$\tilde{\mathcal{P}}_{\pm} = \frac{\lambda^{\frac{1\mp1}{2}}\bar{\xi}_{\pm}}{\rho_{\pm}} \big( \frac{1}{1 + \nu_{1\pm} \Ad_{\tilde{g}}^{-1} + \nu_{2\pm} \Ad_{g,0}^{-1} + \nu_{3\pm} \Ad_{g,0}^{-1}\Ad_{\tilde{g}}^{-1}} {\scriptstyle (1 + \bar{\gamma}_{\pm} \Ad_{g,0}^{-1})}$}
&
$\nu_{3\pm} = -\frac{\lambda^{\mp1}-\chi}{\lambda^{\pm1}-\chi}$
\\
&
\multicolumn{2}{l|}{$\qquad\qquad \quad \ - \frac{1}{1 + \nu_{1\pm} \Ad_{\tilde{g}}^{-1} + \nu_{2\pm} \Ad_{g,1}^{-1} + \nu_{3\pm} \Ad_{g,1}^{-1}\Ad_{\tilde{g}}^{-1}} {\scriptstyle (1 + \bar{\gamma}_{\pm} \Ad_{g,1}^{-1})} \big)$}
&
\\
\hline
\end{tabular}
\caption{The operators $\mathcal{M}_\pm$, $\tilde{\mathcal{M}}_\pm$, $\mathcal{N}_\pm$, $\tilde{\mathcal{N}}_\pm$, $\mathcal{P}_\pm$ and $\tilde{\mathcal{P}}_\pm$ appearing in the Lax connections and actions of the deformed $\Integer_2$-twisted $\eta$-models and $\lambda$-models.}\label{tab:z2}
\end{table}
Note that for all four models we find that
\unskip\footnote{This follows from the fact that the untilded operators can be written as $\mathcal{O} = \frac{1}{2}( \mathscr{O} + \sigma \mathscr{O} \sigma) = P_0 \mathscr{O} P_0 + P_1 \mathscr{O} P_1$, while the tilded operators take the form $\tilde{\mathcal{O}} = \frac12(\mathscr{O} - \sigma \mathscr{O} \sigma) = P_0 \mathscr{O} P_1 + P_1 \mathscr{O} P_0$.
Here we recall that $\tilde{\mathcal{R}}$ commutes with $\sigma$.
Substituting into eq.~\eqref{eq:oos} and using $\sigma^2 =1$, we see that $\mathcal{O}_\pm^{\mathcal{M}} = (\mathscr{M}_\pm \mathscr{N}_\pm^{-1} + \sigma\mathscr{M}_\pm \mathscr{N}_\pm^{-1}\sigma)^{-1} = \frac12 (P_0\mathscr{M}_\pm \mathscr{N}_\pm^{-1}P_0 + P_1\mathscr{M}_\pm \mathscr{N}_\pm^{-1}P_1)^{-1}$ and similarly for $\mathcal{O}_\pm^{\mathcal{N}}$.}
\begin{equation}\begin{aligned}\label{eq:maps}
\{\mathcal{M}_\pm, \mathcal{N}_\pm, \mathcal{P}_\pm,\mathcal{O}_\pm^{\mathcal{M}},\mathcal{O}_\pm^{\mathcal{N}}\} & : \alg{g}_0 \to \alg{g}_0 ~, \qquad
&\{\tilde{\mathcal{M}}_\pm, \tilde{\mathcal{N}}_\pm, \tilde{\mathcal{P}}_\pm\} & : \alg{g}_0 \to \alg{g}_1 ~,
\\
\{\mathcal{M}_\pm, \mathcal{N}_\pm, \mathcal{P}_\pm,\mathcal{O}_\pm^{\mathcal{M}},\mathcal{O}_\pm^{\mathcal{N}}\} & : \alg{g}_1 \to \alg{g}_1 ~, \qquad
&\{\tilde{\mathcal{M}}_\pm, \tilde{\mathcal{N}}_\pm, \tilde{\mathcal{P}}_\pm\} & : \alg{g}_1 \to \alg{g}_0 ~.
\end{aligned}\end{equation}

The Lax connections for the deformed $\Integer_2$-twisted $\eta$-models and $\lambda$-models then take the form
\begin{equation}\label{deformed_lax_connections_general}
\begin{aligned}
\Lax_{\{\mathrm{YB,CC}\}\text{-}\eta\pm}^{(2)} &= \frac{\lambda^{-\frac{1\mp1}{2}}}{2c_0} \Big(c_0+\frac{2c_0}{\lambda^{\pm1}z^{2}-1}+\tilde{\mathcal{R}}\Big) \Big( \mathcal{A}_{\pm}J_{\pm}^{(0)} + \tilde{\mathcal{A}}_{\pm}J_{\pm}^{(1)} \Big) + \frac{\lambda^{-\frac{1\mp1}{2}}z}{\lambda^{\pm1}z^{2}-1} \Big( \tilde{\mathcal{B}}_{\pm}J_{\pm}^{(0)} + \mathcal{B}_{\pm}J_{\pm}^{(1)} \Big) ~, \\
\Lax_{\{\mathrm{YB,CC}\}\text{-}\lambda\pm}^{(2)} &= \Big( 1 + \frac{1-\Ad_{\tilde{g}}^{-1}}{\lambda^{\pm1}z^{2}-1} \Big) \Big( \mathcal{A}_{\pm}J_{\pm}^{(0)} + \tilde{\mathcal{A}}_{\pm}J_{\pm}^{(1)} + \mathcal{C}_{\pm}\tilde{\jmath}_{\pm} \Big)
\\ & \qquad \qquad \qquad \qquad + \frac{\lambda^{-\frac{1\mp1}{2}}z}{\lambda^{\pm1}z^{2}-1} \Big( \tilde{\mathcal{B}}_{\pm}J_{\pm}^{(0)} + \mathcal{B}_{\pm}J_{\pm}^{(1)} + \tilde{\mathcal{C}}_{\pm}\tilde{\jmath}_{\pm} \Big) - \frac{1}{\lambda^{\pm1}z^{2}-1} \tilde{\jmath}_{\pm} ~,
\end{aligned}
\end{equation}
where we have introduced
\begin{equation}\begin{gathered}
\mathcal{A}_{\pm} = \mathcal{O}^{\mathcal{M}}_{\pm} \mathcal{M}_{\pm} ~,
\qquad
\mathcal{B}_{\pm} = \mathcal{O}^{\mathcal{N}}_{\pm} \mathcal{N}_{\pm} ~,
\qquad
\mathcal{C}_{\pm} = \mathcal{O}^{\mathcal{M}}_{\pm} \mathcal{P}_{\pm} ~,
\\ \tilde{\mathcal{A}}_{\pm} = \mathcal{O}^{\mathcal{M}}_{\pm} \tilde{\mathcal{M}}_{\pm} ~, \qquad
\tilde{\mathcal{B}}_{\pm} = \mathcal{O}^{\mathcal{N}}_{\pm} \tilde{\mathcal{N}}_{\pm} ~, \qquad
\tilde{\mathcal{C}}_{\pm} = \mathcal{O}^{\mathcal{N}}_{\pm} \tilde{\mathcal{P}}_{\pm} ~,
\end{gathered}\end{equation}
and the operators $\mathcal{O}^{\mathcal{M}}_\pm$ and $\mathcal{O}^{\mathcal{N}}_\pm$ are defined in eq.~\eqref{eq:oos}.
Using eq.~\eqref{eq:maps} we can check that the part of the Lax connections even in $z$ is valued in $\alg{g}_0$, while the part odd in $z$ is valued in $\alg{g}_1$, in agreement with the discussion beneath eq.~\eqref{eq:laxconnection:zn}.

\paragraph{Actions.}
To find the actions of the deformed $\Integer_2$-twisted models we substitute the Lax connections we have constructed~\eqref{deformed_lax_connections_general} and the levels~\eqref{eq:levelsdef} into the general expressions for the actions~\eqref{eq:action-etageneral} and~\eqref{eq:action-lambdageneral}.
The actions can be written in the form
\begin{equation}\label{eq:genformactionsz2}
\Act_{\{\mathrm{YB,CC}\}\text{-}\{\eta,\lambda\}}^{(2)} = \mathsf{N} \int d^{2}\sigma \, \tr\big( \mathbf{J}_{+}^{T} \mathcal{E} \mathbf{J}_{-} \big) + \Act^{(2)}_{\mathrm{WZ}} ~,
\end{equation}
where $\mathsf{N}$ is a constant, $\mathcal{E}$ is a matrix of operators, $\Act^{(2)}_{\mathrm{WZ}}$ is the Wess-Zumino term, and
\begin{equation}
\text{$\eta$-models}: \quad \mathbf{J} = \begin{pmatrix}J^{(0)}\\J^{(1)}\end{pmatrix} ~, \qquad\qquad \text{$\lambda$-models}: \quad \mathbf{J} = \begin{pmatrix}J^{(0)}\\J^{(1)}\\\tilde{\jmath}\end{pmatrix} ~.
\end{equation}
Recall that the field $\tilde\jmath$ only appears in the $\lambda$-models since we fixed the internal gauge symmetries for the $\eta$-models.
For the YB-deformed $\Integer_{2}$-twisted $\eta$-model we have
\begin{equation}
\mathsf{N} = \frac{c_0\hay}{\lambda(\chi+1)} ~, \qquad \Act^{(2)}_{\mathrm{WZ}} = 0 ~,
\end{equation}
\begin{equation}
\begin{aligned}
\mathcal{E}_{00} &= (\chi+1)(\mathcal{A}_{-}-\lambda\mathcal{A}_{+}^{t}) ~, \qquad
&\mathcal{E}_{01} &= (\chi+1)\tilde{\mathcal{A}}_{-}-\sqrt{\chi}(\lambda+1)\tilde{\mathcal{B}}_{+}^{t} ~, \\
\mathcal{E}_{10} &= \sqrt{\chi}(\lambda+1)\tilde{\mathcal{B}}_{-}-\lambda(\chi+1)\tilde{\mathcal{A}}_{+}^{t} ~, \qquad
&\mathcal{E}_{11} &= \sqrt{\chi}(\lambda+1)(\mathcal{B}_{-}-\mathcal{B}_{+}^{t}) ~.
\end{aligned}
\end{equation}
For the CC-deformed $\Integer_{2}$-twisted $\eta$-model ($\Act_{\mathrm{CC}\text{-}\eta}^{(2)}$) we have
\begin{equation}
\mathsf{N} = \frac{\hay}{2\lambda^{2}(\chi^{2}-1)} ~,
\qquad
\Act^{(2)}_{\mathrm{WZ}} = \frac{c_0\hay(\lambda\chi-1)(\lambda-\chi)}{\lambda(\chi^{2}-1)} \Act_{\mathrm{WZ}}(g) ~,
\end{equation}
\begin{equation}
\begin{aligned}
\mathcal{E}_{00} &= (\lambda\chi-1)(c_0(\lambda+\chi)-(\lambda-\chi)\tilde{\mathcal{R}})\mathcal{A}_{-}
+\lambda(\lambda-\chi)\mathcal{A}_{+}^{t}(c_0(\lambda\chi+1)-(\lambda\chi-1)\tilde{\mathcal{R}}) ~, \\
\mathcal{E}_{01} &= (\lambda\chi-1)(c_0(\lambda+\chi)-(\lambda-\chi)\tilde{\mathcal{R}})\tilde{\mathcal{A}}_{-} + 2c_0\lambda\sqrt{\chi}(\lambda-\chi)\tilde{\mathcal{B}}_{+}^{t} ~, \\
\mathcal{E}_{10} &= \lambda(2c_0\sqrt{\chi}(\lambda\chi-1)\tilde{\mathcal{B}}_{-} +(\lambda-\chi)\tilde{\mathcal{A}}_{+}^{t}(c_0(\lambda\chi+1)-(\lambda\chi-1)\tilde{\mathcal{R}})) ~, \\
\mathcal{E}_{11} &= 2c_0\lambda\sqrt{\chi}((\lambda\chi-1)\mathcal{B}_{-} +(\lambda-\chi)\mathcal{B}_{+}^{t}) ~.
\end{aligned}
\end{equation}
For the YB-deformed $\Integer_{2}$-twisted $\lambda$-model ($\Act_{\mathrm{YB}\text{-}\lambda}^{(2)}$) we have
\begin{equation}
\mathsf{N} = \frac{\hay}{\lambda(\chi+1)} ~, \qquad
\Act^{(2)}_{\mathrm{WZ}} = - \hay \Act_{\mathrm{WZ}}(\tilde g) ~,
\end{equation}
\begin{equation}
\begin{aligned}
\mathcal{E}_{00} &= \lambda(\chi+1)((1-\Ad_{\tilde{g}}^{-1})\mathcal{A}_{-} -\mathcal{A}_{+}^{t}(1-\Ad_{\tilde{g}})) ~, \qquad &
\mathcal{E}_{01} &= \lambda(\chi+1)(1-\Ad_{\tilde{g}}^{-1})\tilde{\mathcal{A}}_{-} -\sqrt{\chi}(\lambda+1)\tilde{\mathcal{B}}_{+}^{t} ~, \\
\mathcal{E}_{10} &= \sqrt{\chi}(\lambda+1)\tilde{\mathcal{B}}_{-} - \lambda(\chi+1)\tilde{\mathcal{A}}_{+}^{t}(1-\Ad_{\tilde{g}}) ~, \qquad &
\mathcal{E}_{11} &= \sqrt{\chi}(\lambda+1)(\mathcal{B}_{-} -\mathcal{B}_{+}^{t}) ~, \\
\mathcal{E}_{02} &= -\lambda(\chi+1)(1 -(1-\Ad_{\tilde{g}}^{-1})\mathcal{C}_{-} +\mathcal{A}_{+}^{t}\Ad_{\tilde{g}}) ~, \qquad
&\mathcal{E}_{12} &= \sqrt{\chi}(\lambda+1)\tilde{\mathcal{C}}_{-} -\lambda(\chi+1)\tilde{\mathcal{A}}_{+}^{t}\Ad_{\tilde{g}}~, \\
\mathcal{E}_{20} &= \lambda(\chi+1)(1 +\Ad_{\tilde{g}}^{-1}\mathcal{A}_{-} -\mathcal{C}_{+}^{t}(1-\Ad_{\tilde{g}})) ~, \qquad
&\mathcal{E}_{21} &= \lambda(\chi+1)\Ad_{\tilde{g}}^{-1}\tilde{\mathcal{A}}_{-} -\sqrt{\chi}(\lambda+1)\tilde{\mathcal{C}}_{+}^{t} ~, \\
\mathcal{E}_{22} &= \lambda(\chi+1)(\Ad_{\tilde{g}}^{-1}\mathcal{C}_{-} -\mathcal{C}_{+}^{t}\Ad_{\tilde{g}}) ~.
\end{aligned}
\end{equation}
For the CC-deformed $\Integer_{2}$-twisted $\lambda$-model ($\Act_{\mathrm{CC}\text{-}\lambda}^{(2)}$) we have
\begin{equation}\begin{split}
\mathsf{N} = \frac{\hay}{\lambda(\chi^{2}-1)} ~,\qquad
\Act^{(2)}_{\mathrm{WZ}} &= - \hay \Act_{\mathrm{WZ}}(\tilde g) + \frac{\hay(\lambda\chi-1)(\lambda-\chi)}{\lambda(\chi^{2}-1)} \Act_{\mathrm{WZ}}(g) ~,
\end{split}\end{equation}
\begin{equation}
\begin{aligned}
\mathcal{E}_{00} &= (\lambda\chi-1)(\chi-\lambda\Ad_{\tilde{g}}^{-1})\mathcal{A}_{-} +(\lambda-\chi)\mathcal{A}_{+}^{t}(\lambda\chi-\Ad_{\tilde{g}}) ~, \\
\mathcal{E}_{01} &= (\lambda\chi-1)(\chi-\lambda\Ad_{\tilde{g}}^{-1})\tilde{\mathcal{A}}_{-} +\sqrt{\chi}(\lambda-\chi)\tilde{\mathcal{B}}_{+}^{t} ~, \\
\mathcal{E}_{02} &= -\lambda(\lambda\chi-1) +(\lambda\chi-1)(\chi-\lambda\Ad_{\tilde{g}}^{-1})\mathcal{C}_{-} -\lambda(\chi^{2}-1)\mathcal{A}_{+}^{t}\Ad_{\tilde{g}} ~, \\
\mathcal{E}_{10} &= \sqrt{\chi}(\lambda\chi-1)\tilde{\mathcal{B}}_{-} +(\lambda-\chi)\tilde{\mathcal{A}}_{+}^{t}(\lambda\chi-\Ad_{\tilde{g}}) ~, \\
\mathcal{E}_{11} &= \sqrt{\chi}((\lambda\chi-1)\mathcal{B}_{-} +(\lambda-\chi)\mathcal{B}_{+}^{t}) ~, \\
\mathcal{E}_{12} &= \sqrt{\chi}(\lambda\chi-1)\tilde{\mathcal{C}}_{-} -\lambda(\chi^{2}-1)\tilde{\mathcal{A}}_{+}^{t}\Ad_{\tilde{g}} ~, \\
\mathcal{E}_{20} &= -(\lambda-\chi) +\lambda(\chi^{2}-1)\Ad_{\tilde{g}}^{-1}\mathcal{A}_{-} +(\lambda-\chi)\mathcal{C}_{+}^{t}(\lambda\chi-\Ad_{\tilde{g}}) ~, \\
\mathcal{E}_{21} &= \lambda(\chi^{2}-1)\Ad_{\tilde{g}}^{-1}\tilde{\mathcal{A}}_{-} +\sqrt{\chi}(\lambda-\chi)\tilde{\mathcal{C}}_{+}^{t} ~, \\
\mathcal{E}_{22} &= \lambda(\chi^{2}-1)(\Ad_{\tilde{g}}^{-1}\mathcal{C}_{-} -\mathcal{C}_{+}^{t}\Ad_{\tilde{g}}) ~.
\end{aligned}
\end{equation}

\paragraph{Undeformed and untwisted limits.}
In the undeformed limit of the YB-deformed $\Integer_2$-twisted $\eta$-model and $\lambda$-model, which corresponds to taking $\chi \to 1$, or equivalently $\beta \to 0$, we recover the undeformed $\Integer_2$-twisted $\eta$-model~\eqref{eq:actionetatwist} and $\lambda$-model~\eqref{eq:actionlambdatwist} as expected, with
\begin{equation}
c_0 \eta = \frac{\lambda - 1}{\lambda + 1} ~.
\end{equation}
It is also possible to take the $\chi \to 1$ limit in the CC-deformed models, but doing so requires simultaneously taking $g \to 1$ in a non-abelian dual type limit~\cite{Sfetsos:2013wia} and is expected to give the non-abelian duals of the undeformed models.

To recover the untwisted models, we can simply set $\sigma$ to be the identity automorphism in the $\Integer_2$-twisted models.
This means that $\mathcal{R}_{g,0} = \mathcal{R}_{g,1} = \mathcal{R}_g$ and $\Ad_{g,0} = \Ad_{g,1} = \Ad_g$.
It follows from the expressions in \tabref{tab:z2} that $\tilde{\mathcal{M}}_\pm = \tilde{\mathcal{N}}_\pm = \tilde{\mathcal{P}}_\pm = 0$, hence from eq.~\eqref{eq:oos} we have that $\mathcal{O}^{\mathcal{M}}_\pm = \frac12\mathcal{N}_\pm\mathcal{M}_\pm^{-1}$ and $\mathcal{O}^{\mathcal{N}}_\pm = \frac12\mathcal{M}_\pm\mathcal{N}_\pm^{-1}$.
We also note that setting $\sigma = 1$ in~\eqref{rotated_currents_2} implies that $J^{(0)} = j$ and $J^{(1)} = 0$.

For the YB-deformed untwisted $\eta$-model ($\Act_{\mathrm{YB}\text{-}\eta}^{(1)}$) we find
\begin{equation}\label{eq:untwist}
\Act^{(1)}_{\mathrm{YB}\text{-}\eta}(g) = -4c_0^2\hay \nu_{1+} \int d^2\sigma \,\tr\big(j_+ \frac{1}{1-\nu_{1+} \tilde{\mathcal{R}} - \nu_{2+} \mathcal{R}_g} j_-\big) ~,
\end{equation}
which we recognise as the bi-Yang-Baxter deformation of the principal chiral model \cite{Klimcik:2008eq}.
For the CC-deformed untwisted $\eta$-model and the YB-deformed untwisted $\lambda$-model we have
\begin{equation}\label{eq:comp1}
\Act^{(1)}_{\mathrm{CC}\text{-}\eta}(g) = -\frac{2c_0^2\hay }{\nu_{1+}^{-1} + \nu_{1-}^{-1}} \int d^2\sigma \, \tr\Big(j_+ \frac{1-\mathcal{Q} \Ad_g}{1+\mathcal{Q} \Ad_g} j_- \Big) -\frac{2c_0^2\hay }{\nu_{1+}^{-1} + \nu_{1-}^{-1}} \Act_{\mathrm{WZ}}(g) ~, \qquad \mathcal{Q} = \frac{\nu_{1+}}{\nu_{1-}} \frac{1 + \nu_{1-} \tilde{\mathcal{R}}}{1 - \nu_{1+} \tilde{\mathcal{R}}} ~,
\end{equation}
and
\begin{equation}\label{eq:comp2}
\Act^{(1)}_{\mathrm{YB}\text{-}\lambda}(\tilde g) = - \hay \int d^2\sigma \, \tr\Big(\tilde \jmath_+ \frac{1-\mathcal{Q} \Ad_{\tilde g}}{1+\mathcal{Q} \Ad_{\tilde g}} \tilde \jmath_- \Big) - \hay \Act_{\mathrm{WZ}}(\tilde{g}) ~, \qquad \mathcal{Q} = \frac{\nu_{2+}}{\nu_{2-}}\frac{1+\nu_{2-} \mathcal{R}}{1-\nu_{2+} \mathcal{R}} ~,
\end{equation}
where we have used the unfixed internal gauge symmetries to set $g = 1$ in $\Act^{(1)}_{\mathrm{YB}\text{-}\lambda}$ since $\grp{G}_0 = \grp{G}$ in the untwisted case.
\unskip\footnote{Recall that for $\eta$-type boundary conditions $g$ transforms as $g\to g v$, $v\in\grp{G}_0$, cf.~eq.~\eqref{eq:intgaugefixbc}.}
As summarised in~\secref{sec:general-theory}, we expect that the CC-deformed untwisted $\eta$-model and the YB-deformed untwisted $\lambda$-model are the same.
Comparing the actions~\eqref{eq:comp1} and~\eqref{eq:comp2} we can explicitly see this is the case if we interchange $g$ and $\tilde g$ and $\mathcal{R}$ and $\tilde{\mathcal{R}}$ and redefine the parameters.
This model was first written down in~\cite{Sfetsos:2015nya}.
Finally, the CC-deformed untwisted $\lambda$-model is a particular double deformation of the WZW model initially constructed in \cite{Sfetsos:2014cea,Sfetsos:2017sep} and its action is
\begin{equation}\begin{split}
\Act^{(1)}_{\mathrm{CC}\text{-}\lambda}(g,\tilde g) & =
\hay \int d^2\sigma \,\tr \Big(\frac{(1+\nu_{1+})(1+\nu_{1-})}{1-\nu_{1+}\nu_{1-}} j_+\tilde{\mathcal{E}}_{00} j_-
+ \tilde{\jmath}_+ \tilde{\mathcal{E}}_{22}\tilde{\jmath}_-
\\ & \hspace{100pt}
- (1+\nu_{1+}) j_+ \tilde{\mathcal{E}}_{02}\tilde{\jmath}_-
+ (1+\nu_{1-}) \tilde\jmath_+ \tilde{\mathcal{E}}_{20} j_- \Big)
\\ & \qquad - \frac{\hay(1+\nu_{1+})(1+\nu_{1-})}{1-\nu_{1+}\nu_{1-}} \Act_{\mathrm{WZ}}(g) - \hay \Act_{\mathrm{WZ}}(\tilde{g}) ~,
\end{split}\end{equation}
where
\begin{equation}\begin{split}
\tilde{\mathcal{E}}_{00} 
& = 1 - \frac{2}{1+\nu_{1+}\Ad_{\tilde g} + \nu_{1-} \nu_{3+} \Ad_g + \nu_{3+}\Ad_{\tilde g}\Ad_g} (1+\nu_{1+}\Ad_{\tilde g}) ~,
\\
\tilde{\mathcal{E}}_{22} 
& = 1 - \frac{2}{1+\nu_{1-}\nu_{3+}\Ad_g + \nu_{1+} \Ad_{\tilde g} + \nu_{3+}\Ad_g\Ad_{\tilde g}}(1+\nu_{1-}\nu_{3+}\Ad_g) ~,
\\
\tilde{\mathcal{E}}_{02} & = \frac{2}{1+\nu_{1+}\Ad_{\tilde g} + \nu_{1-} \nu_{3+} \Ad_g + \nu_{3+}\Ad_{\tilde g}\Ad_g} \Ad_{\tilde g} ~,
\\
\tilde{\mathcal{E}}_{20} & = \frac{2}{1+\nu_{1-}\nu_{3+}\Ad_g + \nu_{1+} \Ad_{\tilde g} + \nu_{3+}\Ad_g\Ad_{\tilde g}}\nu_{3+}\Ad_g ~.
\end{split}\end{equation}

\section{Backgrounds for the \texorpdfstring{$\Integer_2$}{Z2}-twisted \texorpdfstring{$\grp{SU}(2)$}{SU(2)} models}\label{sec:examples}

Having constructed the models, let us give some explicit examples of the backgrounds that follow in the simplest case of $\grp{G} = \grp{SU}(2)$.
We introduce the generators of $\alg{su}(2)$
\begin{equation}\label{su2_generators}
T_{1} = \begin{pmatrix}
0 & i \\
i & 0
\end{pmatrix} ~, \qquad
T_{2} = \begin{pmatrix}
0 & 1 \\
-1 & 0
\end{pmatrix} ~, \qquad
T_{3} = \begin{pmatrix}
i & 0 \\
0 & -i
\end{pmatrix} ~,
\end{equation}
with the usual matrix trace given by $\Tr(T_a T_b) = -2\delta_{ab}$.

We take the $\Integer_2$ automorphism $\sigma$ to act as
\begin{equation}\label{su2z2automorphism}
\sigma(T_3) = T_3 ~, \qquad \sigma(T_{1}) = - T_{1} ~, \qquad \sigma(T_2) = - T_2 ~.
\end{equation}
Here we have chosen an inner automorphism
\begin{equation}\label{eq:53}
\sigma = \Ad_{\exp(\frac{\pi}{2} T_3)} ~,
\end{equation}
since there are no outer automorphisms of $\alg{su}(2)$.
We will return to this in \secref{sec:equivalence}.

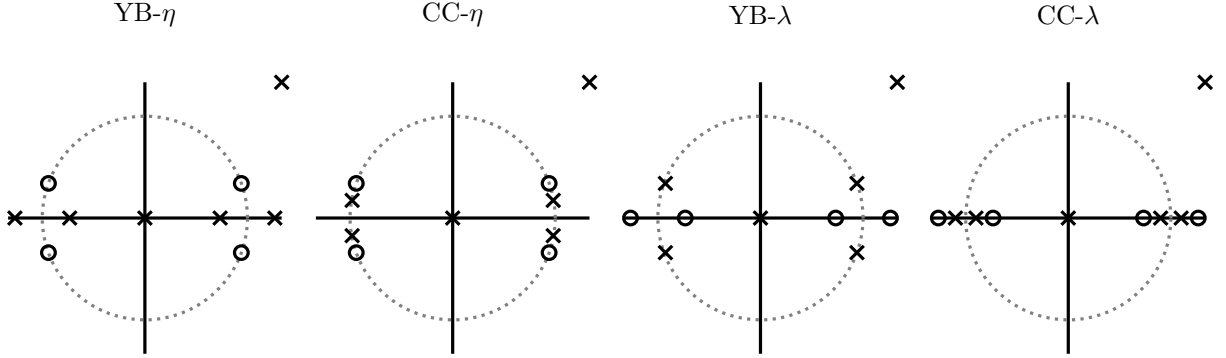
\begin{figure}
\begin{center}
\begin{tikzpicture}[scale=0.9]
\def\n{4.5}
\node at (3,3) {YB-$\eta$};
\draw[-,very thick] (1,0)--(5,0);
\draw[-,very thick] (3,-2)--(3,2);
\draw[dotted,very thick,gray] (3,0) circle (1.5);
\draw[-,very thick] (2.9,-0.1)--(3.1,0.1);
\draw[-,very thick] (2.9,0.1)--(3.1,-0.1);
\draw[-,very thick] (4.9,2.1)--(5.1,1.9);
\draw[-,very thick] (4.9,1.9)--(5.1,2.1);
\draw[-,very thick] (4.00,0.1)--(4.20,-0.1);
\draw[-,very thick] (4.00,-0.1)--(4.20,0.1);
\draw[-,very thick] (4.80,0.1)--(5.00,-0.1);
\draw[-,very thick] (4.80,-0.1)--(5.00,0.1);
\draw[very thick] (4.41,0.51) circle (0.1);
\draw[very thick] (4.41,-0.51) circle (0.1);
\draw[-,very thick] (1.20,0.1)--(1.00,-0.1);
\draw[-,very thick] (1.20,-0.1)--(1.00,0.1);
\draw[-,very thick] (2.00,0.1)--(1.80,-0.1);
\draw[-,very thick] (2.00,-0.1)--(1.80,0.1);
\draw[very thick] (1.59,0.51) circle (0.1);
\draw[very thick] (1.59,-0.51) circle (0.1);
\node at (3+2*\n,3) {YB-$\lambda$};
\draw[-,very thick] (1+\n,0)--(5+\n,0);
\draw[-,very thick] (3+\n,-2)--(3+\n,2);
\draw[dotted,very thick,gray] (3+\n,0) circle (1.5);
\draw[-,very thick] (2.9+\n,-0.1)--(3.1+\n,0.1);
\draw[-,very thick] (2.9+\n,0.1)--(3.1+\n,-0.1);
\draw[-,very thick] (4.9+\n,2.1)--(5.1+\n,1.9);
\draw[-,very thick] (4.9+\n,1.9)--(5.1+\n,2.1);
\draw[-,very thick] (4.37+\n, 0.36)--(4.57+\n, 0.16);
\draw[-,very thick] (4.37+\n, 0.16)--(4.57+\n, 0.36);
\draw[-,very thick] (4.37+\n, -0.36)--(4.57+\n, -0.16);
\draw[-,very thick] (4.37+\n, -0.16)--(4.57+\n, -0.36);
\draw[very thick] (4.41+\n,0.51) circle (0.1);
\draw[very thick] (4.41+\n,-0.51) circle (0.1);
\draw[-,very thick] (1.43+\n, 0.36)--(1.63+\n, 0.16);
\draw[-,very thick] (1.43+\n, 0.16)--(1.63+\n, 0.36);
\draw[-,very thick] (1.43+\n, -0.36)--(1.63+\n, -0.16);
\draw[-,very thick] (1.43+\n, -0.16)--(1.63+\n, -0.36);
\draw[very thick] (1.59+\n,0.51) circle (0.1);
\draw[very thick] (1.59+\n,-0.51) circle (0.1);
\node at (3+\n,3) {CC-$\eta$};
\draw[-,very thick] (1+2*\n,0)--(5+2*\n,0);
\draw[-,very thick] (3+2*\n,-2)--(3+2*\n,2);
\draw[dotted,very thick,gray] (3+2*\n,0) circle (1.5);
\draw[-,very thick] (2.9+2*\n,-0.1)--(3.1+2*\n,0.1);
\draw[-,very thick] (2.9+2*\n,0.1)--(3.1+2*\n,-0.1);
\draw[-,very thick] (4.9+2*\n,2.1)--(5.1+2*\n,1.9);
\draw[-,very thick] (4.9+2*\n,1.9)--(5.1+2*\n,2.1);
\draw[-,very thick] (4.31+2*\n,0.61)--(4.51+2*\n,0.41);
\draw[-,very thick] (4.31+2*\n,0.41)--(4.51+2*\n,0.61);
\draw[-,very thick] (4.31+2*\n,-0.41)--(4.51+2*\n,-0.61);
\draw[-,very thick] (4.31+2*\n,-0.61)--(4.51+2*\n,-0.41);
\draw[very thick] (4.10+2*\n,0) circle (0.1);
\draw[very thick] (4.90+2*\n,0) circle (0.1);
\draw[-,very thick] (1.51+2*\n,0.61)--(1.71+2*\n,0.41);
\draw[-,very thick] (1.51+2*\n,0.41)--(1.71+2*\n,0.61);
\draw[-,very thick] (1.51+2*\n,-0.41)--(1.71+2*\n,-0.61);
\draw[-,very thick] (1.51+2*\n,-0.61)--(1.71+2*\n,-0.41);
\draw[very thick] (1.10+2*\n,0) circle (0.1);
\draw[very thick] (1.90+2*\n,0) circle (0.1);
\node at (3+3*\n,3) {CC-$\lambda$};
\draw[-,very thick] (1+3*\n,0)--(5+3*\n,0);
\draw[-,very thick] (3+3*\n,-2)--(3+3*\n,2);
\draw[dotted,very thick,gray] (3+3*\n,0) circle (1.5);
\draw[-,very thick] (2.9+3*\n,-0.1)--(3.1+3*\n,0.1);
\draw[-,very thick] (2.9+3*\n,0.1)--(3.1+3*\n,-0.1);
\draw[-,very thick] (4.9+3*\n,2.1)--(5.1+3*\n,1.9);
\draw[-,very thick] (4.9+3*\n,1.9)--(5.1+3*\n,2.1);
\draw[-,very thick] (4.55+3*\n,0.1)--(4.75+3*\n,-0.1);
\draw[-,very thick] (4.55+3*\n,-0.1)--(4.75+3*\n,0.1);
\draw[-,very thick] (4.25+3*\n,0.1)--(4.45+3*\n,-0.1);
\draw[-,very thick] (4.25+3*\n,-0.1)--(4.45+3*\n,0.1);
\draw[very thick] (4.10+3*\n,0) circle (0.1);
\draw[very thick] (4.90+3*\n,0) circle (0.1);
\draw[-,very thick] (1.55+3*\n,0.1)--(1.75+3*\n,-0.1);
\draw[-,very thick] (1.55+3*\n,-0.1)--(1.75+3*\n,0.1);
\draw[-,very thick] (1.25+3*\n,0.1)--(1.45+3*\n,-0.1);
\draw[-,very thick] (1.25+3*\n,-0.1)--(1.45+3*\n,0.1);
\draw[very thick] (1.10+3*\n,0) circle (0.1);
\draw[very thick] (1.90+3*\n,0) circle (0.1);
\end{tikzpicture}
\end{center}
\caption{The configurations of poles and zeroes of the $\Integer_2$-equivariant trigonometric 1-forms for, from the left, (i) the YB-$\eta$ model with $(c_0,c)=(i,i)$, the (ii) CC-$\eta$ model with $(c_0,c)=(i,1)$, the (iii) YB-$\lambda$ model with $(c_0,c)=(1,i)$ and (iv) the CC-$\lambda$ model with $(c_0,c)=(1,1)$.}\label{fig3}
\end{figure}

For compatibility with $\Integer_2$-equivariance, the R-matrix $\tilde{\mathcal{R}}$ should commute with $\sigma$.
This condition is satisfied by the familiar Drinfel'd-Jimbo R-matrix~\cite{Drinfeld:1985rx,Jimbo:1985zk}, which acts as
\begin{equation}\label{eq:su2dj}
\tilde{\mathcal{R}}T_{3} = 0 ~, \qquad \tilde{\mathcal{R}} (T_1) = T_2 ~, \qquad \tilde{\mathcal{R}} (T_2) = - T_1 ~.
\end{equation}
Note that this is the case because we have chosen to align the Cartan subalgebra annihilated by $\tilde{\mathcal{R}}$ with the fixed point subalgebra of $\sigma$.
If we had chosen a rotated $\tilde{\mathcal{R}}$ it would not commute with $\sigma$.
Since $\alg{su}(2)$ is a compact Lie algebra, the Drinfel'd-Jimbo R-matrix satisfies the modified classical Yang-Baxter equation with $c_0 = i$.
Therefore, for the $\eta$-models we have $c_0 = i$.
For simplicity, for the YB deformations we also take the associated R-matrix $\mathcal{R}$ to be the same as $\tilde{\mathcal{R}}$, hence we also have $c=i$ for these models.
This means that for the examples below, for $\eta$-type boundary conditions we have $c=i$, while for $\lambda$-type we have $c=1$.
This, and the resulting reality conditions for the parameters $\lambda$ and $\chi$, are summarised in \tabref{tbl:realitysu2}.
\begin{table}
\centering
\begin{tabular}{|c||c|c||c|c||c|c|}
\hline
& $c_0 =$ & $c =$ & $\lambda \in$ & $\chi \in$ & $\alpha \in$ & $\beta \in$ \\ \hline\hline
YB-$\eta$ & $i$ & $i$ & $\Unit$ & $\Real$ & $i\Real$ & $\Real$ \\ \hline
CC-$\eta$ & $i$ & $1$ & $\Unit$ & $\Unit$ & $i\Real$ & $i\Real$ \\ \hline
YB-$\lambda$ & $1$ & $i$ & $\Real$ & $\Unit$ & $\Real$ & $i\Real$ \\ \hline
CC-$\lambda$ & $1$ & $1$ & $\Real$ & $\Real$ & $\Real$ & $\Real$ \\ \hline
\end{tabular}
\caption{The values of $c_0$ and $c$ in the $\grp{SU}(2)$ (YB,CC)-($\eta$,$\lambda$) models and the reality conditions for $\lambda$ and $\chi$.
The parameter $\lambda$ always belongs to the real line $\mathfrak{R}$, which is the real numbers $\Real$ for $c_0=1$ and the unit circle $\Unit$ for $c_0 = i$.}\label{tbl:realitysu2}
\end{table}
The configurations of poles and zeroes are shown in \figref{fig3}

The fields of our models are $g\in \grp{SU}(2)$ and $\tilde g \in \grp{U}(1)$ and we define the Maurer-Cartan 1-forms
\begin{equation}
j_{\pm} = g^{-1}\partial_{\pm}g ~, \qquad
\tilde{\jmath}_{\pm} = \tilde{g}^{-1}\partial_{\pm}\tilde{g} ~,
\end{equation}
and their projections
\begin{equation}
J^{(0)} = P_{0}j ~, \qquad J^{(1)} = P_{1}j ~,
\end{equation}
where
\begin{equation}
P_{0} = \frac{1}{2} (1 + \sigma) ~, \qquad P_{1} = \frac{1}{2} (1 - \sigma) ~.
\end{equation}
We parametrise the group-valued fields as
\begin{equation}
g = \exp\left(\frac{\varphi+\phi}{2} T_{3}\right) \exp\left(\theta T_{1}\right) \exp\left(\frac{\varphi-\phi}{2} T_{3}\right) ~, \label{g_field_general} \qquad
\tilde{g} = \exp\left(\varrho T_{3}\right) ~,
\end{equation}
where $\varrho$, $\theta$, $\varphi$, and $\phi$ are scalar fields

Let us recall that for the $\lambda$-models we have not fixed the internal gauge symmetries.
As explained in \secref{sec:bcgs} we cannot use these to fix $\tilde g = 1$ as we did for the $\eta$-models.
Instead we use them to fix the field $g$ using eq.~\eqref{eq:intgaugefixbc} for the YB-deformed models and eq.~\eqref{eq:intgaugefixbc2} for the CC-deformed models.
Thus, the final parametrisation of the group-valued fields $g$ and $\tilde{g}$ after fixing the internal gauge symmetries is
\begin{equation}\label{g_field_fixed}
\begin{aligned}
& \text{YB-$\eta$:} \qquad &
g &= \exp\big(\frac{\varphi+\phi}{2} T_{3}\big) \exp\big(\theta T_{1}\big) \exp\big(\frac{\varphi-\phi}{2} T_{3}\big) ~, \qquad
& \tilde{g} &= 1 ~,
\\
& \text{CC-$\eta$:} \qquad &
g &= \exp\big(\frac{\varphi+\phi}{2} T_{3}\big) \exp\big(\theta T_{1}\big) \exp\big(\frac{\varphi-\phi}{2} T_{3}\big) ~, \qquad
& \tilde{g} &= 1 ~,
\\
& \text{YB-$\lambda$:} \qquad &
g & = \exp\big(\varphi T_{3}\big) \exp\big(\theta T_{1}\big) ~, \qquad
& \tilde{g} &= \exp\big(\varrho T_{3}\big) ~,
\\
& \text{CC-$\lambda$:} \qquad &
g &= \exp\big(\frac{\varphi}{2} T_{3}\big) \exp\big(\theta T_{1}\big) \exp\big(\frac{\varphi}{2} T_{3}\big) ~, \qquad
& \tilde{g} &= \exp\big(\varrho T_{3}\big) ~.
\end{aligned}
\end{equation}

Substituting these parametrisations into the actions~\eqref{eq:genformactionsz2} and taking the normalised bilinear form $\tr = -2 \Tr$, where $\Tr$ is the matrix trace, we find 2d sigma models taking the standard form
\begin{equation}
\Act_{\mathrm{SM}} = \int d^2 \sigma \, (G_\ind{MN}(\Phi) + B_\ind{MN}(\Phi)) \partial_+ \Phi^\ind{M} \partial_- \Phi^\ind{M} ~,
\end{equation}
where $G_\ind{MN}(\Phi)$ and $B_\ind{MN}(\Phi)$ are interpreted as the components of a background metric and B-field respectively.
We find the following background for the YB-deformed $\Integer_2$-twisted $\grp{SU}(2)$ $\eta$-model
\begin{sequation}\label{eq:ybdefeta}
\begin{aligned}
G_{\theta\theta} &= 4 \Lambda_{\mathrm{YB}\text{-}\eta} (\chi^{\frac{1}{2}}+\chi^{-\frac{1}{2}}) (\lambda-\lambda^{-1})^2 ~, \\
G_{\phi\phi} &= 4 \Lambda_{\mathrm{YB}\text{-}\eta} (\chi^{\frac{1}{2}}+\chi^{-\frac{1}{2}}) ((\lambda-\lambda^{-1})^2 \cos^2\theta +4 (\lambda + \lambda^{-1} - \chi - \chi^{-1}) \sin^2\theta) \sin^2\theta ~, \\
G_{\varphi\varphi} &= 4 \Lambda_{\mathrm{YB}\text{-}\eta} (\chi^{\frac{1}{2}}+\chi^{-\frac{1}{2}}) ((\lambda-\lambda^{-1})^2\sin^2\theta + 4(\lambda+\lambda^{-1}-\chi-\chi^{-1}) \cos^2\theta)\cos^2\theta ~, \\
G_{\phi\varphi} &= 4 \Lambda_{\mathrm{YB}\text{-}\eta} (\chi^{\frac{1}{2}}+\chi^{-\frac{1}{2}}) (\lambda+\lambda^{-1} - 2\chi)(\lambda + \lambda^{-1} - 2 \chi^{-1}) \sin^2\theta \cos^2\theta ~,
\\
G_{\theta\phi} &= G_{\theta\varphi} = 0 ~,
\\
B_{\theta\phi} &= 2i \Lambda_{\mathrm{YB}\text{-}\eta} (\chi^{\frac{1}{2}}-\chi^{-\frac{1}{2}}) (\lambda-\lambda^{-1}) ( (\lambda + \lambda^{-1} - \chi - \chi^{-1}) \cos2\theta + (\chi^{\frac{1}{2}}+\chi^{-\frac{1}{2}})^2) \sin2\theta ~, \\
B_{\theta\varphi} &= 2i \Lambda_{\mathrm{YB}\text{-}\eta} (\chi^{\frac{1}{2}}-\chi^{-\frac{1}{2}}) (\lambda-\lambda^{-1}) ( (\lambda + \lambda^{-1} - \chi - \chi^{-1}) \cos2\theta - (\chi^{\frac{1}{2}}+\chi^{-\frac{1}{2}})^2) \sin2\theta ~,
\\ B_{\phi\varphi} &= 0 ~,
\end{aligned}
\end{sequation}
where
\begin{equation}\label{eq:ybdefeta2}
\Lambda_{\mathrm{YB}\text{-}\eta}(\theta;\lambda,\chi) = \frac{-4 i\hay}{(\chi^{\frac{1}{2}}+\chi^{-\frac{1}{2}}) (\lambda-\lambda^{-1}) (4+(\chi^{\frac{1}{2}}-\chi^{-\frac{1}{2}})^2 \sin^22\theta)} ~.
\end{equation}
The background of the undeformed $\Integer_2$-twisted $\grp{SU}(2)$ $\eta$-model can be found by setting $\chi = 1$ in the above metric and B-field.
Note that the B-field vanishes at this point.
This is in distinction to the untwisted $\grp{SU}(2)$ $\eta$-model, whose background is given below in eq.~\eqref{eq:ybeud1}, again setting $\chi = 1$.

For the CC-deformed $\Integer_2$-twisted $\grp{SU}(2)$ $\eta$-model we find the background
\begin{sequation}
\begin{aligned}
G_{\theta\theta} &=\Lambda_{\mathrm{CC}\text{-}\eta} (\lambda-\lambda^{-1}) \\ & \qquad\qquad ( (\lambda^{\frac{1}{2}}+\lambda^{-\frac{1}{2}})^2 (\chi^{\frac{1}{2}}-\chi^{-\frac{1}{2}})^2 + (\lambda + \lambda^{-1} - \chi - \chi^{-1})
(4 \sin^2\theta\cos^2\varphi - (\chi^{\frac{1}{2}}-\chi^{-\frac{1}{2}})^2 \cos^2\theta )) ~, \\
G_{\phi\phi} &=\Lambda_{\mathrm{CC}\text{-}\eta} (\lambda-\lambda^{-1}) (\chi-\chi^{-1})^2 \sin^2\theta ~, \\
G_{\varphi\varphi} &=\Lambda_{\mathrm{CC}\text{-}\eta} (\lambda-\lambda^{-1}) ((\lambda^{\frac{1}{2}}-\lambda^{-\frac{1}{2}})^2 (\chi^{\frac{1}{2}}+\chi^{-\frac{1}{2}})^2 - 4 (\lambda + \lambda^{-1} - \chi - \chi^{-1}) \cos^2\varphi) \cos^2\theta
~, \\
G_{\theta\varphi} &=\Lambda_{\mathrm{CC}\text{-}\eta} (\lambda-\lambda^{-1}) (\lambda + \lambda^{-1} - \chi - \chi^{-1}) \sin2\theta \sin2\varphi ~,
\\ G_{\theta\phi} &= G_{\phi\varphi} = 0 ~.
\\
B_{\theta\phi} &= -\Lambda_{\mathrm{CC}\text{-}\eta} (\chi-\chi^{-1}) (\lambda + \lambda^{-1} - \chi - \chi^{-1}) \sin2\theta \sin2\varphi ~, \\
B_{\theta\varphi} &= \Lambda_{\mathrm{CC}\text{-}\eta} (\chi-\chi^{-1}) ((\lambda^{\frac{1}{2}}-\lambda^{-\frac{1}{2}})^2 (\chi^{\frac{1}{2}}+\chi^{-\frac{1}{2}})^2 - 4 (\lambda + \lambda^{-1} - \chi - \chi^{-1}) \cos^2\varphi) \cos^2\theta ~,
\\
B_{\phi\varphi} &= 0 ~,
\end{aligned}
\end{sequation}
where
\begin{equation}
\Lambda_{\mathrm{CC}\text{-}\eta}(\theta,\varphi;\lambda,\chi) = \frac{-4i\hay(\lambda + \lambda^{-1} - \chi - \chi^{-1})}{(\chi-\chi^{-1})^2 ( (\lambda-\lambda^{-1})^2 - (\lambda + \lambda^{-1} - \chi - \chi^{-1}) \cos^2\theta ( (\lambda+\lambda^{-1} + 2\cos2\varphi))} ~.
\end{equation}
As mentioned above, taking $\chi \to 1$ in the CC-deformed models requires simultaneously taking $g\to1$, which here corresponds to $\theta, \varphi \to 0$ and is expected to give the non-abelian dual of the undeformed $\Integer_2$-twisted $\grp{SU}(2)$ $\eta$-model.

For the YB-deformed $\Integer_2$-twisted $\grp{SU}(2)$ $\lambda$-model the background is
\begin{sequation}
\begin{aligned}
G_{\varrho\varrho} &=\Lambda_{\mathrm{YB}\text{-}\lambda} (\lambda-\lambda^{-1}) (\chi^{\frac{1}{2}}+\chi^{-\frac{1}{2}}) (4 + (\chi^{\frac{1}{2}}-\chi^{-\frac{1}{2}})^2 \sin^22\theta) ~, \\
G_{\theta\theta} &= 4\Lambda_{\mathrm{YB}\text{-}\lambda} (\lambda-\lambda^{-1}) (\chi^{\frac{1}{2}}+\chi^{-\frac{1}{2}}) (\lambda + \lambda^{-1} - \chi - \chi^{-1}) ~, \\
G_{\varphi\varphi} &= 4\Lambda_{\mathrm{YB}\text{-}\lambda} (\lambda-\lambda^{-1}) (\chi^{\frac{1}{2}}+\chi^{-\frac{1}{2}}) (\lambda + \lambda^{-1} - \chi - \chi^{-1}) \sin^22\theta ~, \\
G_{\varrho\theta} &= 2i\Lambda_{\mathrm{YB}\text{-}\lambda} (\lambda-\lambda^{-1}) (\chi^{\frac{1}{2}}+\chi^{-\frac{1}{2}}) (\chi-\chi^{-1}) \sin2\theta ~,
\\ G_{\varrho\varphi} &= G_{\theta\varphi} = 0 ~,
\\
B_{\varrho\varphi} &= 8\Lambda_{\mathrm{YB}\text{-}\lambda} (\chi^{\frac{1}{2}}+\chi^{-\frac{1}{2}}) (\lambda + \lambda^{-1} - \chi - \chi^{-1}) \cos2\theta ~, \\
B_{\theta\varphi} &= 2i\Lambda_{\mathrm{YB}\text{-}\lambda} (\lambda^{\frac{1}{2}}+\lambda^{-\frac{1}{2}})^2 (\chi^{\frac{1}{2}}-\chi^{-\frac{1}{2}}) (\lambda + \lambda^{-1} - \chi - \chi^{-1}) \sin4\theta ~,
\\
B_{\varrho\theta} &= 0 ~,
\end{aligned} \end{sequation}
where
\begin{equation}
\Lambda_{\mathrm{YB}\text{-}\lambda}(\theta;\lambda,\chi) = -\frac{4\hay}{ (\chi^{\frac{1}{2}}+\chi^{-\frac{1}{2}}) (4(\lambda + \lambda^{-1} - \chi - \chi^{-1}) + (\lambda^{\frac{1}{2}}+\lambda^{-\frac{1}{2}})^2 (\chi^{\frac{1}{2}}-\chi^{-\frac{1}{2}})^2 \sin^22\theta)} ~.
\end{equation}
The background of the undeformed $\Integer_2$-twisted $\grp{SU}(2)$ $\lambda$-model can again be found by setting $\chi = 1$ in the above metric and B-field.

Finally, the background of the CC-deformed $\Integer_2$-twisted $\grp{SU}(2)$ $\lambda$-model is
\begin{sequation}
\begin{aligned}
G_{\varrho\varrho} &=\Lambda_{\mathrm{CC}\text{-}\lambda} (\chi-\chi^{-1})^2 (((\lambda^{\frac{1}{2}}-\lambda^{-\frac{1}{2}})^2 (\chi^{\frac{1}{2}}+\chi^{-\frac{1}{2}})^2 - 4(\lambda+\lambda^{-1} - \chi-\chi^{-1}) \cos^2\varphi) \cot^2\theta
+ (\lambda-\lambda^{-1})^2) ~, \\
G_{\theta\theta} &=\Lambda_{\mathrm{CC}\text{-}\lambda} (\lambda+\lambda^{-1} - \chi-\chi^{-1})^2 ((\lambda^{\frac{1}{2}}+\lambda^{-\frac{1}{2}})^2 (\chi^{\frac{1}{2}}-\chi^{-\frac{1}{2}})^2 + 4 (\lambda+\lambda^{-1} - \chi-\chi^{-1}) \cos^2\varphi) ~, \\
G_{\varphi\varphi} &=\Lambda_{\mathrm{CC}\text{-}\lambda} (\lambda+\lambda^{-1} - \chi-\chi^{-1})^2 ((\lambda^{\frac{1}{2}}-\lambda^{-\frac{1}{2}})^2 (\chi^{\frac{1}{2}}+\chi^{-\frac{1}{2}})^2 - 4 (\lambda+\lambda^{-1} - \chi-\chi^{-1}) \cos^2\varphi) \cot^2\theta ~, \\
G_{\varrho\theta} &= -2\Lambda_{\mathrm{CC}\text{-}\lambda} (\chi-\chi^{-1}) (\lambda+\lambda^{-1} - \chi-\chi^{-1})^2 \sin2\varphi \cot\theta ~, \\
G_{\varrho\varphi} &= -\Lambda_{\mathrm{CC}\text{-}\lambda} (\chi-\chi^{-1}) (\lambda + \lambda^{-1} - \chi - \chi^{-1}) ((\lambda^{\frac{1}{2}}-\lambda^{-\frac{1}{2}})^2 (\chi^{\frac{1}{2}}+\chi^{-\frac{1}{2}})^2
\\ & \hspace{210pt} - 4 (\lambda+\lambda^{-1} - \chi-\chi^{-1}) \cos^2\varphi) \cot^2\theta ~, \\
G_{\theta\varphi} &= 2\Lambda_{\mathrm{CC}\text{-}\lambda} (\lambda+\lambda^{-1} - \chi-\chi^{-1}) \sin2\varphi \cot\theta ~,
\\ B_{\varrho\theta} &= B_{\varrho\varphi} = B_{\theta\varphi} = 0 ~,
\end{aligned}
\end{sequation}
where
\begin{equation}\label{eq:ccdeflambda}
\Lambda_{\mathrm{CC}\text{-}\lambda} (\lambda,\chi) = -\frac{4\hay}{(\lambda + \lambda^{-1} - \chi - \chi^{-1}) (\lambda-\lambda^{-1}) (\chi-\chi^{-1})^2} ~.
\end{equation}
As for the CC-deformed $\Integer_2$-twisted $\eta$-model, taking $\chi \to 1$ requires simultaneously taking $g\to1$, i.e.~$\theta, \varphi \to 0$, and is expected to yield the non-abelian dual of the undeformed $\Integer_2$-twisted $\grp{SU}(2)$ $\lambda$-model.

As expected, and by construction, all four of the backgrounds are real when the parameters $\lambda$ and $\chi$ satisfy the appropriate reality conditions as given in \tabref{tbl:realitysu2}.
It is interesting to note that the background of the YB-deformed $\Integer_2$-twisted $\eta$-model is also real for $\lambda \in \Real$ and $\chi \in \Unit$ with $\hay \to -i \hay$.
Similarly, the background of the CC-deformed $\Integer_2$-twisted $\eta$-model is real for $\lambda \in \Real$ and $\chi \in \Real$ with $\hay \to -i \hay$.
This is not an accident, and follows from the identity $P_0 \tilde{\mathcal{R}} = \tilde{\mathcal{R}} P_0 = 0$, which can be easily checked from eqs.~\eqref{su2z2automorphism,eq:su2dj}.
Since $\tilde{\mathcal{R}}$ only enters through the boundary condition at the simple poles $0$ and $\infty$~\eqref{etalamp0}, which is an equation valued in $\alg{g}_0$ by $\Integer_N$-equivariance, it follows that it can be replaced by $P_0 \tilde{\mathcal{R}} = \tilde{\mathcal{R}} P_0$ without changing the analysis.
Here $P_0 \tilde{\mathcal{R}} = \tilde{\mathcal{R}} P_0$ is an antisymmetric solution to the modified classical Yang-Baxter equation on $\alg{g}_0$.
For the $\grp{SU}(2)$ models $\alg{g}_0$ is abelian, hence $P_0 \tilde{\mathcal{R}} = \tilde{\mathcal{R}} P_0 = 0$ is trivially a solution to both the non-split and split modified classical Yang-Baxter equations on $\alg{g}_0$.
This means that in this special case we can also take $c_0 = 1$ for the $\eta$-models, which corresponds to the analytic continuations described above, and still expect to find real sigma models.
This is in contrast to the untwisted models as can be seen from the background~\eqref{eq:ybeud1,eq:ybeud2} below.

We conclude this section by noting that for each of these backgrounds we have checked that the equations of motion that follow from the sigma model are equivalent to the zero-curvature equation of the Lax connections~\eqref{deformed_lax_connections_general} for all $z$.

\subsection{Equivalence of the untwisted and \texorpdfstring{$\Integer_2$}{Z2}-twisted \texorpdfstring{$\grp{SU}(2)$}{SU(2)} models.}

The YB-deformed $\Integer_2$-twisted $\grp{SU}(2)$ $\eta$-model~\eqref{eq:ybdefeta} has a $\grp{U}(1) \times\grp{U}(1)$ symmetry parametrised by shifts in $\phi$ and $\varphi$.
In addition, its background B-field is closed.
As previously anticipated, taking $\chi \to 1$ and then $\lambda \to 1$ while rescaling $\hay$ we find the background of the $\grp{SU}(2)$ principal chiral model.
Integrable deformations of the $\grp{SU}(2)$ principal chiral model preserving a $\grp{U}(1) \times\grp{U}(1)$ symmetry were studied in~\cite{Lukyanov:2012zt}, and in~\cite{Hoare:2014pna} it was shown that when the B-field is closed the deformed model is the bi-Yang-Baxter deformation~\cite{Klimcik:2008eq} of the $\grp{SU}(2)$ principal chiral model.
However, we have already seen that the bi-Yang-Baxter deformation of the principal chiral model is precisely the YB-deformed untwisted $\eta$-model~\eqref{eq:untwist}.
This suggests that the YB-deformed untwisted and $\Integer_2$-twisted $\grp{SU}(2)$ $\eta$-models are equivalent, where by equivalent we mean that they are the same up to a closed B-field and parameter-dependent field redefinitions.
This is indeed the case.
Again using the parametrisation~\eqref{g_field_fixed} and $\tr = -2\Tr$, the background for the YB-deformed untwisted $\eta$-model~\eqref{eq:untwist} is
\unskip\footnote{If we set $\theta = \arcsin r$, $\hay = \frac{1}{8(\kappa_- - \kappa_+)}$, $\lambda = \frac{1+\kappa_+ \kappa_- + i(\kappa_- - \kappa_+)}{\sqrt{1+\kappa_+^2}\sqrt{1+\kappa_-^2}}$ and $\chi = \frac{\sqrt{1+\kappa_-^2}}{\sqrt{1+\kappa_+^2}}$ in~\eqref{eq:ybdefeta,eq:ybdefeta2} the background metric of the YB-deformed untwisted $\grp{SU}(2)$ $\eta$-model becomes
\begin{equation*}
\frac{1}{1 + \kappa_-^2 (1 - r^2) + \kappa_+^2 r^2}\Big(
\frac{dr^2}{1-r^2} + (1 - r^2) (1 + \kappa_-^2 (1 - r^2))d\varphi^2
+ r^2 (1 + \kappa_+^2 r^2) d\phi^2
+ 2 \kappa_- \kappa_+ r^2 (1 - r^2) d\varphi d\phi \Big) ~,
\end{equation*}
which matches that of the bi-Yang-Baxter deformation of the $\grp{SU}(2)$ principal chiral model from the literature, see e.g.~\cite{Hoare:2014oua}.
Note that the parameters $\kappa_\pm \in \Real$, which is consistent with $\lambda \in \Unit$ and $\chi \in \Real$, and is in agreement with \tabref{tbl:realitysu2} for the YB-deformed $\eta$-model.}
\begin{sequation}\label{eq:ybeud1}
\begin{aligned}
G_{\theta\theta} &= \Delta_{\mathrm{YB}\text{-}\eta} (\lambda-\lambda^{-1})^2 ~, \\
G_{\phi\phi} &= \Delta_{\mathrm{YB}\text{-}\eta} ((\lambda - \lambda^{-1})^2 - (\lambda+\lambda^{-1}-2\chi^{-1})^2\sin^2 \theta)\sin^2\theta ~, \\
G_{\varphi\varphi} &= \Delta_{\mathrm{YB}\text{-}\eta} ((\lambda - \lambda^{-1})^2 - (\lambda+\lambda^{-1}-2\chi)^2\cos^2 \theta)\cos^2\theta ~, \\
G_{\phi\varphi} &= \Delta_{\mathrm{YB}\text{-}\eta}(\lambda+\lambda^{-1}-2\chi)(\lambda+\lambda^{-1} - 2\chi^{-1})\sin^2\theta\cos^2\theta ~, \\
G_{\theta\phi} &= G_{\theta\varphi} = 0 ~,
\\
B_{\theta\phi} &= -i \Delta_{\mathrm{YB}\text{-}\eta} (\lambda - \lambda^{-1}) (\lambda + \lambda^{-1} - 2\chi) \sin\theta\cos\theta ~, \\
B_{\theta\varphi} &= -i \Delta_{\mathrm{YB}\text{-}\eta} (\lambda - \lambda^{-1}) (\lambda + \lambda^{-1} - 2\chi^{-1}) \sin\theta\cos\theta ~, \\
B_{\phi\varphi} &= 0 ~,
\end{aligned}
\end{sequation}
where
\begin{equation}\label{eq:ybeud2}
\Delta_{\mathrm{YB}\text{-}\eta}(\theta;\lambda,\chi) = \frac{-4 i\hay}{(\lambda-\lambda^{-1})(\chi\cos^2\theta+\chi^{-1}\sin^2\theta)} ~.
\end{equation}
If we implement the field redefinition
\begin{equation}\label{eq:untitwsu2}
\sin \theta \to \frac{\sin\theta}{\sqrt{\sin^2\theta + \chi \cos^2\theta}} ~,
\end{equation}
in~\eqref{eq:ybdefeta,eq:ybdefeta2} we find~\eqref{eq:ybeud1,eq:ybeud2} up to a closed B-field demonstrating equivalence as claimed.

Switching from $\eta$-type to $\lambda$-type boundary conditions corresponds~\cite{Vicedo:2015pna} to a 2d duality transformation known as Poisson-Lie duality~\cite{Klimcik:1995ux,Klimcik:1995jn}.
If this involves going from $c=i$ to $c=1$ then an analytic continuation of the parameters is also needed~\cite{Hoare:2015gda}.
Since the fields at the simple poles $0$ and $\infty$ are constrained to lie in $\grp{G}_0$, it follows that the $\eta$-models and $\lambda$-models should be related by Poisson-Lie duality with respect to $\grp{G}_0$ together with an analytic continuation if going between the $c_0 = i$ and $c_0 = 1$ regimes.

For the $\grp{SU}(2)$ models we have that $\grp{G}_0 = \grp{U}(1)$ and the Poisson-Lie duality simply becomes a T-duality.
Explicitly, if we T-dualise $\Act^{(2)}_{\mathrm{YB}\text{-}\lambda}$ in the coordinate $\varrho$, which originated from the field $\tilde g$, with dual coordinate $\vartheta$, then after redefining $\vartheta \to \frac{2\hay\lambda}{\chi}(\varphi-\phi)$ and $\varphi \to \frac12(\varphi+\phi)$ we recover $\Act^{(2)}_{\mathrm{YB}\text{-}\eta}$ in the $c_0 = 1$ regime.
Similarly, if we T-dualise $\Act^{(2)}_{\mathrm{CC}\text{-}\lambda}$ in the coordinate $\varrho$ we recover $\Act^{(2)}_{\mathrm{CC}\text{-}\eta}$ in the $c_0 = 1$ regime after redefining $\vartheta \to - \frac{4\hay\lambda}{\chi} \phi$.
Let us recall that, in contrast to the untwisted models, the deformed $\Integer_2$-twisted $\grp{SU}(2)$ $\eta$-models are also real in the $c_0 = 1$ regime.

We also expect that the YB-deformed and CC-deformed models are related by Poisson-Lie duality with respect to $\grp{G}$ together with an analytic continuation.
This is consistent with the observation that the YB-deformed $\grp{SU}(2)$ models have a $\grp{U}(1)\times\grp{U}(1)$ symmetry while the CC-deformed $\grp{SU}(2)$ models only have a $\grp{U}(1)$ symmetry.

Based on the above discussion it is reasonable to state that the $\Integer_2$-twisted $\grp{SU}(2)$ models do not lead to new integrable sigma models.
The YB-deformed $\Integer_2$-twisted $\grp{SU}(2)$ $\eta$-model is equivalent to the bi-Yang-Baxter deformation of the $\grp{SU}(2)$ principal chiral model, and similarly the CC-deformed $\Integer_2$-twisted $\grp{SU}(2)$ $\eta$-model should be equivalent to its untwisted counterpart~\eqref{eq:comp1}.
These two models are expected to be related by Poisson-Lie duality and analytic continuation.
The deformed $\Integer_2$-twisted $\lambda$-models can then be found by continuing the $\eta$-models to the $c_0 = 1$ regime, which preserves reality, and T-dualising.
An analogous statement holds for the undeformed $\Integer_2$-twisted $\grp{SU}(2)$ $\eta$-model, which is equivalent to the Yang-Baxter deformation of the $\grp{SU}(2)$ principal chiral model~\cite{Fukushima:2020kta}, and can be mapped to the undeformed $\Integer_2$-twisted $\grp{SU}(2)$ $\lambda$-model by T-duality in the $c_0=1$ regime~\cite{Borsato:2024alk}.

\section{Comments on equivalence and outer automorphisms}\label{sec:equivalence}

Based on the discussion of the $\grp{SU}(2)$ models in the previous section, it is natural to ask when do we find inequivalent models from the $\Integer_N$-twisting.
As before, we define equivalence as being the same up to a closed B-field and parameter-dependent field redefinitions.
We will not give a complete answer to this question, but we do give examples where the models are inequivalent.

We focus on the simplest models, which are the untwisted and $\Integer_2$-twisted $\eta$-models
\begin{align}
\Act^{(1)}_\eta & = \int d^{2}\sigma \, \tr\Big( j_{+}\frac{1+\eta^{2}}{1-\eta\tilde{\mathcal{R}}}j_{-} \Big) ~,\label{eq:untwistedS1} \\
\Act^{(2)}_\eta &= \int d^{2}\sigma \, \tr\Big( j_{+}P_0\frac{1+\eta^{2}}{1-\eta\tilde{\mathcal{R}}}j_{-} + j_{+} P_1 j_{-} \Big)~, \label{eq:twistedS2}
\end{align}
where we have set $c_0 = i$, and define the operators
\begin{equation}\label{eq:o1o2}
\mathcal{O}_1 = \frac{1+\eta^{2}}{1-\eta\tilde{\mathcal{R}}} ~, \qquad \text{and} \qquad \mathcal{O}_2 = P_0\frac{1+\eta^{2}}{1-\eta\tilde{\mathcal{R}}} + P_1 ~.
\end{equation}
Thus far we have considered the $\grp{SU}(2)$ models with the Drinfel'd-Jimbo R-matrix~\eqref{eq:su2dj} and the inner $\Integer_2$ automorphism~\eqref{su2z2automorphism}.
It is therefore natural to ask what happens if we consider an automorphism that is not inner.

\subsection{Alternative real forms}

The first type of automorphism that we consider is one that is not inner in the real form, but is inner in the complexification.
For this we need to consider an alternative non-compact real form and we take $\alg{g} = \alg{sl}(2;\Real)$.
We can easily construct the $\grp{SL}(2;\Real)$ models by analytically continuing the $\grp{SU}(2)$ models from \secref{sec:examples}.
If we define
\begin{equation}
S_1 = -i T_1 ~, \qquad S_2 = T_2 ~, \qquad S_3 = -i T_3 ~,
\end{equation}
then $S_a$ are generators of $\alg{sl}(2;\Real)$.
Eq.~\eqref{g_field_general} then implies that the first step in the analytic continuation is to send
\begin{equation}
\theta \to i \theta ~, \qquad \varphi \to i \varphi ~, \qquad \phi \to i \phi ~, \qquad \varrho \to i \varrho ~.
\end{equation}
Now turning to the $\Integer_2$ automorphism~\eqref{su2z2automorphism} and R-matrix~\eqref{eq:su2dj}, we see that $\sigma$ preserves the $\alg{sl}(2;\Real)$ real form, however, we need to analytically continue $\tilde{\mathcal{R}} \to i \tilde{\mathcal{R}}$ and $\mathcal{R} \to i \mathcal{R}$.
This means that these are now split R-matrices, which corresponds to taking $c = c_0 = 1$ for the $\eta$-type boundary conditions.
From \tabref{tbl:realitysu2}, this implies that both $\lambda \in \Real$ and $\chi \in \Real$ for all four models.
Finally, we note that since $c_0$ appears as an overall factor in the twisted trigonometric 1-forms~\eqref{undeftwist,deftwist} we should also analytically continue $\hay \to - i\hay$.
It is possible to check that implementing this analytic continuation in the backgrounds~\eqref{eq:ybdefeta,-,eq:ccdeflambda} yields real sigma models as expected.

The $\Integer_2$ automorphism $\sigma$, which now acts as
\begin{equation}\label{sl2Rz2automorphism}
\sigma(S_3) = S_3 ~,\qquad \sigma(S_{1,2}) = -S_{1,2} ~,
\end{equation}
is no longer inner in the real form.
There is no $h \in \grp{SL}(2;\Real)$ such that $\sigma = \Ad_h$.
As a consistency check, it is straightforward to see that the analytic continuation of~\eqref{eq:53} gives $\sigma = \Ad_{\exp(-\frac{i \pi}{2} S_3)}$, but $\exp(-\frac{i \pi}{2} S_3) \notin \grp{SL}(2;\Real)$.
However, in spite of the automorphism not being inner in the real form, the map between the YB-deformed $\Integer_2$-twisted and untwisted $\eta$-models~\eqref{eq:untitwsu2} remains real after analytically continuing to $\grp{SL}(2;\Complex)$.
Therefore, again we do not find new models.

\subsection{Outer automorphisms for \texorpdfstring{$\grp{SU}(n)$}{SU(n)}}

As a more non-trivial example we consider the $\grp{SU}(n)$ models with $n>2$ and the $\Integer_2$ outer automorphism of $\alg{su}(n)$.
Using the Cartan-Weyl basis for $\alg{sl}(n;\Complex)$ defined in \appref{Cartan-Weyl_su(N)}, the $\alg{su}(n)$ real form is generated by
\begin{equation}\label{eq:sunrealform}
\{i H_i, i(E_i+F_i), (E_i - F_i), i(E_{i,r}+F_{i,r}), (E_{i,r} - F_{i,r}) \} ~.
\end{equation}
The standard Drinfel'd-Jimbo R-matrix~\cite{Drinfeld:1985rx,Jimbo:1985zk} then acts as
\begin{equation}\label{eq:dj}
\tilde{\mathcal{R}}H_{i} = 0 ~, \qquad \tilde{\mathcal{R}}E_{i} = iE_{i} ~, \qquad \tilde{\mathcal{R}}F_{i} = -iF_{i} ~,\qquad \tilde{\mathcal{R}}E_{i,r} = iE_{i,r} ~, \qquad \tilde{\mathcal{R}}F_{i,r} = -iF_{i,r} ~,
\end{equation}
which preserves the real form~\eqref{eq:sunrealform} and is an antisymmetric solution to the non-split modified classical Yang-Baxter equation.
Recalling that the rank of $\alg{su}(n)$ is $n-1$, the Dynkin diagram outer automorphism of $\alg{su}(n)$ can be represented graphically as
\begin{equation}
\begin{tikzpicture}
\foreach \x in {0,1,2,3,4} {
\draw[thick] (0.5 + 1*\x, 1) circle (5pt); 
}
\foreach \x in {0,1,3} {
\draw[thick] (0.7 + 1*\x, 1) -- ++(0.6,0); 
}
\draw[thick,dotted] (0.7 + 1*2, 1) -- ++(0.6,0); 
\foreach \x/\label in {0/1, 1/2, 2/3, 3/{n-2}, 4/{n-1}} {
\node at (0.5 + 1*\x, 1.5) {$\scriptstyle \label$};
}
\draw[thick,->] (0.7 + 1*5, 1) -- ++(0.6,0); 
\end{tikzpicture}
\qquad
\begin{tikzpicture}
\foreach \x in {0,1,2,3,4} {
\draw[thick] (0.5 + 1*\x, 1) circle (5pt); 
}
\foreach \x in {0,1,3} {
\draw[thick] (0.7 + 1*\x, 1) -- ++(0.6,0); 
}
\draw[thick,dotted] (0.7 + 1*2, 1) -- ++(0.6,0); 
\foreach \x/\label in {0/{n-1}, 1/{n-2}, 2/{n-3}, 3/2, 4/1} {
\node at (0.5 + 1*\x, 1.5) {$\scriptstyle \label$};
}
\end{tikzpicture}
\end{equation}
and we take its action on the generators to be
\unskip\footnote{\label{foot:z2autf}Note that eq.~\eqref{eq:z2aut} is related to the $\Integer_2$ outer automorphism
\begin{equation*}
\begin{gathered}
\sigma(H_{i}) = H_{n-i} ~, \qquad \sigma(E_{i}) = E_{n-i} ~, \qquad \sigma(F_{i}) = F_{n-i} ~,
\\
\sigma(E_{i,r}) = (-1)^rE_{n-i-r,r} ~, \qquad \sigma(F_{i,r}) = (-1)^rF_{n-i-r,r} ~,
\end{gathered}
\end{equation*}
by an inner automorphism that acts as $\Ad_{\exp(i h_i H_i)}$.
For example, a set of real numbers $\{h_i\}$ can be found such that $\exp(i h_i H_i) = \diag(1,-1,1,-1,\dots,-(-1)^n)$ in the defining representation, which achieves this.
In general, composing an outer automorphism with an inner automorphism can change the fixed point subalgebra $\alg{g}_0$.
For odd $n$ the two outer automorphisms both have $\alg{g}_0 = \alg{so}(n)$, while for even $n$ the fixed point subalgebra is now $\alg{g}_0 = \alg{sp}(\frac{n}{2})$.
Since $\rank \alg{sp}(\frac{n}{2}) = \rank \alg{so}(n)$ for even $n$, this does not change the symmetry analysis below.}
\begin{equation}\label{eq:z2aut}
\begin{gathered}
\sigma(H_{i}) = H_{n-i} ~, \qquad \sigma(E_{i}) = -E_{n-i} ~, \qquad \sigma(F_{i}) = -F_{n-i} ~,
\\
\sigma(E_{i,r}) = -E_{n-i-r,r} ~, \qquad \sigma(F_{i,r}) = -F_{n-i-r,r} ~.
\end{gathered}
\end{equation}
It is straightforward to see that $\sigma$ commutes with $\tilde{\mathcal{R}}$.
\unskip\footnote{It is also possible to take the Reshetikhin twist~\cite{Reshetikhin:1990ep} of the Drinfel'd-Jimbo R-matrix as our solution to the non-split modified classical Yang-Baxter equation, which acts on the positive and negative roots as in eq.~\eqref{eq:dj}, but as $\tilde{\mathcal{R}} H_i = \beta_i{}^j H_j$ on the Cartan generators, where $\beta_{i}{}^{j}\tr(H_jH_k)$ is an antisymmetric matrix of real numbers.
Adding a Reshetikhin twist corresponds to implementing TsT transformations in the integrable sigma model~\cite{Osten:2016dvf}.
When constructing the $\Integer_2$-twisted $\eta$-models we require that the Reshetikhin twist satisfies the additional constraint $\beta_i{}^j = \beta_{n-i}{}^{n-j}$ for the R-matrix to commute with either of the outer automorphisms in eq.~\eqref{eq:z2aut} or \foottref{foot:z2autf}.
This is consistent with the TsT interpretation since it corresponds to restricting to isometries that survive the $\Integer_2$-twisting.}
The fixed point subalgebra $\alg{g}_0 = \alg{so}(n)$, which has rank $\lfloor\frac{n}{2}\rfloor$.

To test whether the untwisted and $\Integer_2$-twisted $\eta$-models are equivalent or not we compare their global symmetries.
The right-acting symmetry of the principal chiral model
\begin{equation}
g \to g g_\ind{R} ~, \qquad g_\ind{R} \in \grp{G} ~.
\end{equation}
is broken to $\grp{H}_\ind{R}$ by the presence of the deforming operators $\mathcal{O}_1$ and $\mathcal{O}_2$, defined in eq.~\eqref{eq:o1o2}, in the actions~\eqref{eq:untwistedS1} and~\eqref{eq:twistedS2}.
Parametrising $g_\ind{R} = \exp^{X_\ind{R}}$ and considering infinitesimal transformations, we see that for a symmetry generated by $X_\ind{R} \in \alg{g}$ to survive the deformation we require that
\begin{equation}\label{eq:symmetries}
\ad_{X_{\ind{R}}}\mathcal{O}_{1,2} = \mathcal{O}_{1,2}\ad_{X_\ind{R}} ~,
\end{equation}
as an operator equation.

For the untwisted model this is equivalent to simply requiring
\begin{equation}\label{eq:rmatsym}
\ad_{X_{\ind{R}}}\tilde{\mathcal{R}} = \tilde{\mathcal{R}}\ad_{X_\ind{R}} ~,
\end{equation}
while for the twisted model we additionally have
\begin{equation}\label{eq:z2sym}
\ad_{X_{\ind{R}}}\sigma = \sigma\ad_{X_\ind{R}} ~.
\end{equation}
The first condition~\eqref{eq:rmatsym} is satisfied by $X_\ind{R} \in \alg{u}(1)^{\oplus\rank \alg{g}}$, the Cartan subalgebra of $\alg{g}$, while eq.~\eqref{eq:z2sym} is satisfied by $X_\ind{R} \in \alg{g}_0$, the fixed point subalgebra of $\sigma$.

Therefore, the right-acting symmetries of the untwisted and $\Integer_2$-twisted $\grp{SU}(n)$ $\eta$-models, with the R-matrix $\tilde{\mathcal{R}}$ and outer automorphism $\sigma$ defined in eqs.~\eqref{eq:dj} and~\eqref{eq:z2aut} respectively, are $\grp{H}_\ind{R} = \grp{U}(1)^{\rank \alg{g}}$ and $\grp{H}_\ind{R} = \grp{U}(1)^{\rank \alg{g}_0}$ respectively.
These are the Cartan subgroups of $\grp{G}$ and $\grp{G}_0$ generated by $\{iH_i\}$ and $\{i (H_i + H_{n-i})\}$, which have dimensions $\rank \alg{su}(n) = n-1$ and $\rank \alg{so}(n) = \lfloor \frac{n}{2}\rfloor$ respectively.
It follows that the untwisted and $\Integer_2$-twisted $\grp{SU}(n)$ $\eta$-models, with the latter twisted by a $\Integer_2$ outer automorphism of $\alg{su}(n)$, are not equivalent since they have different global symmetries.
\unskip\footnote{Let us note that if we take $\sigma$ to be an inner automorphism that is compatible with the Drinfel'd-Jimbo R-matrix~\eqref{eq:dj} with or without a Reshetikhin twist, as was done in \cite{Fukushima:2020kta}, then the untwisted and $\Integer_2$-twisted models generically have the same symmetry.
Indeed, since $\sigma$ should commute with $\tilde{\mathcal{R}}$, such an inner automorphism acts as $\Ad_{h_\ind{R}}$ with $h_{\ind{R}} \in \grp{H}_\ind{R} = \grp{U}(1)^{\rank \alg{g}}$.
Because the Cartan subgroup is abelian, we conclude that the adjoint action of the full $\grp{H}_\ind{R} = \grp{U}(1)^{\rank \alg{g}}$ still commutes with such a $\sigma$.
The $\grp{SU}(2)$ models considered in \secref{sec:examples} fall into this class.
The two outer $\Integer_2$ automorphisms in eq.~\eqref{eq:z2aut} and \foottref{foot:z2autf} are related by an inner automorphism of this type, hence for similar reasons the resulting $\Integer_2$-twisted models have the same global symmetries as we have already argued in \foottref{foot:z2autf}.}

\medskip

As an explicit example, let us consider $\alg{g} = \alg{su}(2) \oplus \alg{su}(2)$ and the outer automorphism that interchanges the two copies of $\alg{su}(2)$.
While $\alg{su}(2) \oplus \alg{su}(2)$ is not simple, the constructions described above work for such a setup assuming we take the bilinear form to not mix the two copies of $\alg{su}(2)$ and be normalised same for each.
Defining the generators
\begin{equation}
\begin{aligned}
&& T^+_1 &= \begin{pmatrix}
T_1 & 0 \\
0 & T_1
\end{pmatrix} ~,
& T^+_2 &= \begin{pmatrix}
T_2 & 0 \\
0 & T_2
\end{pmatrix} ~,
& T^+_3 &= \begin{pmatrix}
T_3 & 0 \\
0 & T_3
\end{pmatrix} ~, \\
&& T^-_1 &= \begin{pmatrix}
T_1 & 0 \\
0 & -T_1
\end{pmatrix} ~,
& T^-_2 &= \begin{pmatrix}
T_2 & 0 \\
0 & -T_2
\end{pmatrix} ~,
& T^-_3 &= \begin{pmatrix}
T_3 & 0 \\
0 & -T_3
\end{pmatrix} ~,
\end{aligned}
\end{equation}
where the generators $\{T_a\}$ of $\alg{su}(2)$ are defined in \eqref{su2_generators}, the $\Integer_2$ automorphism acts as
\begin{equation}
\sigma(T^+_a) = T^+_a ~, \qquad \sigma(T^-_a) = - T^-_a ~,
\end{equation}
hence the fixed point subalgebra $\alg{g}_0$ is generated by $\{T^+_a\}$.
We take the R-matrix to be the Drinfel'd-Jimbo R-matrix~\eqref{eq:su2dj} on each copy of $\alg{su}(2)$, implying that
\begin{equation}
\tilde{\mathcal{R}}T^+_3 = \tilde{\mathcal{R}}T^-_3 = 0, \qquad \tilde{\mathcal{R}}T^+_1 = -T^+_2, \qquad \tilde{\mathcal{R}}T^-_1 = -T^-_2, \qquad \tilde{\mathcal{R}}T^+_2 = T^+_1, \qquad \tilde{\mathcal{R}}T^-_2 = T^-_1.
\end{equation}

Parametrising the group-valued field $g \in \grp{SU}(2) \times \grp{SU}(2)$ as
\begin{equation}
g = \exp\left( \phi_1 T^+_3 + \phi_2 T^-_3 \right) \exp\left( \theta_1 T^+_1 + \theta_2 T^-_1 \right) \exp\left( \psi_1 T^+_3 + \psi_2 T^-_3 \right) ~,
\end{equation}
and again taking $\tr = -2\Tr$, the non-vanishing components for the metric and B-field of the untwisted and $\Integer_2$-twisted $\grp{SU}(2) \times \grp{SU}(2)$ $\eta$-models are
\begin{align*}
&\textbf{Untwisted model} && \textbf{Twisted model}
\\
&G_{\theta_1 \theta_1} = G_{\theta_2 \theta_2} = 8 ~,
&
&G_{\theta_1 \theta_1} = G_{\theta_2 \theta_2} = 8 ~,
\\
&G_{\phi_1 \phi_1} = G_{\phi_2 \phi_2} = 8 +4\eta^2 (1 + \cos4\theta_1 \cos4\theta_2) ~,
&\qquad
&\begin{aligned}
G_{\phi_1 \phi_1} & = 8 (1 + \eta^2 \cos^22\theta_1\cos^22\theta_2 ) ~, \\
G_{\phi_2 \phi_2} & = 8 (1 + \eta^2 \sin^22\theta_1\sin^22\theta_2 ) ~,
\end{aligned}
\\
&G_{\psi_1 \psi_1} = G_{\psi_2 \psi_2} = 8 (1+\eta^2) ~,
&
&\begin{aligned}
G_{\psi_1 \psi_1} & = 8 (1+\eta^2) ~, \\
G_{\psi_2 \psi_2} & = 8 ~,
\end{aligned}
\\
&G_{\phi_1 \psi_1} = G_{\phi_2 \psi_2} = 8 (1+\eta^2) \cos2\theta_1 \cos2\theta_2 ~,
&
&\begin{aligned}
G_{\phi_1 \psi_1} &= 8 (1+\eta^2) \cos2\theta_1 \cos2\theta_2~, \\
G_{\phi_2 \psi_2} &= 8 \cos2\theta_1 \cos2\theta_2~,
\end{aligned}
\\
&G_{\phi_1 \phi_2} = -4 \eta^2 \sin4\theta_1 \sin4\theta_2 ~,
&&
G_{\phi_1 \phi_2} = -2\eta^2 \sin4\theta_1 \sin4\theta_2 ~,
\refstepcounter{equation} \tag{\theequation}
\\
&G_{\phi_1 \psi_2} = G_{\phi_2 \psi_1} = -8 (1+\eta^2) \sin2\theta_1 \sin2\theta_2 ~,
&&
\begin{aligned}
G_{\phi_1 \psi_2} &= -8 \sin2\theta_1 \sin2\theta_2 ~, \\
G_{\phi_2 \psi_1} &= -8 (1+\eta^2) \sin2\theta_1 \sin2\theta_2 ~,
\end{aligned}
\\
&B_{\theta_1 \phi_1} = B_{\theta_2 \phi_2} = -8\eta \sin2\theta_1 \cos2\theta_2 ~,
&&
\begin{aligned}
B_{\theta_1 \phi_1} & = -8\eta \sin2\theta_1 \cos2\theta_2 \cos^22\psi_2 ~,\\
B_{\theta_2 \phi_2} & = -8\eta \sin2\theta_1 \cos2\theta_2 \sin^22\psi_2 ~,
\end{aligned}
\\
&B_{\theta_1 \phi_2} = B_{\theta_2 \phi_1} = -8\eta \cos2\theta_1 \sin2\theta_2 ~,
&&
\begin{aligned}
B_{\theta_1 \phi_2} & = -8\eta \cos2\theta_1 \sin2\theta_2 \cos^22\psi_2 ~, \\
B_{\theta_2 \phi_1} & = -8\eta \cos2\theta_1 \sin2\theta_2 \sin^22\psi_2 ~,
\end{aligned}
\\
&&& B_{\theta_1 \theta_2} = 4\eta \sin4\psi_2 ~,
\\
&&& B_{\phi_1 \phi_2} = -2\eta (\cos4\theta_1-\cos4\theta_2) \sin4\psi_2 ~.
\end{align*}
The first thing we note is that the untwisted model has a manifest $\grp{U}(1)^4$ global symmetry.
Two of these, shifts in $\phi_1$ and $\phi_2$, come from the unbroken left-acting symmetries, while the other two, shifts in $\psi_1$ and $\psi_2$ correspond to the right-acting Cartan symmetry.
The twisted model only has a manifest $\grp{U}(1)^3$ global symmetry, and we see explicitly that the right-acting Cartan symmetry not generated by an element of $\alg{g}_0$, i.e.~shifts in $\psi_2$, is no longer a symmetry.
A further indication that these two backgrounds are not equivalent is that the B-field for the untwisted model is closed, while for the $\Integer_2$-twisted model it is not.

\section{Conclusions}\label{sec:conclusions}

In this paper we have introduced a new class of integrable sigma models, the $\Integer_N$-twisted trigonometric sigma models.
Starting from 4d Chern-Simons on the product of 2d space-time and a cylinder, these models are constructed by introducing a $\Integer_N$ branch cut along the non-compact direction of the cylinder.
We have described the general procedure for constructing real Lorentz-invariant $\Integer_N$-twisted trigonometric sigma models and implemented this for a number of explicit examples.
Two of these are the $\Integer_N$-twisted $\eta$-model and $\lambda$-model whose actions are given in eqs.~\eqref{eq:actionetatwist} and~\eqref{eq:actionlambdatwist}, while their Lax connections can be found in~\eqref{eq:laxetatwist} and~\eqref{eq:laxlambdatwist}.
These models are expected to be related by Poisson-Lie duality with respect to the fixed-point subgroup of the $\Integer_N$ automorphism $\sigma$, $\grp{G}_0$.
The remaining four, constructed in \secref{sec:double-deformation} are deformations of these models, or their non-abelian duals with respect to the global $\grp{G}$ symmetry, and, in the nomenclature we have introduced, are the:
\begin{itemize}
\item YB-deformed $\Integer_2$-twisted $\eta$-model,
\item CC-deformed $\Integer_2$-twisted $\eta$-model,
\item YB-deformed $\Integer_2$-twisted $\lambda$-model,
\item CC-deformed $\Integer_2$-twisted $\lambda$-model.
\end{itemize}
Here the YB-deformed models have the $\Integer_N$-twisted $\eta$-model and $\lambda$-model as their undeformed limits, while the CC-deformed models have their non-abelian duals.
The specialisation to $N=2$ for these doubly-deformed models is for practical reasons and there is, in principle, no obstruction to generalising to any $N$.

Having introduced this class of models, there are a number of interesting directions to explore, including whether the twisted sigma models are equivalent to untwisted sigma models, their underlying integrable structure, and whether we can generalise the construction to other symmetric sigma models, including those that describe the worldsheet theories of strings.

\paragraph{Equivalence to untwisted models.}
We have started to answer the first of these questions.
In particular, we argued that the $\Integer_2$-twisted $\grp{SU}(2)$ and $\grp{SL}(2,\Real)$ models, with $\sigma$ an inner automorphism (either in the real form or complexified group), are all related to the untwisted models by parameter-dependent field redefinitions, dropping a closed B-field and 2d space-time dualities.
We then showed that the untwisted and $\Integer_2$-twisted $\grp{SU}(n)$ $\eta$-models, with $\sigma$ given by an outer automorphism, have different global symmetries.
Moreover, in the closely related case of $\grp{SU}(2) \times \grp{SU}(2)$ with the automorphism that swaps the two copies, the untwisted model has a closed B-field, while the $\Integer_2$-twisted one does not.
Therefore, in these cases the two models cannot be equivalent.
While we leave a full analysis for future work, it is natural to speculate that the inequivalent $\eta$-models might be enumerated by the outer automorphisms of the Lie algebra $\alg{g}$.

It is also interesting to note that for the $\eta$-models the R-matrix $\tilde{\mathcal{R}}$ is required to commute with $\sigma$, while for the $\lambda$-models there is no analogous restriction.
Therefore, if there are $\Integer_N$ automorphisms with the property that there exists no R-matrix, either split or non-split, that commutes with $\sigma$, the $\Integer_N$-twisted $\eta$-model would not exist, while the $\Integer_N$-twisted $\lambda$-model would.

\paragraph{Underlying integrable structure.}
Much remains to be understood with regards to the integrable structure of these models.
A key next step would be to analyse them in the Hamiltonian formalism, computing the Poisson bracket of the Lax matrix and extracting the conserved charges~\cite{Lacroix:2017isl,Lacroix:2018njs}, and investigating their underlying affine Gaudin models~\cite{Vicedo:2017cge,Vicedo:2019dej,Lacroix:2020flf}.
It is also interesting to note that the Yang-Baxter deformation of the principal chiral model has $q$-deformed symmetries~\cite{Delduc:2016ihq,Delduc:2017brb}.
For the same finite Lie algebra $\alg{g}$, we can consider untwisted or twisted affinisations, each of which can be $q$-deformed.
Inequivalent twistings are again classified by outer automorphisms of $\alg{g}$, hence we may ask if the symmetries of the $\Integer_N$-twisted $\eta$-models are given by the $q$-deformed twisted affine Lie algebras.
Derivations of conserved charges from the monodromy matrix for the untwisted and $\Integer_2$-twisted $\grp{SU}(2)$ $\eta$-models supporting this idea appear in \cite{Appadu:2017bnv}.

It would also be interesting to understand the $\mathcal{E}$-model formulation~\cite{Klimcik:1995dy,Klimcik:1996nq} of the twisted models, which unifies models with the same twist function in a single first-order 2d model on a doubled space.
Different boundary conditions then correspond to integrating out different sets degrees of freedom.
For example, both the Yang-Baxter deformation of the principal chiral model, the untwisted $\eta$-model, and the current-current deformation of the WZW model, the untwisted $\lambda$-model, can be found from the same $\mathcal{E}$-model~\cite{Klimcik:2015gba}.
$\mathcal{E}$-models can be found directly from 4d Chern-Simons \cite{Lacroix:2020flf}, and given the importance of gauge symmetries in our analysis, degenerate $\mathcal{E}$-models may play a central role~\cite{Klimcik:2019kkf,Liniado:2023uoo}.

Let us also briefly comment on quantum integrability and S-matrices.
For compact Lie groups $\grp{G}$, the principal chiral model is asymptotically free and strongly-coupled in the IR.
Their massive S-matrices are tensor products of two $\grp{G}$-invariant R-matrices, corresponding to the left-acting and right-acting symmetries, each with an infinite-dimensional Yangian symmetry~\cite{Wiegmann:1984ec,Ogievetsky:1984pv,Ogievetsky:1987vv}.
The $q$-deformations of these R-matrices are known and are in correspondence with the $q$-deformed untwisted and twisted affine Lie algebras~\cite{Jimbo:1985ua}.
Again, since these are classified by outer automorphisms, we may ask if the quantum S-matrices of the untwisted models are tensor products of a Yangian and a $q$-deformed untwisted R-matrix, while those of the twisted models are tensor products of a Yangian and a $q$-deformed twisted R-matrix.
Both the untwisted and twisted R-matrices for the same Lie group $\grp{G}$ have the same Yangian R-matrix as their undeformed limit, mirroring the fact that the untwisted and twisted sigma models both limit to the principal chiral model.
As a starting point it would be important to compute the renormalisation group flow of the twisted models either directly in 2d~\cite{Friedan:1980jf,Braaten:1985is} or using general results from 4d Chern-Simons~\cite{Derryberry:2021rne,Lacroix:2024wrd}.

\paragraph{Generalisations to other symmetric sigma models.}

There are a number of generalisations to other symmetric sigma models that it would be possible to pursue.
In this paper we have considered disorder defects, however, there is another important class known as order defects.
While disorder defects lead to a non-ultralocal Poisson bracket of the Lax matrix, order defects lead to an ultralocal bracket making their integrability analysis more tractable.
Therefore, it would be interesting to consider $\Integer_N$-twistings of trigonometric integrable models constructed from order defects such as those in~\cite{Caudrelier:2022jco,Costello:2019tri}, or mixed order and disorder defects.

\begin{figure}
\begin{center}
\begin{tikzpicture}
\draw[-,very thick] (-2,0)--(2,0);
\draw[-,very thick] (0,-2)--(0,2);
\draw[dotted,very thick,gray] (0,0) circle (1.5);
\draw[-,very thick] (-0.1,-0.1)--(0.1,0.1);
\draw[-,very thick] (-0.1,0.1)--(0.1,-0.1);
\draw[-,very thick] (1.9,2.1)--(2.1,1.9);
\draw[-,very thick] (1.9,1.9)--(2.1,2.1);
\draw[-,very thick] (1,0.1)--(1.2,-0.1);
\draw[-,very thick] (1,-0.1)--(1.2,0.1);
\draw[-,very thick] (1.8,0.1)--(2,-0.1);
\draw[-,very thick] (1.8,-0.1)--(2,0.1);
\draw[very thick] (1.41,0.51) circle (0.1);
\draw[very thick] (1.41,-0.51) circle (0.1);
\draw[-,very thick] (-1.8,0.1)--(-2,-0.1);
\draw[-,very thick] (-1.8,-0.1)--(-2,0.1);
\draw[-,very thick] (-1,0.1)--(-1.2,-0.1);
\draw[-,very thick] (-1,-0.1)--(-1.2,0.1);
\draw[very thick] (-1.41,0.51) circle (0.1);
\draw[very thick] (-1.41,-0.51) circle (0.1);
\draw[-,very thick] (-0.1,1.2)--(0.1,1.4);
\draw[-,very thick] (-0.1,1.4)--(0.1,1.2);
\draw[-,very thick] (-0.1,1.6)--(0.1,1.8);
\draw[-,very thick] (-0.1,1.8)--(0.1,1.6);
\draw[very thick] (0.41,1.45) circle (0.1);
\draw[very thick] (-0.41,1.45) circle (0.1);
\draw[-,very thick] (-0.1,-1.6)--(0.1,-1.8);
\draw[-,very thick] (-0.1,-1.8)--(0.1,-1.6);
\draw[-,very thick] (-0.1,-1.2)--(0.1,-1.4);
\draw[-,very thick] (-0.1,-1.4)--(0.1,-1.2);
\draw[very thick] (0.41,-1.45) circle (0.1);
\draw[very thick] (-0.41,-1.45) circle (0.1);
\end{tikzpicture}
\end{center}
\caption{An example of a $\Integer_2$-equivariant twist function where the 2d integrable sigma model would be a 2-field model.}\label{fig4}
\end{figure}
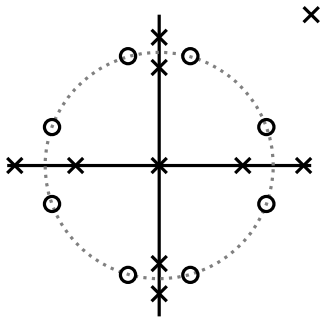

Returning to disorder defects, the simplest extension is to ask how the construction is modified if we allow the residues at the fixed points $z_{\mathrm{f.p.}} = 0,\infty$ to sum to a non-zero value.
In the untwisted case with $\eta$-type boundary conditions this gives the Yang-Baxter deformation of the principal chiral model plus WZ term \cite{Hoare:2020mpv}.
Another direction would be to explore coupled models along the lines of~\cite{Delduc:2018hty,Bassi:2019aaf}.
This would generalise our explicit examples to models with more fields, where on the original sheet we have more than a double pole or pair of simple poles.
If we have $N_{\mathrm{f}}$ double poles and pairs of simple poles, then we expect to find a twisted model with $N_{\mathrm{f}}\dim\grp{G}$ degrees of freedom.
An example of such a twist function on the $N$-fold cover is given in \figref{fig4}.

Finally, let us comment on the generalisation to coset models, including the symmetric space sigma model and the $\Integer_4$-coset models that describe the worldsheet theories of strings.
Such models have been studied from the perspective of 4d Chern-Simons in, e.g., \cite{Costello:2019tri,Tian:2020ryu,Fukushima:2020dcp,Tian:2020pub,Costello:2020lpi,Berkovits:2024reg}.
Coset models are constructed from 4d Chern-Simons by introducing a branch cut between two marked points that are not poles or zeroes.
As an example, if we start from a twist function with $N_{\mathrm{f}}+1$ double poles and pairs of simple poles on $\CP^1$, following the standard construction, the resulting 2d integrable sigma model will have target space $\grp{G}^{N_{\mathrm{f}}+1} /\grp{G}$ \cite{Bassi:2019aaf}.
Introducing a $\Integer_N$ branch cut between two marked points and imposing $\Integer_N$-equivariance then yields an integrable sigma model on $\grp{G}^{N_{\mathrm{f}}+1}/\grp{G}_0$.
This class of sigma models have been constructed from affine Gaudin models in~\cite{Arutyunov:2020sdo}.
Note that for a $\Integer_N$ branch cut between marked points, the branch points become zeroes of order $N-1$ on the $N$-fold cover, in contrast with the construction described in this paper where the branch points remain simple poles.

In principle, to construct a $\Integer_N$-twisted $\Integer_M$-coset model we would need to introduce two branch cuts on the original sheet, a $\Integer_N$ branch cut between two simple poles and a $\Integer_M$ branch cut between two marked points.
The simplest example would be the $\Integer_2$-twisted $\Integer_2$-coset $\eta$-model, for which the untwisted model is the Yang-Baxter deformation of the symmetric space sigma model~\cite{Delduc:2013fga}.
Since this involves introducing more than one branch cut it appears that the Riemann surface is higher genus and the resulting model, if it exists, would be elliptic.
Disorder defects in elliptic models have been investigated, e.g. in~\cite{Lacroix:2023qlz}, and it would be interesting to see if it is possible to construct such a $\Integer_N$-twisted $\Integer_M$-coset model.

The elliptic sigma models of~\cite{Lacroix:2023qlz} are related to the trigonometric sigma models in this paper.
Working with $\grp{G} = \grp{SL}(n,\Real)$, the Riemann surface of 4d Chern-Simons is taken to be a torus and $\Integer_n \times \Integer_n$-equivariance is imposed.
The $\Integer_n \times \Integer_n$ action on the torus is along its two cycles, similar to the $\Integer_N$ action we have considered, which acts along the cycle of the cylinder, while the $\Integer_n$ automorphisms of the Lie algebra are particular inner automorphisms.
The trigonometric limit corresponds to decompactifying one of the cycles and yields a special case of the $\Integer_N$-twisted $\eta$-model.
For $\grp{G} = \grp{SL}(n,\Real)$ this is a $\Integer_n$-twisted $\eta$-model, where the R-matrix $\tilde{\mathcal{R}}$ is constructed from the $\Integer_n$ automorphism that previously implemented the $\Integer_n$-equivariance along the now decompactified cycle and coincides with a particular Reshetikhin twist of the split Drinfel'd-Jimbo R-matrix.
For $n=2$ there is no Reshetikhin twist and the whole construction can be analytically-continued to the compact real form $\grp{SU}(2)$ while preserving reality.
In this case the resulting sigma model is the elliptic deformation of the $\grp{SU}(2)$ principal chiral model~\cite{Cherednik:1981df} whose trigonometric limit is the $\Integer_2$-twisted $\grp{SU}(2)$ $\eta$-model.

If it is possible to construct a $\Integer_N$-twisted $\Integer_M$-coset model, this may be a way to bypass the presence of a curvature singularity in the supergravity backgrounds for the $\eta$ deformations of $\AdS$ superstrings described by $\Integer_4$ supercosets~\cite{Hoare:2018ngg,Seibold:2019dvf}.

\subsection*{Acknowledgements}
We would like to thank R.~Borsato, L.T.~Cole, R.A.~Cullinan, S.~Lacroix, J.L.~Miramontes, A.L.~Retore, E.~Viana and B.~Vicedo for related discussions, and the authors of \cite{Borsato:2024alk} for sharing an early version of their draft.
We also thank E.~Viana for bringing to our attention the possibility of missing coefficients in the $\Integer_N$-twisted $\eta$-model action in the first version of this preprint.
The work of BH was supported by a UKRI Future Leaders Fellowship (grant number MR/T018909/1).
The work of RH was supported by an STFC PhD Studentship (grant number ST/X508342/1 (2713401)).

\appendix

\section{Conventions}\label{app:conv}

\subsection{Lie bracket and symmetric ad-invariant bilinear form}\label{app:conv1}

We consider a simple Lie algebra $\alg{g}^{\Complex}$ over $\Complex$ with antisymmetric Lie bracket $[,]$ and symmetric ad-invariant bilinear form $\tr()$.
We extend these to take arguments that are forms on a $d$-dimensional manifold with coordinates $x^i$ in the standard way.
Let $A_m$ and $A_n$ be an $m$-form and an $n$-form.
We have that $[A_m, A_n]$ is an $m+n$-form valued in $\alg{g}^{\Complex}$ explicitly given by
\begin{equation}
[A_m, A_n] = \frac{1}{m!n!} [(A_m)_{i_1i_2\dots i_m}, (A_n)_{j_1j_2\dots j_n}] dx^{i_1} \wedge dx^{i_2} \wedge \dots \wedge dx^{i_m} \wedge dx^{j_1} \wedge dx^{j_2} \wedge \dots \wedge dx^{j_n} ~,
\end{equation}
and with the symmetry property
\begin{equation}
[A_m,A_n] = - (-1)^{mn}[A_n,A_m] ~.
\end{equation}
Moreover, $\tr(A_m A_n)$ is an $m+n$-form valued in $\Complex$ explicitly given by
\begin{equation}
\tr( A_m A_n ) = \frac{1}{m!n!} \tr((A_m)_{i_1i_2\dots i_m} (A_n)_{j_1j_2\dots j_n}) dx^{i_1} \wedge dx^{i_2} \wedge \dots \wedge dx^{i_m} \wedge dx^{j_1} \wedge dx^{j_2} \wedge \dots \wedge dx^{j_n} ~,
\end{equation}
and with the symmetry property
\begin{equation}
\tr(A_m A_n)= (-1)^{mn}\tr(A_nA_m) ~.
\end{equation}
With these definitions we have the following generalisation of ad-invariance
\begin{equation}
\tr(A_m[A_n,A_p]) = \tr([A_m,A_n]A_p) ~,
\end{equation}
where $A_p$ is a $p$-form.

\subsection{Cartan-Weyl basis for \texorpdfstring{$\alg{sl}(n;\Complex)$}{sl(n;C)}}\label{Cartan-Weyl_su(N)}

Defining $\mathbf{e}_{i,j}$ to be the matrix unit with a 1 in the $i$th row and $j$th column, the Cartan-Weyl generators for $\alg{sl}(n;\Complex)$ are given by
\begin{equation}\begin{gathered}
H_{i} = \mathbf{e}_{i,i} - \mathbf{e}_{i+1,i+1} ~, \qquad E_{i} = \mathbf{e}_{i,i+1} ~, \qquad F_{i} = \mathbf{e}_{i+1,i} ~, \qquad i = 1,\dots n-1 ~, \\ E_{i,r} = \mathbf{e}_{i,i+r+1} ~, \qquad F_{i,r} = \mathbf{e}_{i+r+1,i} ~, \qquad r = 1,\dots n-1-i ~.
\end{gathered}\end{equation}
The Cartan-Weyl basis is divided into primary and descendent generators, see, e.g.,~\cite{Serre2000}.
The primary generators are divided into three groups: the diagonal $H$ generators, the upper triangular $E$ generators, and the lower triangular $F$ generators, also known as the Cartan generators, the positive simple roots and the negative simple roots respectively.
The number of each type of primary generator is equal to the rank of the Lie algebra, which for $\alg{sl}(n;\Complex)$ is $\rank \alg{sl}(n;\Complex) = n-1$.
The descendent generators are of two types: the descendant $E$ generators, which are formed using commutators of the primary $E$ generators, and the descendant $F$ generators, formed using the commutators of the primary $F$ generators.
These are, respectively, the remaining positive and negative non-simple roots.

The primary generators of the Cartan-Weyl basis $\{H_{i},E_{i},F_{i}\}$ satisfy the Chevalley-Serre relations
\begin{equation}\begin{gathered}
[H_{i},H_{j}] = 0 ~, \qquad [E_{i},F_{j}] = \delta_{ij}H_{i} ~, \qquad [H_{i},E_{j}] = A_{ij}E_{j} ~, \qquad [H_{i},F_{j}] = -A_{ij}F_{j} \\
\ad_{E_{i}}^{1-A_{ij}} E_{j} = 0, \qquad \ad_{F_{i}}^{1-A_{ij}} F_{j} = 0,
\end{gathered}\end{equation}
where $A_{ij}$ is the Cartan matrix.

\begin{bibtex}[\jobname]

@article{Arutyunov:2020sdo,
author = "Arutyunov, Gleb and Bassi, Cristian and Lacroix, Sylvain",
title = "{New integrable coset sigma models}",
eprint = "2010.05573",
archivePrefix = "arXiv",
primaryClass = "hep-th",
reportNumber = "ZMP-HH/20-19",
doi = "10.1007/JHEP03(2021)062",
journal = "JHEP",
volume = "03",
pages = "062",
year = "2021"
}

@article{Ashwinkumar:2023zbu,
author = "Ashwinkumar, Meer and Sakamoto, Jun-ichi and Yamazaki, Masahito",
title = "{Dualities and Discretizations of Integrable Quantum Field Theories from 4d Chern-Simons Theory}",
eprint = "2309.14412",
archivePrefix = "arXiv",
primaryClass = "hep-th",
month = "9",
year = "2023"
}

@article{Bassi:2019aaf,
author = "Bassi, Cristian and Lacroix, Sylvain",
title = "{Integrable deformations of coupled $\sigma$-models}",
eprint = "1912.06157",
archivePrefix = "arXiv",
primaryClass = "hep-th",
reportNumber = "ZMP-HH/19-26",
doi = "10.1007/JHEP05(2020)059",
journal = "JHEP",
volume = "05",
pages = "059",
year = "2020"
}

@article{Benini:2020skc,
author = "Benini, Marco and Schenkel, Alexander and Vicedo, Benoit",
title = "{Homotopical Analysis of 4d Chern-Simons Theory and Integrable Field Theories}",
eprint = "2008.01829",
archivePrefix = "arXiv",
primaryClass = "hep-th",
doi = "10.1007/s00220-021-04304-7",
journal = "Commun. Math. Phys.",
volume = "389",
number = "3",
pages = "1417--1443",
year = "2022"
}

@article{Berkovits:2024reg,
author = "Berkovits, Nathan and Pitombo, Rodrigo S.",
title = "{4D Chern-Simons and the pure spinor AdS5\texttimes{}S5 superstring}",
eprint = "2401.03976",
archivePrefix = "arXiv",
primaryClass = "hep-th",
doi = "10.1103/PhysRevD.109.106015",
journal = "Phys. Rev. D",
volume = "109",
number = "10",
pages = "106015",
year = "2024"
}

@article{Borsato:2024alk,
author = "Borsato, Riccardo and Itsios, Georgios and Miramontes, J. Luis and Siampos, Konstantinos",
title = "{Integrability of the \ensuremath{\lambda}-deformation of the PCM with spectators}",
eprint = "2407.20323",
archivePrefix = "arXiv",
primaryClass = "hep-th",
reportNumber = "HU-EP-24/24",
doi = "10.1007/JHEP03(2025)112",
journal = "JHEP",
volume = "03",
pages = "112",
year = "2025"
}

@article{Braaten:1985is,
author = "Braaten, Eric and Curtright, Thomas L. and Zachos, Cosmas K.",
title = "{Torsion and Geometrostasis in Nonlinear Sigma Models}",
reportNumber = "UFTP-85-01, ANL-HEP-PR-85-03",
doi = "10.1016/0550-3213(86)90196-3",
journal = "Nucl. Phys. B",
volume = "260",
pages = "630",
year = "1985",
note = "[Erratum: Nucl.Phys.B 266, 748--748 (1986)]"
}

@article{Cherednik:1981df,
author = "Cherednik, I. V.",
title = "{Relativistically Invariant Quasiclassical Limits of Integrable Two-dimensional Quantum Models}",
doi = "10.1007/BF01086395",
journal = "Theor. Math. Phys.",
volume = "47",
pages = "422--425",
year = "1981"
}

@article{Cole:2023umd,
author = "Cole, Lewis T. and Cullinan, Ryan A. and Hoare, Ben and Liniado, Joaquin and Thompson, Daniel C.",
title = "{Integrable deformations from twistor space}",
eprint = "2311.17551",
archivePrefix = "arXiv",
primaryClass = "hep-th",
doi = "10.21468/SciPostPhys.17.1.008",
journal = "SciPost Phys.",
volume = "17",
number = "1",
pages = "008",
year = "2024"
}

@article{Costello:2013sla,
author = "Costello, Kevin",
editor = "Donagi, Ron and Douglas, Michael R. and Kamenova, Ljudmila and Rocek, Martin",
title = "{Integrable lattice models from four-dimensional field theories}",
eprint = "1308.0370",
archivePrefix = "arXiv",
primaryClass = "hep-th",
doi = "10.1090/pspum/088/01483",
journal = "Proc. Symp. Pure Math.",
volume = "88",
pages = "3--24",
year = "2014"
}

@article{Costello:2013zra,
author = "Costello, Kevin",
title = "{Supersymmetric gauge theory and the Yangian}",
eprint = "1303.2632",
archivePrefix = "arXiv",
primaryClass = "hep-th",
month = "3",
year = "2013"
}

@article{Costello:2017dso,
author = "Costello, Kevin and Witten, Edward and Yamazaki, Masahito",
title = "{Gauge Theory and Integrability, I}",
eprint = "1709.09993",
archivePrefix = "arXiv",
primaryClass = "hep-th",
reportNumber = "IPMU17-0136",
doi = "10.4310/ICCM.2018.v6.n1.a6",
journal = "ICCM Not.",
volume = "06",
number = "1",
pages = "46--119",
year = "2018"
}

@article{Costello:2018gyb,
author = "Costello, Kevin and Witten, Edward and Yamazaki, Masahito",
title = "{Gauge Theory and Integrability, II}",
eprint = "1802.01579",
archivePrefix = "arXiv",
primaryClass = "hep-th",
reportNumber = "IPMU18-0025",
doi = "10.4310/ICCM.2018.v6.n1.a7",
journal = "ICCM Not.",
volume = "06",
number = "1",
pages = "120--146",
year = "2018"
}

@article{Costello:2019tri,
author = "Costello, Kevin and Yamazaki, Masahito",
title = "{Gauge Theory And Integrability, III}",
eprint = "1908.02289",
archivePrefix = "arXiv",
primaryClass = "hep-th",
reportNumber = "IPMU19-0110",
month = "8",
year = "2019"
}

@article{Costello:2020lpi,
author = "Costello, Kevin and Stefa\'nski, Bogdan",
title = "{Chern-Simons Origin of Superstring Integrability}",
eprint = "2005.03064",
archivePrefix = "arXiv",
primaryClass = "hep-th",
doi = "10.1103/PhysRevLett.125.121602",
journal = "Phys. Rev. Lett.",
volume = "125",
number = "12",
pages = "121602",
year = "2020"
}

@article{Delduc:2013fga,
author = "Delduc, Francois and Magro, Marc and Vicedo, Benoit",
title = "{On classical $q$-deformations of integrable sigma-models}",
eprint = "1308.3581",
archivePrefix = "arXiv",
primaryClass = "hep-th",
doi = "10.1007/JHEP11(2013)192",
journal = "JHEP",
volume = "11",
pages = "192",
year = "2013"
}

@article{Delduc:2016ihq,
author = "Delduc, Francois and Lacroix, Sylvain and Magro, Marc and Vicedo, Benoit",
title = "{On q-deformed symmetries as Poisson\textendash{}Lie symmetries and application to Yang\textendash{}Baxter type models}",
eprint = "1606.01712",
archivePrefix = "arXiv",
primaryClass = "hep-th",
doi = "10.1088/1751-8113/49/41/415402",
journal = "J. Phys. A",
volume = "49",
number = "41",
pages = "415402",
year = "2016"
}

@article{Delduc:2017brb,
author = "Delduc, Francois and Kameyama, Takashi and Magro, Marc and Vicedo, Benoit",
title = "{Affine $q$-deformed symmetry and the classical Yang-Baxter $\sigma$-model}",
eprint = "1701.03691",
archivePrefix = "arXiv",
primaryClass = "hep-th",
doi = "10.1007/JHEP03(2017)126",
journal = "JHEP",
volume = "03",
pages = "126",
year = "2017"
}

@article{Delduc:2018hty,
author = "Delduc, F. and Lacroix, S. and Magro, M. and Vicedo, B.",
title = "{Integrable Coupled $\sigma$ Models}",
eprint = "1811.12316",
archivePrefix = "arXiv",
primaryClass = "hep-th",
reportNumber = "ZMP-HH/18-26",
doi = "10.1103/PhysRevLett.122.041601",
journal = "Phys. Rev. Lett.",
volume = "122",
number = "4",
pages = "041601",
year = "2019"
}

@article{Delduc:2019whp,
author = "Delduc, Francois and Lacroix, Sylvain and Magro, Marc and Vicedo, Benoit",
title = "{A unifying 2D action for integrable $\sigma$-models from 4D Chern\textendash{}Simons theory}",
eprint = "1909.13824",
archivePrefix = "arXiv",
primaryClass = "hep-th",
doi = "10.1007/s11005-020-01268-y",
journal = "Lett. Math. Phys.",
volume = "110",
number = "7",
pages = "1645--1687",
year = "2020"
}

@article{Derryberry:2021rne,
author = "Derryberry, Richard",
title = "{Lax formulation for harmonic maps to a moduli of bundles}",
eprint = "2106.09781",
archivePrefix = "arXiv",
primaryClass = "math.AG",
month = "6",
year = "2021"
}

@article{Drinfeld:1985rx,
author = "Drinfeld, V.G.",
title = "{Hopf algebras and the quantum Yang-Baxter equation}",
journal = "Sov. Math. Dokl.",
volume = "32",
pages = "254--258",
year = "1985"
}

@article{Figueroa-OFarrill:1994vwl,
author = "Figueroa-O'Farrill, Jose M. and Stanciu, Sonia",
title = "{Gauged Wess-Zumino terms and equivariant cohomology}",
eprint = "hep-th/9407196",
archivePrefix = "arXiv",
reportNumber = "QMW-PH-94-22",
doi = "10.1016/0370-2693(94)90304-2",
journal = "Phys. Lett. B",
volume = "341",
pages = "153--159",
year = "1994"
}

@article{Friedan:1980jf,
author = "Friedan, D.",
title = "{Nonlinear Models in Two Epsilon Dimensions}",
reportNumber = "LBL-10981",
doi = "10.1103/PhysRevLett.45.1057",
journal = "Phys. Rev. Lett.",
volume = "45",
pages = "1057",
year = "1980"
}

@article{Fukushima:2020dcp,
author = "Fukushima, Osamu and Sakamoto, Jun-ichi and Yoshida, Kentaroh",
title = "{Yang-Baxter deformations of the AdS$_5\times$S$^5$ supercoset sigma model from 4D Chern-Simons theory}",
eprint = "2005.04950",
archivePrefix = "arXiv",
primaryClass = "hep-th",
reportNumber = "KUNS-2817",
doi = "10.1007/JHEP09(2020)100",
journal = "JHEP",
volume = "09",
pages = "100",
year = "2020"
}

@article{Fukushima:2020kta,
author = "Fukushima, Osamu and Sakamoto, Jun-ichi and Yoshida, Kentaroh",
title = "{Comments on $\eta$-deformed principal chiral model from 4D Chern-Simons theory}",
eprint = "2003.07309",
archivePrefix = "arXiv",
primaryClass = "hep-th",
reportNumber = "KUNS-2802",
doi = "10.1016/j.nuclphysb.2020.115080",
journal = "Nucl. Phys. B",
volume = "957",
pages = "115080",
year = "2020"
}

@article{Hoare:2014oua,
author = "Hoare, Ben",
title = "{Towards a two-parameter q-deformation of AdS$_3 \times S^3 \times M^4$ superstrings}",
eprint = "1411.1266",
archivePrefix = "arXiv",
primaryClass = "hep-th",
reportNumber = "HU-EP-14-44",
doi = "10.1016/j.nuclphysb.2014.12.012",
journal = "Nucl. Phys. B",
volume = "891",
pages = "259--295",
year = "2015"
}

@article{Hoare:2014pna,
author = "Hoare, B. and Roiban, R. and Tseytlin, A. A.",
title = "{On deformations of $AdS_n$ x $S^n$ supercosets}",
eprint = "1403.5517",
archivePrefix = "arXiv",
primaryClass = "hep-th",
reportNumber = "IMPERIAL-TP-AT-2014-02, HU-EP-14-10",
doi = "10.1007/JHEP06(2014)002",
journal = "JHEP",
volume = "06",
pages = "002",
year = "2014"
}

@article{Hoare:2015gda,
author = "Hoare, B. and Tseytlin, A. A.",
title = "{On integrable deformations of superstring sigma models related to $AdS_n \times S^n$ supercosets}",
eprint = "1504.07213",
archivePrefix = "arXiv",
primaryClass = "hep-th",
reportNumber = "IMPERIAL-TP-AT-2015-02, HU-EP-15-21",
doi = "10.1016/j.nuclphysb.2015.06.001",
journal = "Nucl. Phys. B",
volume = "897",
pages = "448--478",
year = "2015"
}

@article{Hoare:2018ngg,
author = "Hoare, Ben and Seibold, Fiona K.",
title = "{Supergravity backgrounds of the $\eta$-deformed AdS$_2 \times S^2 \times T^6 $ and AdS$_5 \times S^5$ superstrings}",
eprint = "1811.07841",
archivePrefix = "arXiv",
primaryClass = "hep-th",
doi = "10.1007/JHEP01(2019)125",
journal = "JHEP",
volume = "01",
pages = "125",
year = "2019"
}

@article{Hoare:2020mpv,
author = "Hoare, B. and Lacroix, S.",
title = "{Yang\textendash{}Baxter deformations of the principal chiral model plus Wess\textendash{}Zumino term}",
eprint = "2009.00341",
archivePrefix = "arXiv",
primaryClass = "hep-th",
reportNumber = "ZMP-HH/20-17",
doi = "10.1088/1751-8121/abc43d",
journal = "J. Phys. A",
volume = "53",
number = "50",
pages = "505401",
year = "2020"
}

@article{Hoare:2021dix,
author = "Hoare, Ben",
title = "{Integrable deformations of sigma models}",
eprint = "2109.14284",
archivePrefix = "arXiv",
primaryClass = "hep-th",
doi = "10.1088/1751-8121/ac4a1e",
journal = "J. Phys. A",
volume = "55",
number = "9",
pages = "093001",
year = "2022"
}

@article{Hoare:2022vnw,
author = "Hoare, Ben and Levine, Nat and Seibold, Fiona K.",
title = "{Bi-\ensuremath{\eta} and bi-\ensuremath{\lambda} deformations of \ensuremath{\mathbb{Z}}$_{4}$ permutation supercosets}",
eprint = "2212.08625",
archivePrefix = "arXiv",
primaryClass = "hep-th",
reportNumber = "Imperial-TP-FS-2022-03",
doi = "10.1007/JHEP04(2023)024",
journal = "JHEP",
volume = "04",
pages = "024",
year = "2023"
}

@article{Jimbo:1985ua,
author = "Jimbo, Michio",
title = "{Quantum r Matrix for the Generalized Toda System}",
reportNumber = "RIMS-506",
doi = "10.1007/BF01221646",
journal = "Commun. Math. Phys.",
volume = "102",
pages = "537--547",
year = "1986"
}

@article{Jimbo:1985zk,
author = "Jimbo, Michio",
title = "{A q-Difference Analogue of U(g) and the Yang-Baxter Equation}",
doi = "10.1007/BF00704588",
journal = "Lett. Math. Phys.",
volume = "10",
pages = "63--69",
year = "1985"
}

@article{Kawaguchi:2011pf,
author = "Kawaguchi, Io and Yoshida, Kentaroh",
title = "{Hybrid classical integrability in squashed sigma models}",
eprint = "1107.3662",
archivePrefix = "arXiv",
primaryClass = "hep-th",
reportNumber = "KUNS-2350",
doi = "10.1016/j.physletb.2011.09.117",
journal = "Phys. Lett. B",
volume = "705",
pages = "251--254",
year = "2011"
}

@article{Kawaguchi:2012ve,
author = "Kawaguchi, Io and Matsumoto, Takuya and Yoshida, Kentaroh",
title = "{The classical origin of quantum affine algebra in squashed sigma models}",
eprint = "1201.3058",
archivePrefix = "arXiv",
primaryClass = "hep-th",
reportNumber = "KUNS-2379",
doi = "10.1007/JHEP04(2012)115",
journal = "JHEP",
volume = "04",
pages = "115",
year = "2012"
}

@article{Klimcik:1995dy,
author = "Klim\v{c}\'i{}k, C. and \v{S}evera, P.",
title = "{Poisson-Lie T duality and loop groups of Drinfeld doubles}",
eprint = "hep-th/9512040",
archivePrefix = "arXiv",
reportNumber = "CERN-TH-95-330",
doi = "10.1016/0370-2693(96)00025-1",
journal = "Phys. Lett. B",
volume = "372",
pages = "65--71",
year = "1996"
}

@article{Klimcik:1995jn,
author = "Klim\v{c}\'i{}k, C.",
editor = "Gava, E. and Narain, K. S. and Vafa, C.",
title = "{Poisson-Lie T duality}",
eprint = "hep-th/9509095",
archivePrefix = "arXiv",
reportNumber = "CERN-TH-95-248",
doi = "10.1016/0920-5632(96)00013-8",
journal = "Nucl. Phys. B Proc. Suppl.",
volume = "46",
pages = "116--121",
year = "1996"
}

@article{Klimcik:1995ux,
author = "Klim\v{c}\'i{}k, C. and \v{S}evera, P.",
title = "{Dual nonAbelian duality and the Drinfeld double}",
eprint = "hep-th/9502122",
archivePrefix = "arXiv",
reportNumber = "CERN-TH-95-39, CERN-TH-95-039",
doi = "10.1016/0370-2693(95)00451-P",
journal = "Phys. Lett. B",
volume = "351",
pages = "455--462",
year = "1995"
}

@article{Klimcik:1996nq,
author = "Klim\v{c}\'i{}k, C. and \v{S}evera, P.",
title = "{NonAbelian momentum winding exchange}",
eprint = "hep-th/9605212",
archivePrefix = "arXiv",
reportNumber = "CERN-TH-96-142",
doi = "10.1016/0370-2693(96)00755-1",
journal = "Phys. Lett. B",
volume = "383",
pages = "281--286",
year = "1996"
}

@article{Klimcik:2002zj,
author = "Klim\v{c}\'i{}k, Ctirad",
title = "{Yang-Baxter sigma models and dS/AdS T duality}",
eprint = "hep-th/0210095",
archivePrefix = "arXiv",
reportNumber = "IML-02-XY",
doi = "10.1088/1126-6708/2002/12/051",
journal = "JHEP",
volume = "12",
pages = "051",
year = "2002"
}

@article{Klimcik:2008eq,
author = "Klim\v{c}\'i{}k, Ctirad",
title = "{On integrability of the Yang-Baxter sigma-model}",
eprint = "0802.3518",
archivePrefix = "arXiv",
primaryClass = "hep-th",
doi = "10.1063/1.3116242",
journal = "J. Math. Phys.",
volume = "50",
pages = "043508",
year = "2009"
}

@article{Klimcik:2015gba,
author = "Klim\v{c}\'i{}k, Ctirad",
title = "{\ensuremath{\eta} and \ensuremath{\lambda} deformations as E -models}",
eprint = "1508.05832",
archivePrefix = "arXiv",
primaryClass = "hep-th",
doi = "10.1016/j.nuclphysb.2015.09.011",
journal = "Nucl. Phys. B",
volume = "900",
pages = "259--272",
year = "2015"
}

@article{Klimcik:2019kkf,
author = "Klim\v{c}\'\i{}k, Ctirad",
title = "{Dressing cosets and multi-parametric integrable deformations}",
eprint = "1903.00439",
archivePrefix = "arXiv",
primaryClass = "hep-th",
doi = "10.1007/JHEP07(2019)176",
journal = "JHEP",
volume = "07",
pages = "176",
year = "2019"
}

@article{Lacroix:2017isl,
author = "Lacroix, Sylvain and Magro, Marc and Vicedo, Benoit",
title = "{Local charges in involution and hierarchies in integrable sigma-models}",
eprint = "1703.01951",
archivePrefix = "arXiv",
primaryClass = "hep-th",
doi = "10.1007/JHEP09(2017)117",
journal = "JHEP",
volume = "09",
pages = "117",
year = "2017"
}

@article{Lacroix:2020flf,
author = "Lacroix, Sylvain and Vicedo, Benoit",
title = "{Integrable $\mathcal{E}$-Models, 4d Chern-Simons Theory and Affine Gaudin Models. I.~Lagrangian Aspects}",
eprint = "2011.13809",
archivePrefix = "arXiv",
primaryClass = "hep-th",
reportNumber = "ZMP-HH/20-22",
doi = "10.3842/SIGMA.2021.058",
journal = "SIGMA",
volume = "17",
pages = "058",
year = "2021"
}

@article{Lacroix:2021iit,
author = "Lacroix, Sylvain",
title = "{Four-dimensional Chern\textendash{}Simons theory and integrable field theories}",
eprint = "2109.14278",
archivePrefix = "arXiv",
primaryClass = "hep-th",
doi = "10.1088/1751-8121/ac48ed",
journal = "J. Phys. A",
volume = "55",
number = "8",
pages = "083001",
year = "2022"
}

@article{Lacroix:2023gig,
author = "Lacroix, Sylvain",
title = "{Lectures on classical Affine Gaudin models}",
eprint = "2312.13849",
archivePrefix = "arXiv",
primaryClass = "hep-th",
month = "12",
year = "2023"
}

@article{Lacroix:2023qlz,
author = "Lacroix, Sylvain and Wallberg, Anders",
title = "{An elliptic integrable deformation of the Principal Chiral Model}",
eprint = "2311.09301",
archivePrefix = "arXiv",
primaryClass = "hep-th",
reportNumber = "CERN-TH-2023-205",
doi = "10.1007/JHEP05(2024)006",
journal = "JHEP",
volume = "05",
pages = "006",
year = "2024"
}

@article{Lacroix:2024wrd,
author = "Lacroix, Sylvain and Wallberg, Anders",
title = "{Geometry of the spectral parameter and renormalisation of integrable sigma-models}",
eprint = "2401.13741",
archivePrefix = "arXiv",
primaryClass = "hep-th",
reportNumber = "CERN-TH-2024-008",
doi = "10.1007/JHEP05(2024)108",
journal = "JHEP",
volume = "05",
pages = "108",
year = "2024"
}

@article{Liniado:2023uoo,
author = "Liniado, Joaquin and Vicedo, Benoit",
title = "{Integrable Degenerate $\mathcal {E}$-Models from 4d Chern\textendash{}Simons Theory}",
eprint = "2301.09583",
archivePrefix = "arXiv",
primaryClass = "hep-th",
doi = "10.1007/s00023-023-01317-x",
journal = "Annales Henri Poincare",
volume = "24",
number = "10",
pages = "3421--3459",
year = "2023"
}

@article{Maillet:1985ek,
author = "Maillet, Jean Michel",
title = "{New Integrable Canonical Structures in Two-dimensional Models}",
reportNumber = "PAR/LPTHE-85-32",
doi = "10.1016/0550-3213(86)90365-2",
journal = "Nucl. Phys. B",
volume = "269",
pages = "54--76",
year = "1986"
}

@article{Maillet:1985fn,
author = "Maillet, Jean Michel",
title = "{Kac-moody Algebra and Extended {Yang-Baxter} Relations in the O($N$) Nonlinear $\sigma$ Model}",
reportNumber = "PAR LPTHE 85-22",
doi = "10.1016/0370-2693(85)91075-5",
journal = "Phys. Lett. B",
volume = "162",
pages = "137--142",
year = "1985"
}

@article{Lukyanov:2012zt,
author = "Lukyanov, Sergei L.",
title = "{The integrable harmonic map problem versus Ricci flow}",
eprint = "1205.3201",
archivePrefix = "arXiv",
primaryClass = "hep-th",
reportNumber = "RUNHETC-2012-10",
doi = "10.1016/j.nuclphysb.2012.08.002",
journal = "Nucl. Phys. B",
volume = "865",
pages = "308--329",
year = "2012"
}

@article{Ogievetsky:1984pv,
author = "Ogievetsky, E. and Reshetikhin, N. and Wiegmann, P.",
title = "{The principal chiral field in two-dimension and classical Lie algebra}",
reportNumber = "NORDITA-84/38",
month = "10",
year = "1984"
}

@article{Ogievetsky:1987vv,
author = "Ogievetsky, E. and Reshetikhin, N. and Wiegmann, P.",
title = "{The principal chiral field in two dimensions on classical lie algebras: The Bethe-ansatz solution and factorized theory of scattering}",
doi = "10.1016/0550-3213(87)90138-6",
journal = "Nucl. Phys. B",
volume = "280",
pages = "45--96",
year = "1987"
}

@article{Seibold:2019dvf,
author = "Seibold, Fiona K.",
title = "{Two-parameter integrable deformations of the $AdS_3 \times S^3 \times T^4$ superstring}",
eprint = "1907.05430",
archivePrefix = "arXiv",
primaryClass = "hep-th",
doi = "10.1007/JHEP10(2019)049",
journal = "JHEP",
volume = "10",
pages = "049",
year = "2019"
}

@article{Sfetsos:2013wia,
author = "Sfetsos, Konstadinos",
title = "{Integrable interpolations: From exact CFTs to non-Abelian T-duals}",
eprint = "1312.4560",
archivePrefix = "arXiv",
primaryClass = "hep-th",
reportNumber = "DMUS-MP-13-23, DMUS--MP--13-23",
doi = "10.1016/j.nuclphysb.2014.01.004",
journal = "Nucl. Phys. B",
volume = "880",
pages = "225--246",
year = "2014"
}

@article{Sfetsos:2014cea,
author = "Sfetsos, Konstantinos and Thompson, Daniel C.",
title = "{Spacetimes for $\lambda$-deformations}",
eprint = "1410.1886",
archivePrefix = "arXiv",
primaryClass = "hep-th",
doi = "10.1007/JHEP12(2014)164",
journal = "JHEP",
volume = "12",
pages = "164",
year = "2014"
}

@article{Sfetsos:2015nya,
author = "Sfetsos, Konstantinos and Siampos, Konstantinos and Thompson, Daniel C.",
title = "{Generalised integrable \ensuremath{\lambda} - and \ensuremath{\eta}-deformations and their relation}",
eprint = "1506.05784",
archivePrefix = "arXiv",
primaryClass = "hep-th",
doi = "10.1016/j.nuclphysb.2015.08.015",
journal = "Nucl. Phys. B",
volume = "899",
pages = "489--512",
year = "2015"
}

@article{Sfetsos:2017sep,
author = "Sfetsos, Konstantinos and Siampos, Konstantinos",
title = "{Integrable deformations of the $G_{k_1} \times G_{k_2}/G_{k_1+k_2}$ coset CFTs}",
eprint = "1710.02515",
archivePrefix = "arXiv",
primaryClass = "hep-th",
reportNumber = "CERN-TH-2017-199",
doi = "10.1016/j.nuclphysb.2017.12.011",
journal = "Nucl. Phys. B",
volume = "927",
pages = "124--139",
year = "2018"
}

@article{Tian:2020pub,
author = "Tian, Jia and He, Yi-Jun and Chen, Bin",
title = "{$\lambda$-Deformed AdS$_5 \times$ S$^5$ superstring from 4D Chern-Simons theory}",
eprint = "2007.00422",
archivePrefix = "arXiv",
primaryClass = "hep-th",
doi = "10.1016/j.nuclphysb.2021.115545",
journal = "Nucl. Phys. B",
volume = "972",
pages = "115545",
year = "2021"
}

@article{Tian:2020ryu,
author = "Tian, Jia",
title = "{Comments on $\lambda$--deformed models from 4D Chern-Simons theory}",
eprint = "2005.14554",
archivePrefix = "arXiv",
primaryClass = "hep-th",
month = "5",
year = "2020"
}

@article{Vicedo:2015pna,
author = "Vicedo, Benoit",
title = "{Deformed integrable \ensuremath{\sigma}-models, classical R-matrices and classical exchange algebra on Drinfel\textquoteright{}d doubles}",
eprint = "1504.06303",
archivePrefix = "arXiv",
primaryClass = "hep-th",
doi = "10.1088/1751-8113/48/35/355203",
journal = "J. Phys. A",
volume = "48",
number = "35",
pages = "355203",
year = "2015"
}

@article{Vicedo:2017cge,
author = "Vicedo, Benoit",
title = "{On integrable field theories as dihedral affine Gaudin models}",
eprint = "1701.04856",
archivePrefix = "arXiv",
primaryClass = "hep-th",
doi = "10.1093/imrn/rny128",
journal = "Int. Math. Res. Not.",
volume = "2020",
number = "15",
pages = "4513--4601",
year = "2020"
}

@article{Vicedo:2019dej,
author = "Vicedo, Benoit",
title = "{4D Chern\textendash{}Simons theory and affine Gaudin models}",
eprint = "1908.07511",
archivePrefix = "arXiv",
primaryClass = "hep-th",
doi = "10.1007/s11005-021-01354-9",
journal = "Lett. Math. Phys.",
volume = "111",
number = "1",
pages = "24",
year = "2021"
}

@article{Wiegmann:1984ec,
author = "Wiegmann, P.",
title = "{Exact factorized S matrix of the chiral field in two-dimensions}",
doi = "10.1016/0370-2693(84)91256-5",
journal = "Phys. Lett. B",
volume = "142",
pages = "173--176",
year = "1984"
}

@article{Witten:1983ar,
author = "Witten, Edward",
editor = "Stone, M.",
title = "{Nonabelian Bosonization in Two-Dimensions}",
reportNumber = "PRINT-83-0934 (PRINCETON)",
doi = "10.1007/BF01215276",
journal = "Commun. Math. Phys.",
volume = "92",
pages = "455--472",
year = "1984"
}

@article{Witten:1991mm,
author = "Witten, Edward",
title = "{On Holomorphic factorization of WZW and coset models}",
reportNumber = "IASSNS-HEP-91-25",
doi = "10.1007/BF02099196",
journal = "Commun. Math. Phys.",
volume = "144",
pages = "189--212",
year = "1992"
}

@book{Serre2000,
author = {Jean-Pierre Serre},
title = {Complex Semisimple Lie Algebras},
publisher = {Springer},
year = {1987},
series = {Springer Monographs in Mathematics},
address = {New York},
isbn = {978-0-387-96569-7}
}

@phdthesis{Lacroix:2018njs,
author = "Lacroix, Sylvain",
title = "{Integrable models with twist function and affine Gaudin models}",
eprint = "1809.06811",
archivePrefix = "arXiv",
primaryClass = "hep-th",
reportNumber = "tel-01900498, 2018LYSEN014",
school = "Lyon, Ecole Normale Superieure",
year = "2018"
}

@article{Appadu:2017bnv,
author = "Appadu, Calan and Hollowood, Timothy J. and Price, Dafydd and Thompson, Daniel C.",
title = "{Yang Baxter and Anisotropic Sigma and Lambda Models, Cyclic RG and Exact S-Matrices}",
eprint = "1706.05322",
archivePrefix = "arXiv",
primaryClass = "hep-th",
doi = "10.1007/JHEP09(2017)035",
journal = "JHEP",
volume = "09",
pages = "035",
year = "2017"
}

@article{Caudrelier:2022jco,
author = "Caudrelier, Vincent and Stoppato, Matteo and Vicedo, Benoit",
title = "{Classical Yang\textendash{}Baxter Equation, Lagrangian Multiforms and Ultralocal Integrable Hierarchies}",
eprint = "2201.08286",
archivePrefix = "arXiv",
primaryClass = "nlin.SI",
doi = "10.1007/s00220-023-04871-x",
journal = "Commun. Math. Phys.",
volume = "405",
number = "1",
pages = "12",
year = "2024"
}

@article{Cole:2024skp,
author = "Cole, Lewis T. and Weck, Peter",
title = "{Integrability in gravity from Chern-Simons theory}",
eprint = "2407.08782",
archivePrefix = "arXiv",
primaryClass = "hep-th",
doi = "10.1007/JHEP10(2024)080",
journal = "JHEP",
volume = "10",
pages = "080",
year = "2024"
}

@article{Reshetikhin:1990ep,
author = "Reshetikhin, N.",
title = "{Multiparameter quantum groups and twisted quasitriangular Hopf algebras}",
doi = "10.1007/BF00626530",
journal = "Lett. Math. Phys.",
volume = "20",
pages = "331--335",
year = "1990"
}

@article{Osten:2016dvf,
author = "Osten, David and van Tongeren, Stijn J.",
title = "{Abelian Yang\textendash{}Baxter deformations and TsT transformations}",
eprint = "1608.08504",
archivePrefix = "arXiv",
primaryClass = "hep-th",
reportNumber = "HU-EP-16-29",
doi = "10.1016/j.nuclphysb.2016.12.007",
journal = "Nucl. Phys. B",
volume = "915",
pages = "184--205",
year = "2017"
}

\end{bibtex}

\bibliographystyle{nb}
\bibliography{\jobname}

\end{document}